\documentclass[fleqn,usenatbib]{mnras}

\usepackage{newtxtext,newtxmath}
\usepackage[T1]{fontenc}

\DeclareRobustCommand{\VAN}[3]{#2}
\let\VANthebibliography\thebibliography
\def\thebibliography{\DeclareRobustCommand{\VAN}[3]{##3}\VANthebibliography}

\usepackage{graphicx}	% Including figure files
\usepackage{amsmath}	% Advanced maths commands

\usepackage{subcaption}
\usepackage{float}
\usepackage{lscape}
\usepackage{booktabs}
\usepackage{multirow}
\usepackage{array} % required for text wrapping in tables
\usepackage{tablefootnote}
\usepackage{xcolor}
\DeclareMathOperator{\arctantwo}{arctan2}
\title[Magnetic fields from GMCs to IRDCs]{Characterising magnetic fields at the onset of star cluster formation:\\
From giant molecular clouds to infrared dark clumps}

\author[R. Ramkumar et al.]{
Ria Ramkumar,$^{1}$\thanks{E-mail: RamkumarR@cardiff.ac.uk}
Nicolas Peretto$^{1}$, Gary A. Fuller$^{2,3}$, Patrick M. Koch$^{4}$, Ya-Wen Tang$^{4}$
\\
$^{1}$Cardiff Hub for Astrophysics Research \& Technology, School of Physics \& Astronomy, Cardiff University, Queen's Buildings, The Parade, Cardiff CF24 3AA, UK\\
$^{2}$Jodrell Bank Centre for Astrophysics, School of Physics and Astronomy, University of Manchester, Oxford Road, Manchester, M13 9PL, UK\\
$^{3}$Physikalisches Institut, University of Cologne, Zülpicher Str. 77, D-50937 Köln, Germany\\
$^{4}$Academia Sinica Institute of Astronomy and Astrophysics, No. 1, Sec. 4, Roosevelt Road, Taipei 10617, Taiwan\\
}

\date{Accepted 2026 June 12. Received 2026 June 4; in original form 2026 February 13}

\pubyear{2026}

\begin{document}
\label{firstpage}
\pagerange{\pageref{firstpage}--\pageref{lastpage}}
\maketitle

% Abstract of the paper
\begin{abstract}
The role of magnetic fields in the observed inefficiency of star formation in Galactic molecular clouds is a widely debated topic, with the past decade seeing an explosion of observational characterisation of magnetic fields in star-forming regions. However, few have studied the spatial evolution of magnetic fields from entire molecular clouds down to parsec-size cluster-forming clumps. In this work, the plane-of-sky morphology of the magnetic fields of eight infrared dark clumps and their parent molecular clouds are derived from {\it Planck} and JCMT POL-2 polarisation data (including some from the BISTRO survey). We also use this data to test multiple methods of calculating $B$-field strengths. Our study shows that the morphologies of magnetic fields in clumps and their parent molecular clouds systematically, and significantly, differ, supported by a line-of-sight correction of the cloud-scale magnetic fields using the velocity gradient technique. We find a strong correlation between gas velocity dispersion and the alignment of magnetic field lines with column density gradients from scales of tens of parsecs to a few parsecs. This correlation is clear evidence of a link between the kinematic properties of the gas and the dynamical importance of magnetic fields. Conversely, the higher magnetic field strengths we measure on cloud scale compared to clump scale contradict magnetic flux conservation and thus highlight the unreliability of such measurements. Altogether, our analysis supports a picture in which magnetic fields have little impact on the dynamical evolution of cluster-forming clumps but do play a role in providing support on larger scales.
\end{abstract}

% Select between one and six entries from the list of approved keywords.
% Don't make up new ones.

%%%%%%%%%%%%%%%%%%%%%%%%%%%%%%%%%%%%%%%%%%%%%%%%%%

%%%%%%%%%%%%%%%%% BODY OF PAPER %%%%%%%%%%%%%%%%%%

\section{Introduction}

Uniform density free-fall models predict a rate of star formation that is two orders of magnitude higher than the current $\sim2$ M$_\odot$/yr observed in the Milky Way \citep{1974ARA&A..12..279Z, 2010ApJ...710L..11R, Elia_2022}. This discrepancy has led to the development of a number of models explaining this inefficiency by the presence of strong magnetic fields \citep[e.g.][]{2015MNRAS.451.3340K, 10.3389/fspas.2019.00007}, interstellar turbulence \citep[e.g.][]{1974ApJ...192L.149Z, 2005ApJ...630..250K} or internal stellar feedback \citep[e.g.][]{2014prpl.conf..243K, 2017ApJ...845..133S, 2018MNRAS.480..800H} within molecular clouds. All of these ingredients contribute to moderating the rate at which, or what fraction of, a molecular cloud collapses. 

Magnetic fields - also referred to as $B$-fields - are both theoretically complex and observationally challenging to detect. As a result, their importance in molecular cloud evolution is not yet fully understood. Magnetic fields dynamically manifest themselves through the Lorentz force, which exclusively acts in a perpendicular direction to the fields and therefore adds anisotropy to the system. In a {\it static} picture with strong $B$-fields, the Lorentz force is able to stop any gravity-driven gas contraction perpendicular to the field, but would allow significant contraction along the field lines, even though at a reduced rate \citep{2008MNRAS.385.1820P, 2021MNRAS.507.5641G}. In a more realistic and {\it dynamic} picture, interstellar turbulence would perturb the field lines, generating transverse magnetic waves (also known as Alfv\'en waves) as a consequence of the magnetic tension of the lines. Those waves would then propagate along the field lines, creating an extra magnetic pressure term that prevents the gas from collapsing along the field lines as well \citep{2003ApJ...597..970F}. One implicit assumption that has been made so far is that the gas is sufficiently ionised so that all particles feel (directly or via collisions) the impact of the $B$-fields. In the molecular gas, ionisation fractions can vary from being on the order of 10$^{-4}$ in the diffuse outer regions of a molecular cloud to 10$^{-9}$ in the dense, shielded regions \citep{1979ApJ...232..729E, 1999ApJ...512..724B, 2009A&A...498..771G, 2025A&A...702A.205B}; molecular clouds are thus largely made up of weakly ionised plasma that departs from the expectations of ideal magnetohydrodynamics (MHD) and perfect flux-freezing of magnetic field lines. Ambipolar diffusion is the foremost non-ideal MHD effect that could explain how star formation may occur in the presence of a dominant magnetic field \citep{1956MNRAS.116..503M, 1991ApJ...371..296M, Desch_2001, Kunz_2009}. In self-gravitating clouds with low ionisation fractions, neutrals and ions feel the pull of gravity towards the clouds' centre of mass. However, due to the Lorentz force ions remain frozen to the magnetic field lines while neutrals drift through them. This differential motion between ions and neutrals creates friction between these two fluids, considerably slowing down the collapse of the neutrals. In star-forming regions, this then means that magnetic field lines that would have been pulled in by gravity in the ideal MHD case are not dragged in to the extent that would have taken place with perfect flux-freezing, causing a redistribution of magnetic flux. This reduces the amount of magnetic flux in the dense region, eventually allowing it to freely collapse. For clouds where the magnetic energy is much larger than its gravitational energy, the ambipolar diffusion timescale $\tau_{ad}$ - i.e. the time it takes for neutrals to drift through a cloud/clump/core - is typically 10 times slower than the free-fall time, implying $\tau_{ad}$ values of the order of several tens of Myr in molecular clouds  \citep{2004ApJ...616..283T, 2019FrASS...6....5H}. Such long timescales could explain the low star formation efficiency of molecular clouds \citep[e.g.][]{2008ApJ...687..354N}. However, the collapse time of clouds strongly depends on their initial ratio of gravitational-to-magnetic energy, or equivalently cloud mass to magnetic flux ratio \citep[e.g.][]{1956MNRAS.116..503M, 1995ApJ...453..271B}, and it is possible that this initial value is one of energy equipartition with gravity, favouring shorter collapse timescales \citep[e.g.][]{2007ApJ...668.1064E}.

Only in the past decade or so, sensitive dust polarisation measurements of star-forming regions have become routinely available, leading to a wealth of magnetic field studies \citep[e.g.][]{2016ApJ...824..132N, 2016SPIE.9914E..03F, 2018JAI.....740008H, 2019FrASS...6...15P, 2020pase.conf..117B}; these are now the primary method of deducing the plane-of-the-sky direction of the magnetic field in molecular clouds. Measurements of the polarisation of light from dust grains allow for the inference of magnetic field direction since non-spherical dust grains tend to spin on their short axes, which align approximately parallel to the direction of the magnetic field. The most accepted theory providing an explanation as to why this occurs is the Radiative Torques - RATs - Alignment Theory \citep{1996ApJ...470..551D, 1997ApJ...480..633D, 2003ApJ...589..289W, 2007MNRAS.378..910L, 2015ARA&A..53..501A}, which explains the alignment of the dust grains in this way as due to the interaction between the paramagnetic dust grains and a non-isotropic radiation field. Thermal (sub-)millimetre emission of spinning dust grains in molecular clouds is polarised; the direction of polarisation is parallel to the long axes of the dust grains, and therefore is perpendicular to the direction of the magnetic field.

Dust polarisation measurements have been extensively reported in the recent literature on star formation research, the main goal often being to quantify the relative importance of gravity, turbulence, and magnetic fields in star-forming regions. Recent observational studies seem to show that the diffuse regions of molecular clouds are magnetically subcritical (supported against collapse by magnetic fields) and sub-to-trans-Alfv\'enic (magnetic field dominates or is in equipartition with turbulence). As density increases, these potentially transition towards magnetically supercritical, trans-to-super-Alfv\'enic (turbulence being equivalent to or dominating the magnetic field) structures \citep{2019FrASS...6....5H}, although there is much uncertainty regarding the energy balance in the denser regions of molecular clouds. 

Most polarisation studies concentrate on the denser structures within molecular clouds that are on the scale of a few parsecs (i.e. clump scale) down to $\sim0.1$~pc or lower (i.e. core scale). In these regions, using variations of the Davis-Chandrasekhar-Fermi (DCF) method \citep{1951PhRv...81..890D,1953ApJ...118..113C},  it is often found that gravity dominates over the magnetic field and turbulence \citep[e.g.][]{2020ApJ...905..158W, 2024MNRAS.528.1460R}, however there have also been findings that gravity is in approximate equipartition with the kinetic and magnetic field energies \citep[e.g.][]{2024ApJ...966..120L, 2024ApJ...976..249G}, and even in some cases that magnetic field dominates over gravity \citep[e.g.][]{2025A&A...696A.163L}. Some studies that split these regions into separate sections find that stability varies across the structure \citep[e.g.][]{2018ApJ...859..151L, 2019ApJ...878...10T, 2024ApJ...962..136W}. These mixed observational results display the challenges in drawing a consistent, observationally-driven picture of the role of magnetic fields in the dynamical evolution of clumps and their sub-structures. While the diversity of situations could be genuine and linked to different initial conditions or evolutionary stages, the large uncertainties linked to the DCF method and the observationally-derived parameters used to derive magnetic field strengths (i.e. gas volume densities, turbulent velocity dispersion and magnetic field direction dispersion) may also be responsible for the emergence of this noisy picture. Methods that focus only on the analysis of the observed morphology of the magnetic field are likely to be more robust \citep[e.g.][]{2012ApJ...747...79K, 2013ApJ...774..128S}

There have been suggestions of a critical density at which there is a transition between the magnetically dominated regime of a molecular cloud to the gravity/turbulence dominated regime. \citet{2010ApJ...725..466C} notes this transition occurring at a hydrogen number density $n_{H}$ of 300~cm$^{-3}$, based on Zeeman measurements of the line-of-sight magnetic field strength. They observe that while the maximum magnetic field strength up to this density is relatively constant at 10~$\mu$G, above this density the magnetic field strength scales with $n_{H}^{2/3}$, marking a transition towards gravity dominating over the magnetic field. However, \citet{2015MNRAS.451.4384T} reanalyse this data and find $n_{H}^{1/2}$. The index of 2/3 is what would be expected for isotropic collapse with a weak magnetic field and flux-freezing, while an index of 1/2 is obtained for anisotropic collapse with a strong magnetic field model and flux-freezing \citep{1966MNRAS.133..265M}. Taking into account ambipolar diffusion, the index is reduced even further \citep[e.g.][]{1993ApJ...415..680F, 2010MNRAS.408..322K}. Indirect evidence of the existence of such a transition has been supported by a change of the relative orientations of magnetic fields and density gradients at a H$_2$ column density of $\sim0.5-1\times10^{22}$cm$^{-2}$ \citep[e.g.][]{2016A&A...586A.135P, 2019ApJ...878..110F, 2019A&A...629A..96S}. This change can be understood in the context of a switch in the internal gas dynamics of a cloud where departure from orthogonality between density gradients and magnetic fields is the consequence of a switch between random flows below the threshold and converging flows (e.g. collapse) above the threshold \citep{2017A&A...607A...2S}. Those relative orientation studies, however, have mostly focussed on a few nearby clouds ($d<1$~kpc) for which the spatial resolution provided by instruments like those on board of {\it Planck} was good enough to resolve their  magnetic field morphology. In the past decade, though, new instruments such as the POL-2 sub-millimetre polarimeter on the James Clerk Maxwell Telescope \citep[JCMT - e.g][]{2025ApJ...983..184Y}, the High-resolution Airborne Wideband Camera Plus (HAWC+) far-infrared camera and polarimeter on the Stratospheric Observatory for Infrared Astronomy \citep[SOFIA - e.g][]{2023AAS...24130802C, 2025ApJ...988..252Z} and the polarimeter on the Atacama Large Millimeter/submillimeter Array \citep[ALMA - e.g][]{2019A&A...630A..54B, 2020MNRAS.496.2790L} have allowed sub-pc resolution observations of magnetic fields up to heliocentric distances of $\sim10$~kpc. 

In this paper, we present a study of the magnetic field properties estimated towards a sample of infrared dark clumps (IRDCs) and their parent molecular clouds. Infrared dark clumps are  over-densities within molecular clouds on the scale of parsecs across, and are thought to represent the earliest stages of star cluster formation \citep[e.g.][]{2006ApJ...641..389R, 2009A&A...505..405P}. They have typical averaged H$_2$ number densities of $10^{3}-10^{4}$ cm$^{-3}$ and typical masses in the range  of $10^2$ to $10^4 {\rm {M}_\odot}$. On the other hand, molecular clouds have lower densities in the range of $10^{2}-10^{3}$ cm$^{-3}$, and larger masses in the range  of $10^4$ to $10^6 {\rm {M}_\odot}$ for sizes in the range of 10 to 100 pc. \citet{2023MNRAS.525.2935P} investigated the radial profiles of the velocity dispersions and virial ratios of a sample of 27 IRDCs ($3<d<5$ kpc) and their parent molecular clouds. In that study, the authors found that the velocity dispersion radial profile of the clumps were systematically flat ($\sigma \propto r^0$) and would only steepen ($\sigma \propto r^{0.5}$), for some of the clouds,  on larger scales. This is in contrast with a universal velocity dispersion - size relationship as proposed by \citet{1981MNRAS.194..809L} and \cite{1987ApJ...319..730S}. It is the change of the velocity dispersion profile's slope from clump to cloud that we refer to as {\it clump dynamical decoupling}. While no magnetic field measurements were made, this dynamical decoupling is reminiscent of the (column) density threshold for magnetic field realignment mentioned above. The research presented in this paper aims to further the \citet{2023MNRAS.525.2935P} study by the inclusion of magnetic field measurements in 8 out of the 27 IRDCs. Following their nomenclature, we refer to the IRDCs as {\it clumps}, and their parent molecular clouds as {\it clouds}.

The paper is organised as follows: in Section \ref{sec:sources+data} the sample and observations are presented. Section \ref{sec:magfielddirection} describes the morphology of the magnetic field in the clumps and clouds in our sample, and in Section \ref{sec:confusion} we address the issue of how the magnetic field morphology in our clouds may be affected by contaminating structures along the line-of-sight. The results of an investigation into the relative alignment between density structures and the magnetic field are shown in Section \ref{sec:hro}. We attempt to constrain the magnetic field strength within our clouds and clumps employing multiple methods which are discussed in Section \ref{sec:Bfieldstrength}, and then use the results from one of these methods to quantify the relative importance of gravity, turbulence, and magnetic fields in our sources in Section \ref{sec:energybalance}. The implications of our results regarding the importance of the magnetic field in star formation are discussed in Section \ref{sec:discussion}. Section \ref{sec:summary} summarises the results of our investigation.

\section{Sample and observations}
\label{sec:sources+data}

\subsection{Sample}

In this study we aim to characterise the magnetic field properties of eight infrared dark clumps and their parent molecular clouds: SDC18.624, SDC24.489, SDC25.166, SDC28.333, SDC34.370, SDC35.527, SDC35.745, and SDC40.283. These eight clump/cloud pairs are taken from the sample of \citet{2023MNRAS.525.2935P}, in which mass, velocity dispersion, and virial ratio radial profiles have been derived for these sources using a combination of {\it Herschel} H$_2$ column density maps, GRS $^{13}$CO(1-0) data, and IRAM 30m N$_{2}$H$^{+}$(1-0) data. These clouds lie close to the Galactic plane and at heliocentric distances within the range 3-5 kpc. Table \ref{tab:SamplePropertiesTable} presents the properties of the clump/cloud pairs measured at two different scales \citep[for more details, see][]{2023MNRAS.525.2935P}.

\begin{table}
\caption{Properties of the IRDCs and their parent clouds, from \citet{2023MNRAS.525.2935P}. $M$ is the mass, $R$ is the effective radius, and $\alpha_{\rm vir}=\frac{2E_{\rm K}}{|E_{\rm G}|}$ is the virial parameter, where $E_{\rm K}$ is the kinetic energy and $E_{\rm G}$ is the gravitational energy.}
\label{tab:SamplePropertiesTable}
\resizebox{\columnwidth}{!}{%
\begin{tabular}{@{}ccclccl@{}}
\toprule
Source    & $M_{\rm clump}$    & $R_{\rm clump}$ & $\alpha_{\rm vir}^{\rm clump}$ & $M_{\rm cloud}$               & $R_{\rm cloud}$ & $\alpha_{\rm vir}^{\rm cloud}$ \\
          & ${\rm ({M}_\odot)}$  & ${\rm (pc)}$      &                           & ${\rm (\times10^{4}{M}_\odot)}$ & ${\rm (pc)}$      &                           \\ \midrule
SDC18.624 & 3239$^{+365}_{-546}$    & 1.66                 &              0.62             & 14.52                              & 12.62                &              3.35             \\
SDC24.489 & 831$^{+167}_{-167}$     & 1.12                 &                2.08           & 2.18                               & 9.19                 &              1.39             \\
SDC25.166 & 3622$^{+374}_{-374}$    & 1.68                 &              0.53             & 3.23                               & 8.65                 &               1.00            \\
SDC28.333 & 13954$^{+1863}_{-1863}$ & 2.65                 &             0.46              & 103.70                             & 30.85                &             0.97              \\
SDC34.370 & 12048$^{+1223}_{-1528}$ & 2.15                 &               0.39            & 18.13                              & 11.33                &             1.05              \\
SDC35.527 & 1499$^{+245}_{-367}$    & 1.36                 &               0.68            & 3.49                               & 8.22                 &             0.99              \\
SDC35.745 & 3490$^{+536}_{-268}$    & 2.01                 &                0.80           & 7.66                               & 11.58                &             2.47              \\
SDC40.283 & 4044$^{+432}_{-432}$    & 1.81                 &              1.15             & 5.00                               & 11.06                &              0.46             \\ \bottomrule
\end{tabular}%
}
\end{table}

\subsection{Observations}

\subsubsection{Dust Polarisation Observations}
\label{sec:dustpolobs}

This work makes use of JCMT POL-2 and {\it Planck} dust polarisation data to probe the direction of the magnetic field at both clump and cloud scales respectively. These telescopes measure Stokes parameters $I$, $Q$, and $U$, where $I$ is the total intensity, $Q$ is the horizontal and vertical polarisation, and $U$ is the diagonal polarisation. Since both of these telescopes only measure linear polarisation, they do not measure Stokes' parameter $V$ (circular polarisation). These quantities allow for the inference of magnetic field pseudovectors from the polarisation pseudovectors - termed so since there is an ambiguity of 180$^\circ$ in the angles, e.g. a pseudovector with angle 20$^\circ$ is equivalent to a pseudovector with angle 200$^\circ$.

\paragraph{JCMT POL-2}

JCMT is a ground-based 15m-diameter telescope. POL-2 (the linear polarimeter) on the JCMT works in conjunction with SCUBA-2 (the detector) to obtain dust polarisation measurements. It operates simultaneously at 450$\mu$m (resolution of $9.8''$) and 850$\mu$m (resolution of $14.6''$). In this work we use the 850$\mu$m data, which has a higher signal-to-noise ratio (SNR). The scanning pattern used for the observations is a Daisy mode scan, and the pixel size of the images is $6''$. Observations of SDC24.489, SDC25.166, SDC34.370, and SDC40.283 were carried out from 2019-08-15 to 2019-08-29 (PID: M19BP058, PI: Nicolas Peretto). SDC18.624 was observed on 2017-07-08 (PID: M17AP019, PI: Patrick Koch), SDC28.333 was observed on 2020-09-25 (PID: M20AL018, PI: Derek Ward-Thompson), SDC35.527 was observed on 2017-08-31 (PID: M17BP050, PI: Tie Liu), and SDC35.745 was observed on 2020-02-16 (PID: M20AP040, PI: Nicolas Peretto). The data for SDC18.624 \citep{2025A&A...696A.163L}, SDC28.333 \citep{2025ApJ...985..222H} - as part of the BISTRO programme \citep{2017ApJ...842...66W} - and SDC35.527 \citep{2018ApJ...859..151L, 2018A&A...620A..26J} have already been published. The POL-2 data for the remaining 5 clumps are published for the first time here.

The JCMT POL-2 data has been reduced using the standard POL-2 data reduction procedure\footnote{https://starlink.eao.hawaii.edu/docs/sc22.htx/sc22.html}. The flux conversion factor is 668.25 Jy beam$^{-1}$ pW$^{-1}$ for all clumps except SDC18.624 and SDC35.527, for which it is 696.6 Jy beam$^{-1}$ pW$^{-1}$.

\paragraph{{\it Planck}} 

{\it Planck} was a space-based telescope positioned at the Earth-Sun L2 point, with the primary aim of mapping the Cosmological Microwave Background \citep{2003NewAR..47.1017L}. In the process, it also measured polarisation data. {\it Planck} 353GHz polarisation data - obtained from the {\it Planck} Legacy Archive\footnote{https://pla.esac.esa.int/} - is used here to calculate the directions of the magnetic field pseudovectors in the molecular clouds being investigated. This data has an angular resolution of $5'$, and the pixel size of the images is $1.5'$. 

During data processing, the {\it Planck} polarisation data had to be corrected for bandpass mismatch (BPM) leakage in order to account for instrumental differences between the detectors. This mismatch led to `fake' polarisation signals even from unpolarised regions \citep{2014A&A...571A...9P, 2015A&A...576A.104P}. The {\it Planck} Collaboration tested two methods to correct for BPM: the first using the ground-measured bandpasses, and the second comparing the response of each detector in regions of the sky where the output of each detector could be measured individually. The second method is the one which is used, since this method results in alignment of the magnetic field pseudovectors preferentially more parallel to the Galactic plane. Since this bandpass mismatch took the form of leakage from the Stokes $I$ into $Q$ and $U$, this issue was more prevalent in the Galactic plane, where the larger $I$ led to a greater leakage into the $Q$ and $U$ parameters. This means that the polarisation measurements in the Galactic plane suffer from larger uncertainties \citep[see][for details on how the bandpass mismatch correction was carried out]{2015A&A...576A.104P}. From \citet{2015A&A...576A.104P}, the uncertainty on the polarisation angles are $<10^\circ$ towards the inner Galactic plane. The assumption made here that the magnetic field is statistically expected to be aligned with the Galactic plane could potentially bias the {\it Planck} measurements.

\subsubsection{Molecular line data}

The $^{13}$CO(1-0) and N$_{2}$H$^{+}$(1-0) data from \citet{2023MNRAS.525.2935P} were used in this work to define the boundaries of the cloud and clump respectively. The $^{13}$CO(1-0) emission line typically traces gas densities of a few 100~cm$^{-3}$ and above, and is therefore a good tracer of molecular clouds \citep{2016ApJ...818..144R}. On the other hand, at densities of a few $\sim10^4$~cm$^{-3}$ $^{13}$CO(1-0) tends to be optically thick, and CO also freezes out onto dust grains \citep{2014MNRAS.438L..56H}. To trace the gas at densities of $10^4$~cm$^{-3}$ and above we use N$_{2}$H$^{+}$(1-0) - a reliable tracer of dense and cold gas \citep{2015PASP..127..299S}.

The $^{13}$CO velocity data was also used to probe overlapping structures along the line-of-sight towards each cloud and for the calculation of the large-scale magnetic field direction pseudovectors in the cloud through the velocity gradient technique (see Section \ref{sec:confusion}).

\paragraph{$^{13}$CO(1-0)}

The $^{13}$CO(1-0) data used is from the Galactic Ring Survey \citep[GRS - ][]{2010ApJ...723..492R}, measured with the FCRAO. This data has an angular resolution of $44''$ (pixel size $\sim22''$), a velocity resolution of 0.21km/s and a one $\sigma$ noise of 0.13 K in T$_A^*$ scale. The H$_{2}$ column density cubes produced by \citet{2023MNRAS.525.2935P} from this data have been used in this work as well; these have an angular resolution of $\sim3'$.

\paragraph{N$_{2}$H$^{+}$(1-0)}

The N$_{2}$H$^{+}$(1-0) molecular line data was obtained using the IRAM 30m telescope. The data has an angular resolution of $28''$ (with a pixel size of $9''$) and a velocity resolution of 0.16 km/s. The final noise ranges from 0.09 K to 0.2 K per velocity channel and pixel \citep{2023MNRAS.525.2935P}. This data is not used directly in this work extensively, but it was utilised by \citet{2023MNRAS.525.2935P} in calculating the boundary values of H$_{2}$ column density for the clumps - these column density values are the ones used in this paper to define the edges of the clumps - along with the velocity dispersions and virial ratios on clump scale.

\subsubsection{\textit{Herschel} H$_2$ column density maps}

The \textit{Herschel} telescope was a space-based telescope that launched alongside {\it Planck}, and observed the Universe at sub-millimetre and far-infrared wavelengths. The multi-wavelength continuum observation of the sky allowed the characterisation of cold thermal dust emission, ideal for the study of star-forming regions. Here, we use the \textit{Herschel} H$_2$ column density maps derived from the Hi-GAL galactic plane survey \citep{2010PASP..122..314M} and presented in   \citet{2016A&A...590A..72P}. The data has a pixel size of $4.5''$ and a native resolution of $18''$, which has been degraded to $28''$ for the images of the clumps in order to match the N$_{2}$H$^{+}$(1-0) data.

\begin{figure*}
	\begin{subfigure}[H]{0.462\textwidth}
	\includegraphics[width=\textwidth]{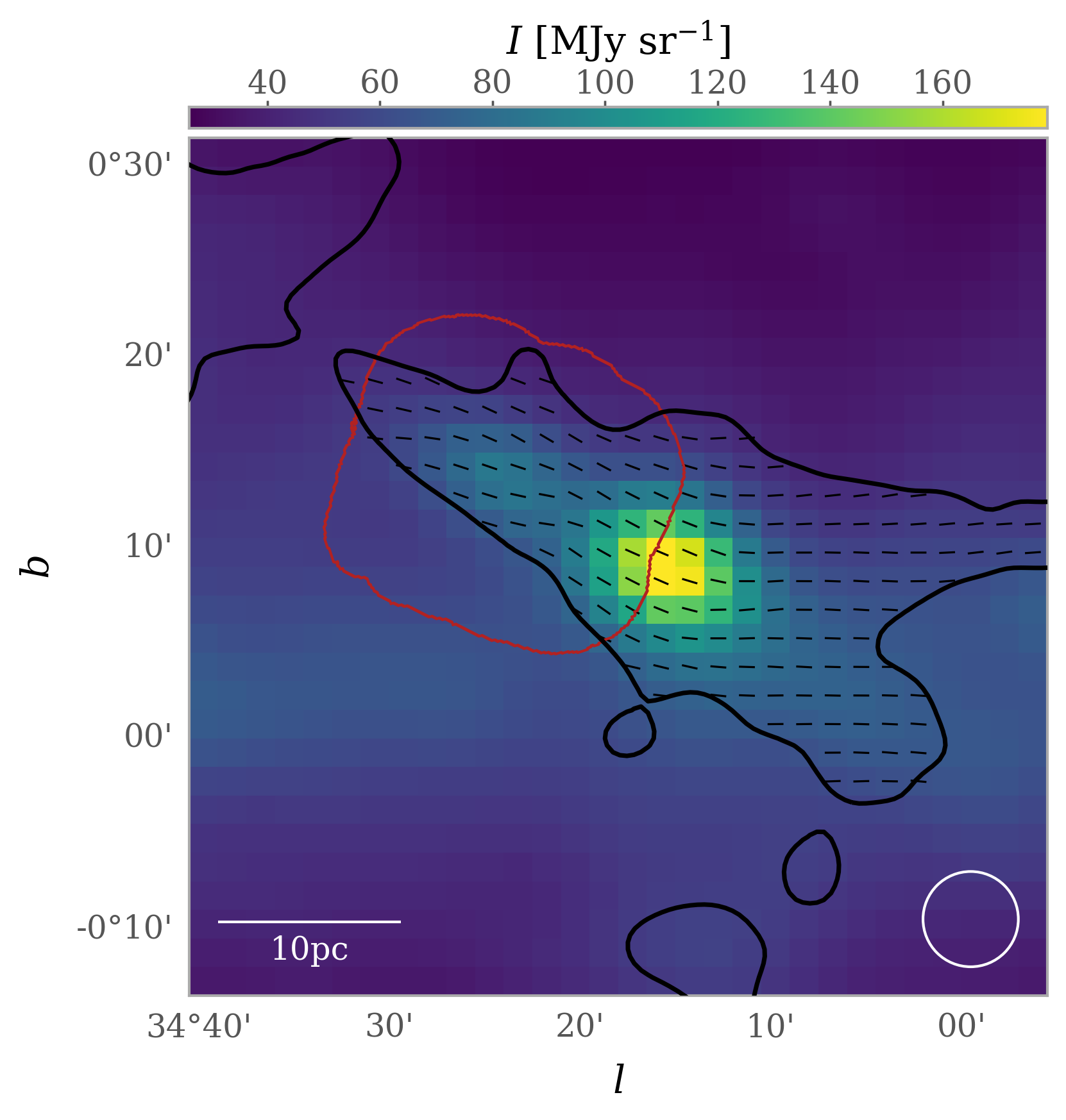}
	\caption{}
	\label{fig:SDC34p370_Planck}
	\end{subfigure}
	\begin{subfigure}[H]{0.528\textwidth}
	\includegraphics[width=\textwidth]{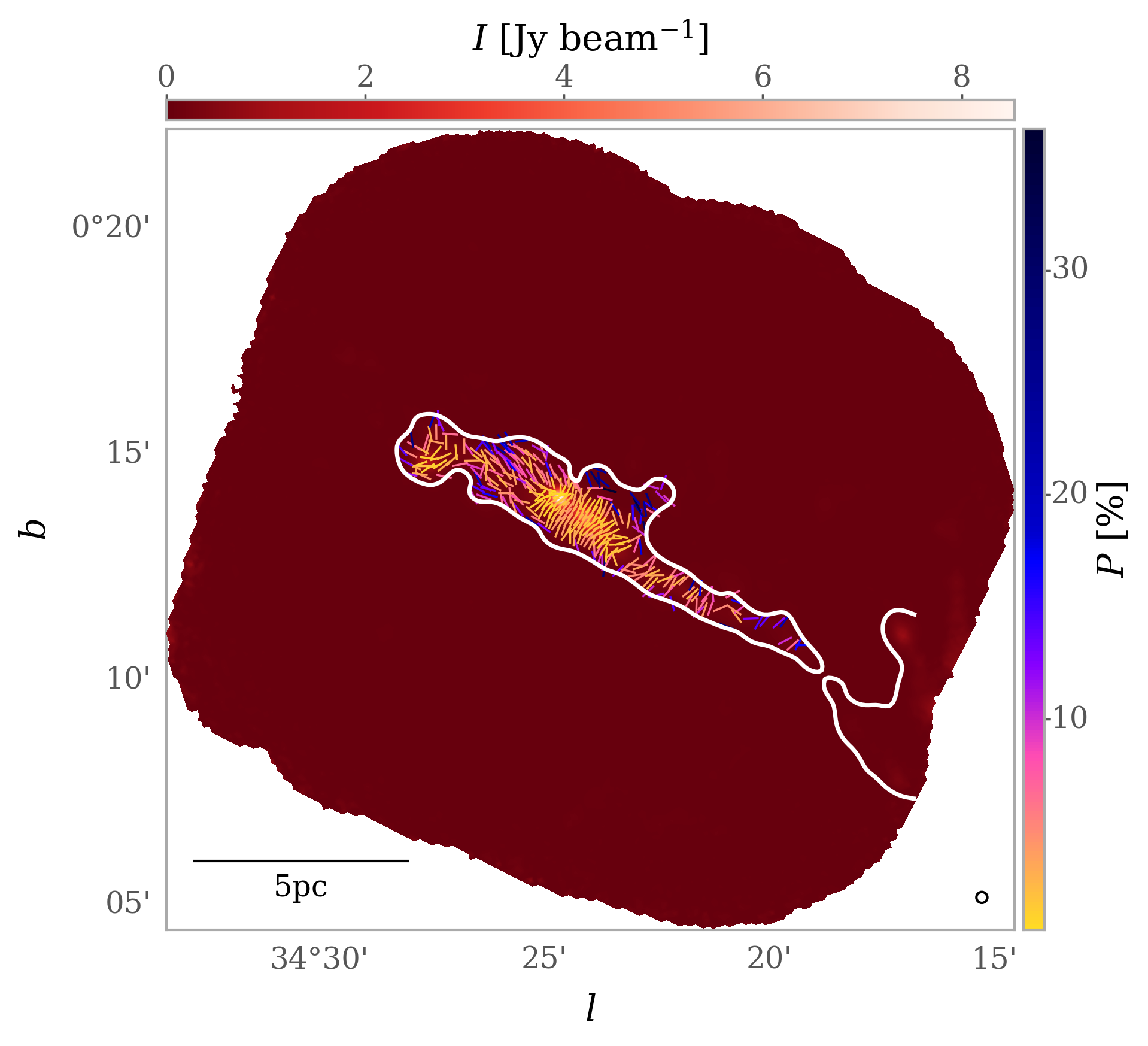}
	\caption{}
	\label{fig:SDC34p370_POL-2}
	\end{subfigure}
	\begin{subfigure}[H]{0.9\textwidth}
    \centering
	\includegraphics[height=0.4\textheight]{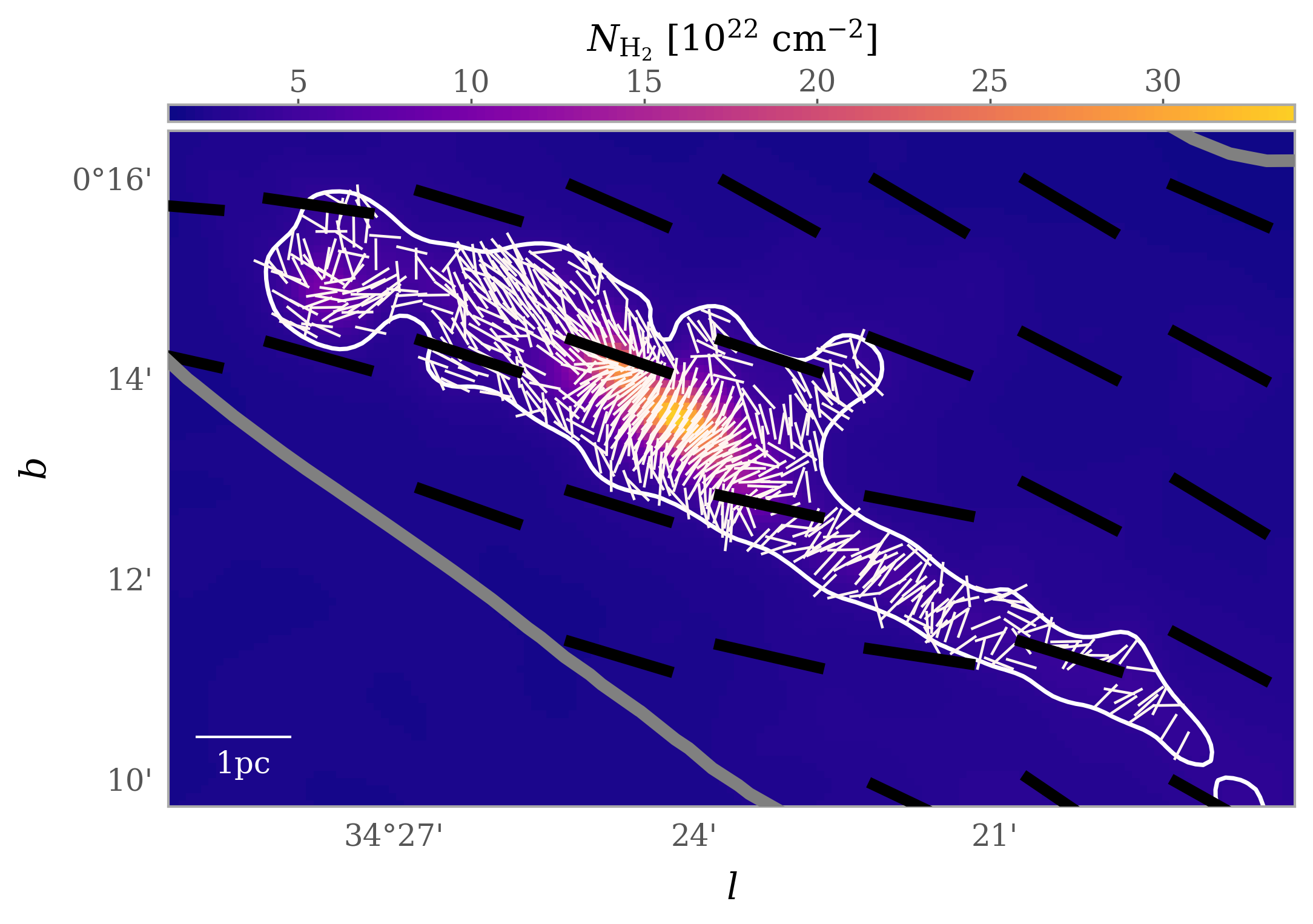}
	\caption{}
	\label{fig:SDC34p370_zoomed}
	\end{subfigure}
    \caption{a) {\it Planck} magnetic field pseudovectors (in black and all of the same arbitrary length; not colour-coded or scaled) within the parent molecular cloud of SDC34.370 (black contour). The outline of the POL-2 intensity image (red contour) of SDC34.370 is overlaid on the {\it Planck} intensity image to show the relative position of the JCMT scanning area. The {\it Planck} beam size is depicted as a white circle in the bottom right corner of the image. b) POL-2 intensity map and magnetic field pseudovectors within SDC34.370, colour-coded by polarisation percentage (colorbar to the right of the image) and all of the same arbitrary length. Only every second pseudovector has been shown for visibility. The white contour marks the edge of the clump \citep[see][for more information]{2023MNRAS.525.2935P}. The JCMT beam size is depicted as a black circle in the bottom right corner of the image. c) A zoom-in of the clump region. A grey contour marks the cloud boundary and black segments show the {\it Planck} magnetic field pseudovectors, while the white contour and segments mark the clump boundary and JCMT POL-2 magnetic field pseudovectors respectively. The background is the {\it Herschel} H$_2$ column density map.}
    \label{fig:magfieldpseudovectors}
\end{figure*}

\section{Magnetic Field Direction from Dust Polarisation}
\label{sec:magfielddirection}

The directions of the polarisation pseudovectors are calculated from the Stokes' $Q$ and $U$ parameters according to the equation:
\begin{equation} 
\phi=\frac{1}{2} \arctantwo (\frac{U}{Q}), 
\end{equation}
where $\phi$ is the angle taken anticlockwise from Galactic north, towards Galactic east. The function $\arctantwo$ ensures that the output angle is in the correct quadrant. {\it Planck} convention however differs, in that this equation instead gives the angle clockwise from north. In order to calculate the angle anticlockwise from north as is IAU convention,  $\phi=\frac{1}{2} \arctantwo (\frac{-U}{Q})$ is used instead with {\it Planck} data \citep{2015A&A...576A.104P}. The directions of the magnetic field pseudovectors are found by rotating the polarisation pseudovectors by 90$^\circ$. Due to the $180^\circ$ ambiguity, the calculated angles are restricted to the range -90$^\circ$ to 90$^\circ$.

The polarisation fraction $P$ is defined as the fraction of the total intensity $I$ that is polarised:

\begin{equation} 
(PI)^2=Q^2+U^2
\end{equation}

Figure \ref{fig:magfieldpseudovectors} shows the plane-of-the-sky magnetic field direction for SDC34.370 as obtained from {\it Planck} on cloud scale and from POL-2 on clump scale. The POL-2 pseudovectors have been converted from equatorial to galactic angles by adding the conversion angle $\theta_{\rm eq-gal}$:

\begin{equation}
  \theta_{\rm eq-gal} = { \rm tan}^{-1} \left( \frac{{\rm cos}(l-32.9^\circ)}{{\rm cos}(b){\rm cot}(62.9^\circ) - {\rm sin}(b){\rm sin}(l-32.9^\circ)} \right)
\end{equation}

\noindent where $l$ is the Galactic longitude and $b$ is the Galactic latitude.

\begin{figure*}
	\begin{subfigure}[H]{0.45\textwidth}
	\includegraphics[width=\columnwidth, height=0.22\textheight]{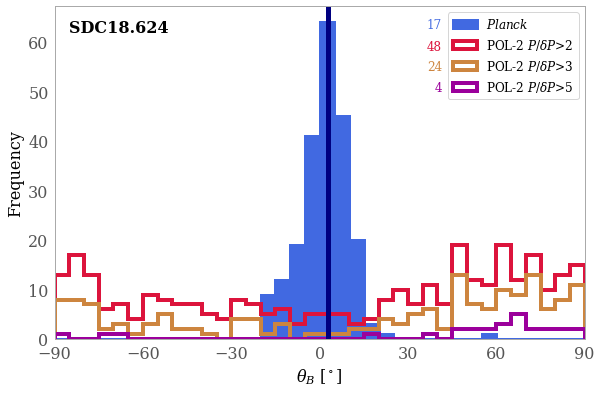}
	\caption{}
	\end{subfigure}
	\begin{subfigure}[H]{0.45\textwidth}
	\includegraphics[width=\columnwidth, height=0.22\textheight]{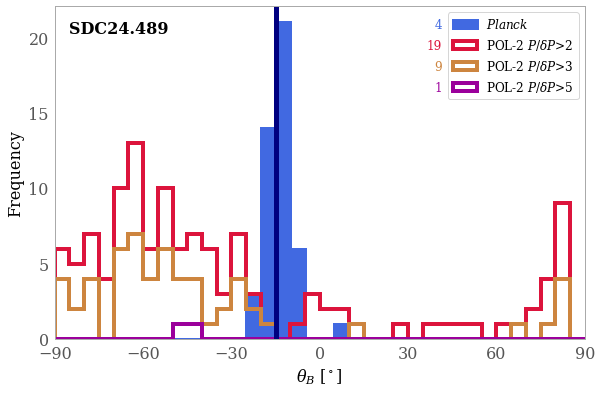}
	\caption{}
	\end{subfigure}
	\begin{subfigure}[H]{0.45\textwidth}
	\includegraphics[width=\columnwidth, height=0.22\textheight]{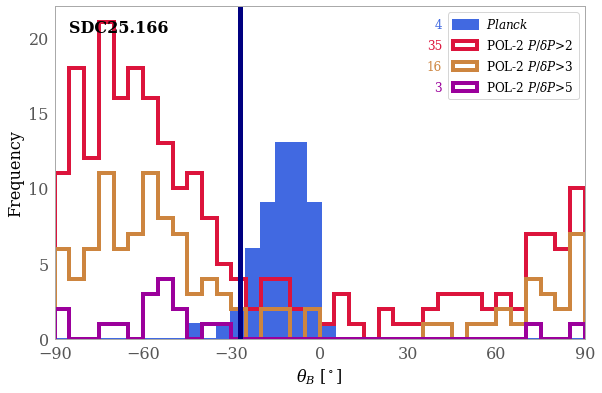}
	\caption{}
	\end{subfigure}
	\begin{subfigure}[H]{0.45\textwidth}
	\includegraphics[width=\columnwidth, height=0.22\textheight]{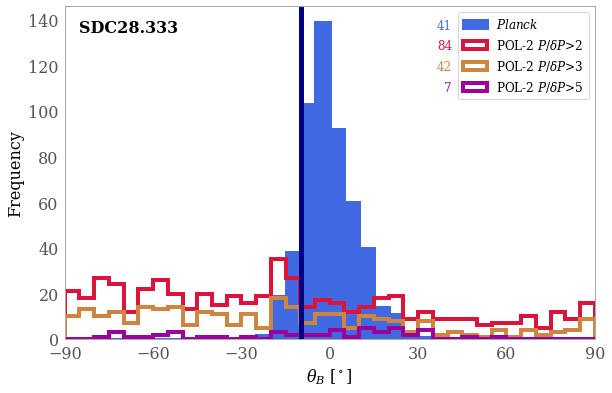}
	\caption{}
	\end{subfigure}
	\begin{subfigure}[H]{0.45\textwidth}
	\includegraphics[width=\columnwidth, height=0.22\textheight]{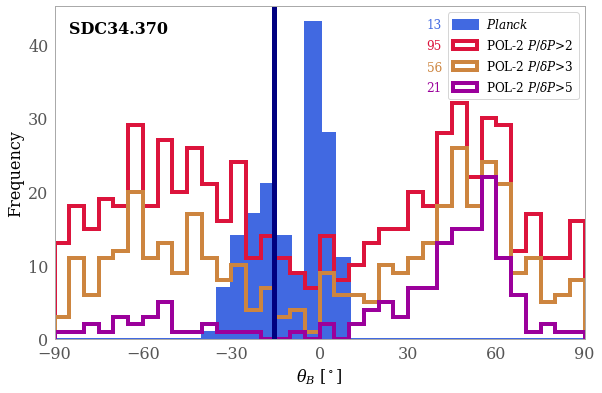}
	\caption{}
	\end{subfigure}
	\begin{subfigure}[H]{0.45\textwidth}
	\includegraphics[width=\columnwidth, height=0.22\textheight]{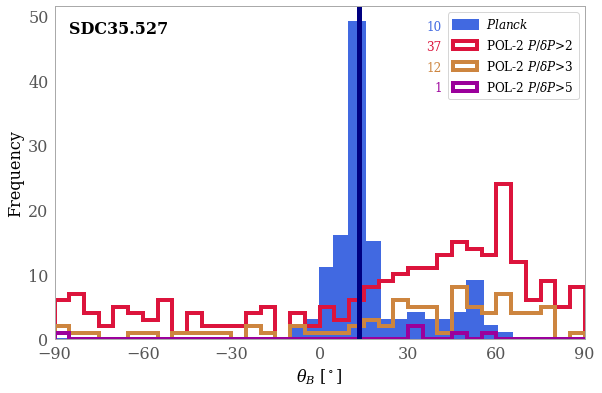}
	\caption{}
	\label{fig:SDC35p527vecthist}
	\end{subfigure}
	\begin{subfigure}[H]{0.45\textwidth}
	\includegraphics[width=\columnwidth, height=0.22\textheight]{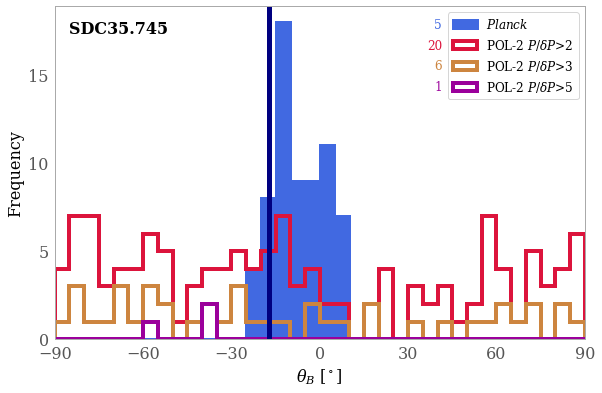}
	\caption{}
    \label{fig:SDC35p745pseudovectorhist}
	\end{subfigure}
	\begin{subfigure}[H]{0.45\textwidth}
	\includegraphics[width=\columnwidth, height=0.22\textheight]{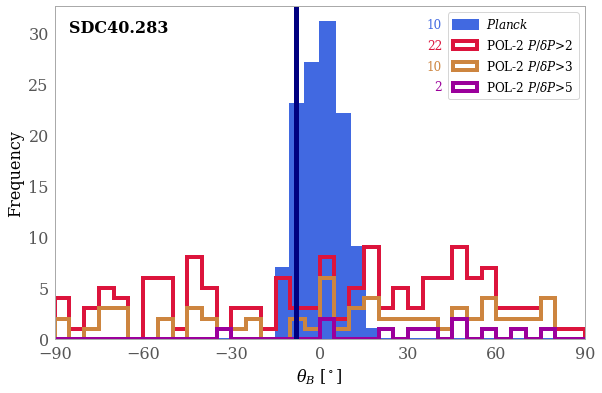}
	\caption{}
    \label{fig:SDC40p283pseudovectorhist}
	\end{subfigure}
    \caption{Histograms of $\theta_{B}$ for each clump and its parent cloud. A vertical blue line marks the average angle of {\it Planck} pseudovectors falling within the boundary of the clump. Numbers in corresponding colours to the histograms give the minimum number of independent measurements that contribute to the given histogram.}
    \label{fig:pseudovectorhists}
\end{figure*}

Similar images for the remaining clump/cloud pairs of the sample are given in Appendix \ref{sec:appendixmagfieldpseudovectorimages}. For the rest of the analysis, we only considered the pseudovectors falling within the cloud and clump boundaries as defined in \citet{2023MNRAS.525.2935P}. Also, only POL-2 and {\it Planck} pseudovectors with signal-to-noise ratios $P/\delta P>2$ and $I/\delta I>10$ have been retained, where $\delta P$ and $\delta I$ represent the uncertainties in $P$ and $I$ respectively. We use these filters for all analyses presented in this paper. We choose to use pseudovectors with $2< P/\delta P<3$ - which are often used in magnetic field studies -  to allow for greater spatial coverage across our clumps (magnetic field pseudovectors in the clouds mostly have $P/\delta P>3$ so are not largely affected). We checked that, statistically, the morphology of the magnetic field inferred from pseudovectors with $P/\delta P>2$ and with $P/\delta P>3$ is similar (see Figure \ref{fig:pseudovectorhists}), and also show how our results would vary if we were to apply $P/\delta P>3$ instead of $P/\delta P>2$ (see Appendix \ref{sec:appendixsnr>3}). $\delta P$ for the {\it Planck} measurements are calculated using Equation (B.2) from \citet{2015A&A...576A.104P}, and $P = \sqrt{\frac{Q^2 + U^2}{I^2} - (\delta P)^2}$. It can already be seen from these images that the general directions of the magnetic field pseudovectors in the clump are very different compared to the directions of the magnetic field pseudovectors in the parent cloud. 

Histograms of the distribution of angles of the magnetic field pseudovectors, which we call $\theta_{B}$, allow for a comparison between the large-scale cloud magnetic field observed with {\it Planck} and the small-scale clump magnetic field observed with JCMT POL-2. Figure \ref{fig:pseudovectorhists} shows the distribution of $\theta_{B}$ for each investigated cloud and its clump. It is apparent that the magnetic field direction on clump scale is distinct from the magnetic field direction in the larger scale cloud. The direction of the magnetic field in the diffuse part of the clouds is usually ordered, with the angles of the pseudovectors strongly peaking at 0$^\circ$ from the horizontal. This is likely due to the cloud scale magnetic field being inherited from the Galactic magnetic field, which tends to align approximately parallel to the plane of the Galaxy - which was the assumption made when the {\it Planck} polarisation data was corrected for bandpass mismatch leakage (see Section \ref{sec:dustpolobs}). The distribution of the directions of the magnetic field pseudovectors in the clumps however is more spread out, with any peaks present often being offset from 0$^\circ$. The vertical blue lines in the plots in Figure \ref{fig:pseudovectorhists} mark the circular mean angle of the {\it Planck} magnetic field pseudovectors that fall within the boundary of the clump, as projected in 2D; these for the most part lie close to the peak of the {\it Planck} histograms. For clouds such as SDC34.370 and SDC35.527 where the {\it Planck} histogram displays a bimodal distribution, the mean direction of the cloud-scale magnetic field in the region of the clump can be seen to coincide with one of the peaks. The numbers in the upper-right corner of each plot in Figure \ref{fig:pseudovectorhists} give the minimum number of independent measurements that contribute to the corresponding histogram. These numbers are minimum values since they quantify the number of independent measurements that would be obtained if all of the pseudovectors were adjacent. This means that for {\it Planck} this number is expected to be close to the true number of independent measurements, but for POL-2 the true number of independent measurements contributing to the histograms is likely to be slightly higher.

\subsection{The effect of resolution}
\label{sec:resolutioneffects}

The measurements of the large-scale magnetic field by {\it Planck} and of the small-scale magnetic field by JCMT POL-2 have extremely different resolutions; it must therefore be checked that the differences in the distributions of the magnetic field pseudovector directions seen in Figures~\ref{fig:magfieldpseudovectors} and \ref{fig:SDC18p624magfieldpseudovectors}-\ref{fig:SDC40p283magfieldpseudovectors} between these two scales are not a consequence of the differing resolutions. Since the JCMT POL-2 data has a much finer resolution than {\it Planck}, we thus might anticipate the dispersion of the magnetic field direction distribution to be larger for POL-2 measurements. Matching the resolutions of the {\it Planck} and POL-2 data allows us to infer what would theoretically have been observed by JCMT on clump scale at {\it Planck}'s resolution.
This was achieved by convolving the JCMT POL-2 $Q$ and $U$ polarisation data to the same resolution as the {\it Planck} polarisation data with a Gaussian kernel, and then recalculating the POL-2 magnetic field pseudovectors with the smoothed data. The maximum extent of the POL-2 maps is $\sim 20'\times 20'$ - these angular scales are small enough that rotation of the reference frame from pixel-to-pixel is negligible and so we do not apply any correction for this when smoothing the maps. As before, only pseudovectors with $P/\delta P>2$ and $I/\delta I>10$ were retained after smoothing, with the polarisation percentage and intensity signal-to-noise ratios calculated from the smoothed $I$, $Q$, and $U$ maps. $P$ and $\delta P$ were calculated in the same way as for the {\it Planck} data, but setting the cross-terms $\delta I\delta Q=\delta I\delta U=\delta Q\delta U=0$ when calculating $\delta P$ using Equation (B.2) from \citet{2015A&A...576A.104P}. Here it should be noted that this method is not suitable for comparing the dispersions of the angles of the large-scale and small-scale magnetic fields, but it is sufficient as a probe for which direction the small-scale magnetic field converges to. 

\begin{figure}
	\includegraphics[width=\columnwidth]{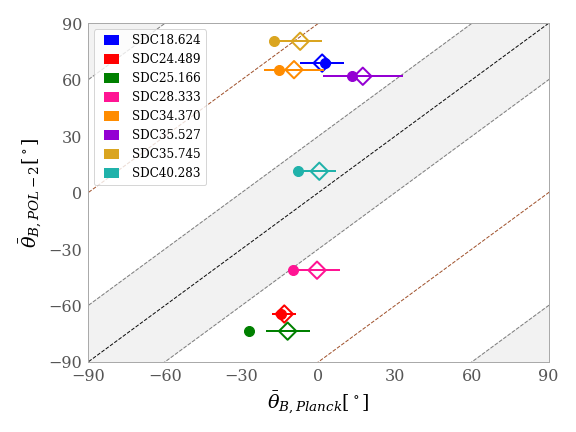}
	\caption{Circular means of the {\it Planck} ($x-$axis) and smoothed POL-2 ($y-$axis) magnetic field directions. Diamond symbols represent the average $\theta_B$ for the whole cloud from {\it Planck} measurements and filled circular symbols represent the average of the {\it Planck} pseudovectors falling within the clump boundary. Dashed lines depict a 1:1 relation (opaque black) between the two (i.e. the {\it Planck} and smoothed POL-2 mean directions are parallel), $\pm30^\circ$ (translucent black) from parallel, and perpendicular (opaque brown). The shaded grey regions highlight where a cloud would lie if the mean directions of the magnetic field on clump and cloud scale were within 30$^\circ$ of each other. Horizontal lines through each point show $\pm$ the circular standard deviations of the {\it Planck} magnetic field distributions. The standard deviations of the smoothed POL-2 distributions have not been included due to the low number of independent measurements.}
    \label{fig:smoothedresults}
\end{figure}

Figure \ref{fig:smoothedresults} shows the circular mean magnetic field angles obtained from {\it Planck} on cloud scale versus the circular mean magnetic field angles on clump scale derived from POL-2 data convolved to {\it Planck} resolution. For all bar one of the samples, the mean direction of the magnetic field within the clump is more than $30^\circ$ different to the mean direction of the magnetic field within its parent molecular cloud - the magnetic field on clump scale is not consistent with the magnetic field on cloud scale even taking into account the resolution differences. This is the case for the cloud-scale mean magnetic field direction considered both over the entirety of the cloud and also only for pseudovectors falling within the boundary of the clump. For most clouds, the range of angles encompassed by the circular standard deviation of the {\it Planck} distribution does not fall within the region of $\pm30^\circ$ from parallel either, highlighting the difference of directions even further. As mentioned previously, the standard deviation of the convolved POL-2 distribution cannot be reliably used due to the smoothing process, and so has been omitted from Figure~\ref{fig:smoothedresults}.

\section{Confusion along the line-of-sight}
\label{sec:confusion}

All clouds in this sample lie in the Galactic plane, making the presence of several, spatially unrelated, clouds along their lines-of-sight (LOS) likely. These contaminants may in particular affect the {\it Planck} large-scale polarised intensity maps. To check whether such contamination may be an issue, we created an average $^{13}$CO(1-0) spectrum for each cloud (see Appendix \ref{sec:appendixvelspec}). These spectra indeed show that while some clouds have only one clear emission peak and are therefore mostly uncontaminated (e.g. SDC34.370), others are clearly not (e.g. SDC25.166).  It is thus important to quantify to what extent the polarisation measurements obtained towards our sample of clouds may have been affected by these foreground/background clouds. Note that for the clumps, the N$_{2}$H$^+$(1-0) spectra do not display multiple components along their lines-of-sight, and are small and dense enough that it can be assumed that POL-2 measurements will not be affected by the same issue, i.e. POL-2 emission is clearly confined to the clumps.

In the following, we describe an alternative method to derive the plane-of-sky magnetic field morphology that only targets the emission from the cloud of interest. This method is known as the velocity gradient technique \citep[VGT - ][]{2017ApJ...835...41G}. 

The VGT is based on the theory of MHD turbulence; fluid motions in the magnetised and turbulent ISM are eddy-like, and these eddies rotate around an axis aligned with the magnetic field, along which they are elongated. Due to magnetic reconnection, these eddies cause the magnetic field and matter to be moved in the direction perpendicular to the local direction of the magnetic field. The elongated eddies have a maximum velocity gradient that is perpendicular to their longest axis - i.e. perpendicular to the magnetic field. This leads to the basis of the VGT - the maximum velocity gradient is expected to be perpendicular to the local direction of the magnetic field in regions that are not dominated by gravity. Multiple versions of the VGT have been developed, each implementing different ways to derive the direction of polarisation pseudovectors, often either through velocity centroid gradients (VCGs) or velocity channel gradients (VChGs) \citep{2018ApJ...853...96L, 2018MNRAS.480.1333H, 2019NatAs...3..776H}. 

VChGs is the method used here to derive the expected magnetic field direction in the diffuse ISM. This particular method uses the idea that the intensity in thin velocity channels - defined as having a width less than the square root of the turbulent velocity dispersion - of a position-position-velocity (PPV) data cube traces the turbulent velocity of the fluid, therefore allowing for the calculation of the velocity gradient through the intensity gradients of the thin velocity slices and consequently the direction of the magnetic field pseudovectors as detailed above.

To briefly summarise the VChGs process, this method involves first calculating intensity gradients in each thin velocity channel of the PPV cube. Subblock averaging is then applied to each pixel in each velocity slice; this requires computing the distribution of the angles of the pseudovectors that are within a block centred on the given pixel, and taking the peak of the Gaussian fitted to that distribution as the value of the angle for that pixel. Pseudo $Q$ and $U$ Stokes' parameters are calculated from the subblock-averaged angles, using the equations
\begin{equation}
Q(x, y) = \sum_i^{n_v} I_i(x, y) \cos(2\psi_i(x, y))
\end{equation}
\begin{equation}
U(x, y) = \sum_i^{n_v} I_i(x, y) \sin(2\psi_i(x, y))
\end{equation}
where $I$ is the intensity, $\psi$ is the subblock-averaged angle and the sum is over the $n_v$ velocity channels of the PPV cube. This results in 2D $Q$ and $U$ maps, from which the directions of the polarisation pseudovectors - and therefore the magnetic field pseudovectors - are calculated.

We use an Alignment Measure (AM) to compare the magnetic field gradients obtained from the VGT and those from the {\it Planck} dust polarisation observations, defined as:

\begin{equation}
\rm{AM} = 2\left(\langle \cos^2 \left(\theta_{\rm{diff}}\right) \rangle - \frac{1}{2}\right) \label{eq:AM}
\end{equation}
where $\theta_{\rm{diff}}$ is the difference in angle between the magnetic field pseudovectors calculated from the VGT and those from {\it Planck}. A value AM=1 indicates perfect alignment between the VGT and {\it Planck} magnetic field pseudovectors, while AM=-1 shows that they are perpendicular.

In the context of the VGT analysis employed here, we use the GRS $^{13}$CO(1-0) cubes.
We apply the VGT to the whole area covered by the {\it Planck} image of each cloud (as seen in Figures \ref{fig:SDC34p370_Planck} and \ref{fig:SDC18p624_Planck}-\ref{fig:SDC40p283_Planck}), first using the full velocity range of each $^{13}$CO(1-0) cube, and then repeat the VGT process using only the velocity range of the cube that corresponds to the velocity of the cloud of interest \citep[as defined by][]{2023MNRAS.525.2935P}. The difference in the AMs within the cloud between the full-velocity and cloud-velocity applications can help to quantify the level of line-of-sight confusion towards a given cloud. We use a block of size 13$\times$13 pixels of the $^{13}$CO(1-0) data (pixel size $22''$) in the subblock averaging step, resulting in a block that has dimensions similar to the resolution of the {\it Planck} data (the effect of block size on the results of the VGT are shown in Appendix \ref{sec:appendixVGTblocksize}). The magnetic field angles derived from the VGT are then rebinned to match the {\it Planck} pixel grid to allow for comparison between the two datasets.

\begin{figure*}
	\begin{subfigure}[H]{0.465\textwidth}
	\includegraphics[width=\columnwidth]{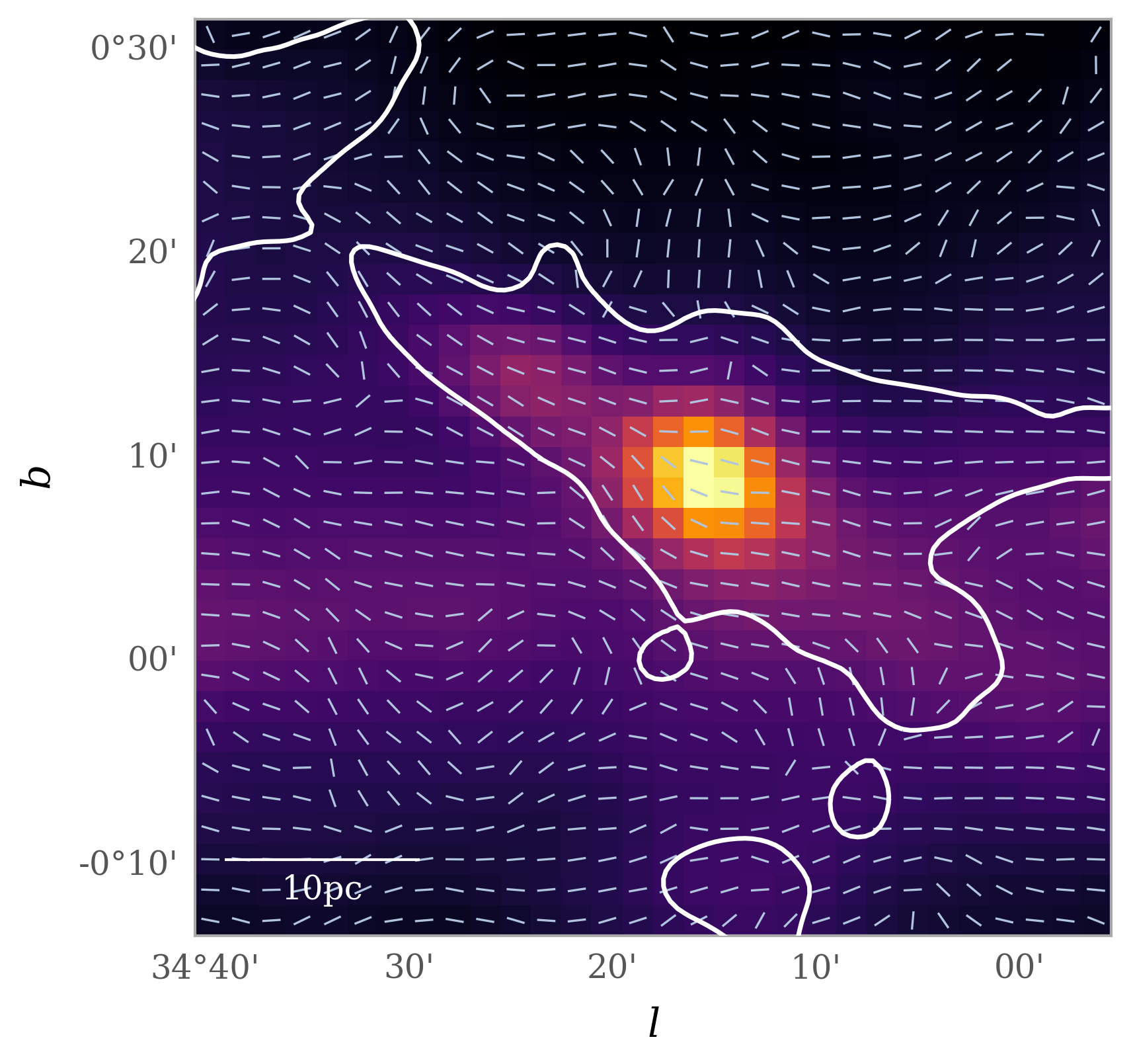}
	\caption{}
	\label{fig:VGT-a}
	\end{subfigure}
	\begin{subfigure}[H]{0.525\textwidth}
	\includegraphics[width=\columnwidth]{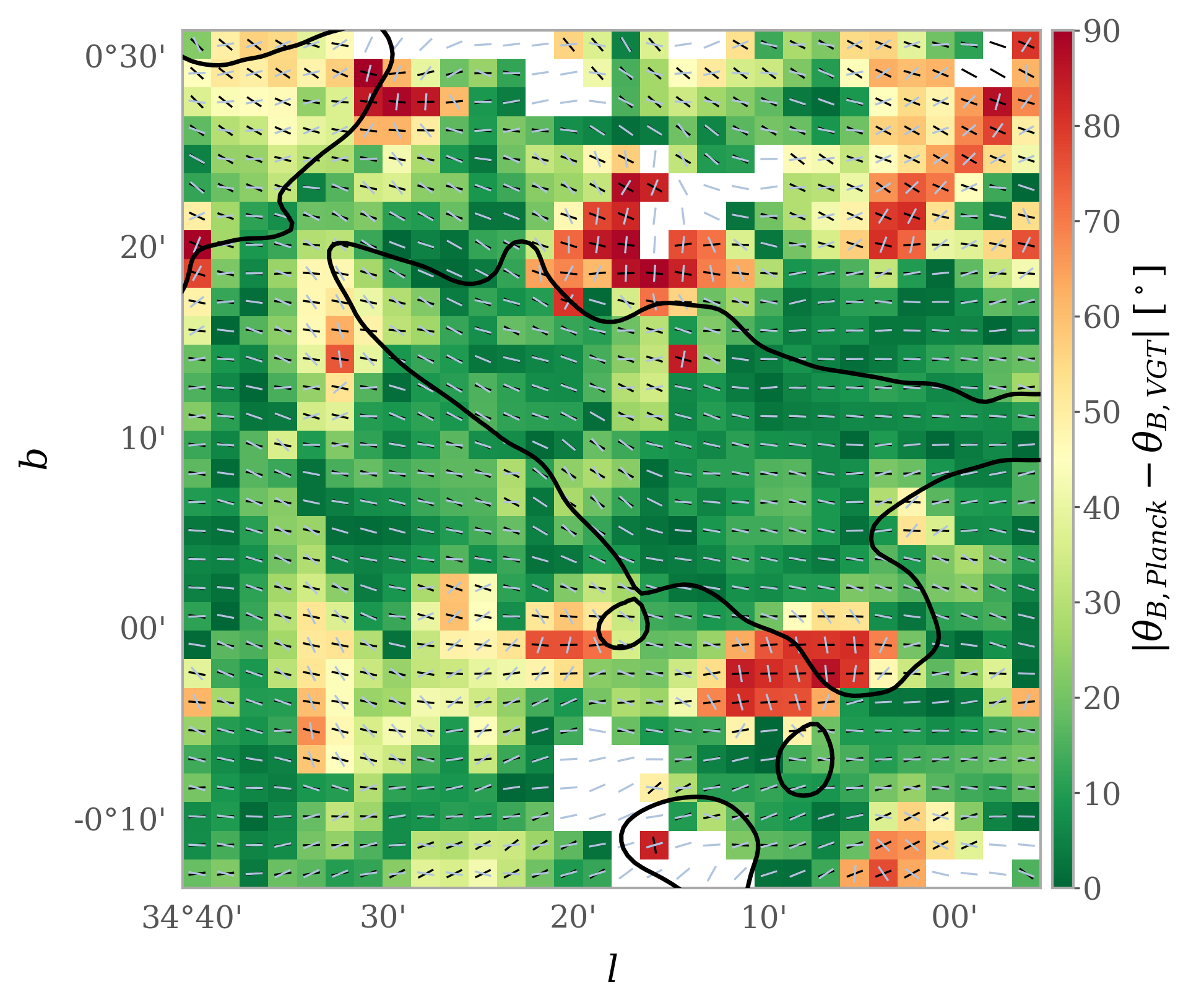}
	\caption{}
	\label{fig:VGT-b}
	\end{subfigure}
    \caption{a) The magnetic field pseudovectors (in pale blue) obtained from applying the VChGs method on SDC34.370, including the full velocity range of the $^{13}$CO(1-0) PPV cube. b) The match of the VChGs pseudovectors (pale blue segments) shown in Figure \ref{fig:VGT-a} to the {\it Planck} pseudovectors (black segments), where the colorbar shows the difference in angle between the VGT pseudovector and the {\it Planck} pseudovector in that pixel.}
    \label{fig:VGT}
\end{figure*}

Testing of this method was carried out on the cloud of SDC34.370, since this cloud was found to be the least contaminated cloud (see Appendix \ref{sec:appendixvelspec}). The VGT results on SDC34.370 are shown in Figure~\ref{fig:VGT}, where the full velocity range of the $^{13}$CO(1-0) cube was included. There are areas of the image where magnetic field pseudovectors have not been able to be calculated from the VGT due to the SNR and fitting error filters applied during the process. In areas where we have been able to obtain magnetic field pseudovectors from the VGT, the generally good alignment between the VGT and {\it Planck} magnetic field pseudovectors provides confidence that this method may be used to quantify the line-of-sight confusion of the {\it Planck} measurements for the other clouds too. 

\begin{table}
\caption{Alignment Measures obtained from applying the VGT on each cloud. The median values, along with the 16th percentile (bottom) and 84th percentile (top) are given in brackets.}
\label{tab:vgttable}
\resizebox{\columnwidth}{!}{%
\begin{tabular}{@{}ccccc@{}}
\toprule
Source    & Whole image,    & Within cloud, & Whole image,               & Within cloud, \\
          & full velocity range  & full velocity range      & cloud velocity range & cloud velocity range      \\ \midrule
SDC18.624 &  0.60 ($0.85^{0.99}_{0.10}$)&        0.78 ($0.91^{0.99}_{0.60}$)&              0.44 ($0.70^{0.98}_{-0.27}$)&         0.68 ($0.89^{0.99}_{0.28}$)\\
SDC24.489 &   0.54 ($0.79^{0.98}_{-0.06}$)&        0.50 ($0.79^{0.96}_{-0.18}$)&                 0.50 ($0.69^{0.96}_{-0.05}$)&          0.24 ($0.38^{0.90}_{-0.41}$)\\
SDC25.166 &   0.45 ($0.77^{0.99}_{-0.46}$)&        0.55 ($0.92^{0.99}_{-0.41}$)&                0.37 ($0.73^{0.99}_{-0.72}$)&          0.52 ($0.97^{1.00}_{-0.53}$)\\
SDC28.333 & 0.49 ($0.82^{0.99}_{-0.33}$)&        0.60 ($0.90^{0.99}_{-0.07}$)&               0.43 ($0.80^{0.98}_{-0.56}$)&         0.49 ($0.83^{0.99}_{-0.40}$)\\
SDC34.370 & 0.55 ($0.83^{0.98}_{-0.11}$)&        0.77 ($0.94^{0.99}_{0.61}$)&                  0.38 ($0.67^{0.97}_{-0.59}$)&         0.74 ($0.89^{0.98}_{0.57}$)\\
SDC35.527 &  0.36 ($0.68^{0.98}_{-0.60}$)&        0.29 ($0.47^{0.92}_{-0.42}$)&                  0.04 ($0.08^{0.88}_{-0.83}$)&          -0.03 ($0.03^{0.65}_{-0.81}$)\\
SDC35.745 &  0.52 ($0.83^{0.98}_{-0.30}$)&         0.92 ($0.97^{0.99}_{0.87}$)&                 0.49 ($0.83^{0.98}_{-0.41}$)&         0.78 ($0.96^{0.99}_{0.56}$)\\
SDC40.283 &  0.53 ($0.85^{0.99}_{-0.23}$)&        0.57 ($0.87^{0.99}_{-0.21}$)&                 0.39 ($0.80^{0.99}_{-0.69}$)&        0.26 ($0.58^{0.99}_{-0.89}$)\\ \bottomrule
\end{tabular}%
}
\end{table}

Table \ref{tab:vgttable} provides the AMs obtained from the VGT, where all values that contribute to the mean have equal weighting. Since means have a tendency to be skewed by outliers and distributions with large spread, we also include the median, and 16th and 84th percentiles of the values of  $2\left( \cos^2 \left(\theta_{\rm{diff}}\right) -  \frac{1}{2}\right)$. As mentioned previously, the VGT has been run twice for each cloud, once using the full velocity range of the PPV cube, and once using only channels within the velocity range of the cloud. For each of these two applications AM values are given for both the whole {\it Planck} image and for only within the outer boundary of the cloud. Images of the results for the VGT such as those given in Figure \ref{fig:VGT} for the other clouds in the sample are provided in Appendix \ref{sec:appendixVGTimages}. 

 Since {\it Planck} traces thermal dust emission, it is sensitive to emission from the diffuse ISM as well, while $^{13}$CO, as a molecular line tracer, observes the emission from molecular gas and so is localised to denser regions of the ISM such as molecular clouds. This means that perfect alignment (i.e. AM=1) cannot be expected due to the different phases of the ISM being traced by {\it Planck} and $^{13}$CO, as well as the fact that means are largely biased by outliers and trailing values. Taking these factors into consideration, column 3 of Table \ref{tab:vgttable} (within the cloud, for the full-velocity range application) is where we expect to see the best agreement between the VGT and {\it Planck}, since this is the region where $^{13}$CO likely traces a larger fraction of the column density traced by {\it Planck}. We do indeed see that this is the column with the best AMs for most of the clouds, although the level of alignment does vary from cloud to cloud. Column 5 shows the main quantity that we are interested in - here the AMs, when compared to those in column 3, should allow us to get an idea of how much the {\it Planck} measurements have been affected by LOS confusion. It can be seen that within the outer boundary of the cloud, the AM when calculating the VGT magnetic field using only the velocity range of the cloud is always lower than when using the full velocity range of the cube, as expected. The AM values in column 5 are also all above 0.2 - which corresponds to a difference in angle of $<40^\circ$ between {\it Planck} and VGT on average - except for SDC35.527.

\begin{figure}
	\includegraphics[width=\columnwidth]{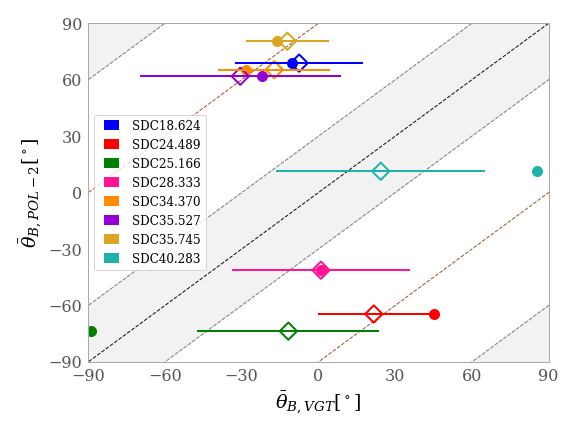}
	\caption{The same as Figure \ref{fig:smoothedresults}, but with the average directions of the magnetic fields in the clouds calculated from the results of the VGT instead of {\it Planck} dust polarisation.}
    \label{fig:vgtsmoothedresults}
\end{figure}

Figure \ref{fig:vgtsmoothedresults} shows the equivalent plot to Figure~\ref{fig:smoothedresults}, but this time the mean directions of the magnetic field on cloud scale are derived from the VGT magnetic field angle distributions within the cloud for the case where the VGT was applied only over the cloud's velocity range. The clump average directions are the same as in Figure~\ref{fig:smoothedresults}. According to the magnetic field directions derived from the VGT, the average directions of the magnetic field within the clouds tends to either stay in a similar position or move towards being more perpendicular to the average magnetic field direction in the clump compared to {\it Planck}, albeit while largely increasing the spread in the distribution of cloud-scale magnetic field pseudovector angles. For SDC40.283, the cloud's mean field direction changes by about 20$^\circ$, but the difference between the average cloud and clump magnetic field directions remains similar.
In terms of the mean direction of the VGT pseudovectors within the boundary of the clump projected in 2D, these are all also outside $\pm30^\circ$ from parallel to the clump mean direction with the exception of SDC25.166. 

\begin{figure}
	\includegraphics[width=\columnwidth]{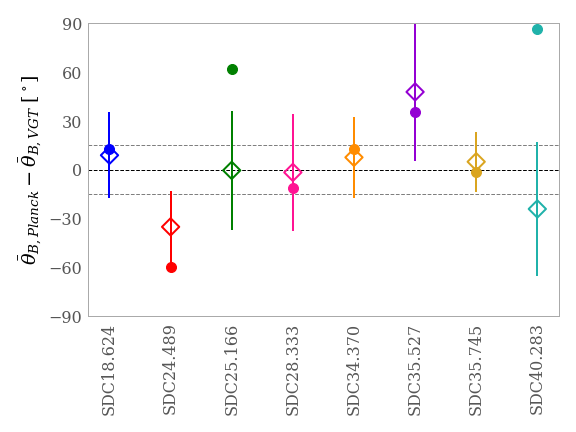}
	\caption{The difference in the average magnetic field angles within the clouds from {\it Planck} and the VGT. A dashed horizontal black line shows a difference of 0$^\circ$, fainter dashed black lines show $\pm$15$^\circ$ difference. Diamond symbols represent the difference in circular mean angle of the magnetic field within the whole cloud, while filled circles represent the difference between the circular mean of the vectors falling within the clump boundary.}
    \label{fig:planckvgtdiff}
\end{figure}

Figure \ref{fig:planckvgtdiff} shows the comparison between the average directions of the magnetic field within the clouds gained from {\it Planck} and the VGT. Five out of the eight clouds have differences in the mean magnetic field direction of the whole cloud between these two methods of $<15^\circ$. Three clouds - SDC24.489, SDC35.527, and SDC40.283 - have differences greater than $15^\circ$. Half of the clouds have differences between the VGT and {\it Planck} mean directions within the clump boundary of $<15^\circ$.

These results from the VGT reinforce our observation that the morphology of the magnetic field varies substantially between cloud and clump scales.

\section{Histogram of Relative Orientations (HRO)}
\label{sec:hro}

The Histogram of Relative Orientations (HRO) was proposed by \citet{2013ApJ...774..128S} as a method to quantify the alignment between the magnetic field and density structures. A common finding in that paper and many others since then is that the orientation of the magnetic field is parallel to density structures at low densities but becomes perpendicular at high densities \citep[e.g.][]{2022FrASS...9.3556L}. This has been interpreted as evidence for a switch from magnetic to gravity-driven gas dynamics \citep{2017A&A...607A...2S}. In this section we show the results of applying the HRO analysis to the clouds and clumps in our sample. For the purposes of this analysis, the $^{13}$CO-based H$_{2}$ column density maps were smoothed to $5'$ to match the resolution of the {\it Planck} polarisation data, so that the intensity gradients probe the same spatial scales as the magnetic field pseudovectors. On clump scale, we use the JCMT POL-2 polarisation data and the \textit{Herschel} column density maps at their native resolutions of $14.6''$ and $18''$, respectively. We choose to not smooth the POL-2 polarisation to match the resolution of \textit{Herschel} in order to preserve the quantity of pseudovectors included in the analysis, since smoothing the POL-2 polarisation data to $18''$ results in the loss of a substantial number of pseudovectors in most samples - the mismatch in resolutions at clump scale is considered trivial due to small difference in physical resolution, i.e. approximately 0.05 to 0.08 pc at the source distances.

We have attempted to correct for any differences arising from using different tracers for column density on cloud and clump scale - till now, on cloud scale the $^{13}$CO-based H$_{2}$ column density maps were used to define the boundaries of the clouds, while on clump scale the \textit{Herschel} dust emission column density maps were used to define the boundaries of the clumps, as was done in \citet{2023MNRAS.525.2935P}. For the purposes of ensuring that the column density is comparable between clump and cloud in the HRO analysis and so can be consistently plotted on the same axis, the \textit{Herschel} column density maps have been corrected for line-of-sight confusion using the ratio of $^{13}$CO column density within the velocity range of the cloud to the $^{13}$CO column density over the full velocity range of the data cube. This allows the column density contour levels of the clouds to be expressed as a corrected \textit{Herschel} column density. Correction factors range from 0.14 to 0.63.

The level of alignment between the magnetic field pseudovector and the pseudovector parallel to the column density contour in each pixel can be quantified by the Alignment Measure (AM), as was defined in Section \ref{sec:confusion} (Equation (\ref{eq:AM})). Here, an AM=1 indicates that the magnetic field pseudovectors are parallel to the column density isocontours (perpendicular to the intensity gradient), while an AM=-1 shows that they are perpendicular to the isocontours (parallel to the intensity gradient). The column density ranges within which an AM is calculated have been obtained by splitting the total number of pixels within the cloud or clump into 3 approximately equal groups, and then calculating the corresponding column density boundaries.

\begin{figure*}
	\begin{subfigure}[H]{0.45\textwidth}
	\includegraphics[width=\columnwidth, height=0.22\textheight]{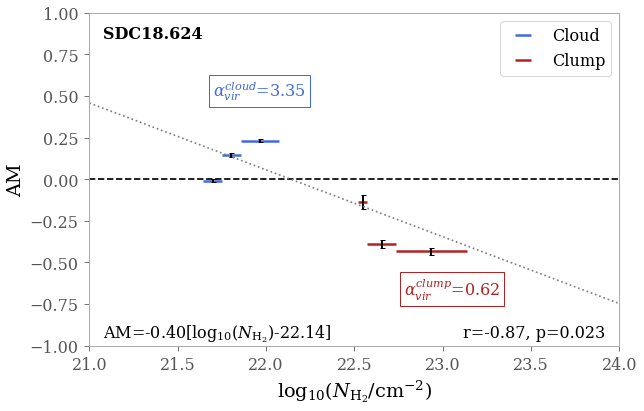}
	\caption{}
	\end{subfigure}
	\begin{subfigure}[H]{0.45\textwidth}
	\includegraphics[width=\columnwidth, height=0.22\textheight]{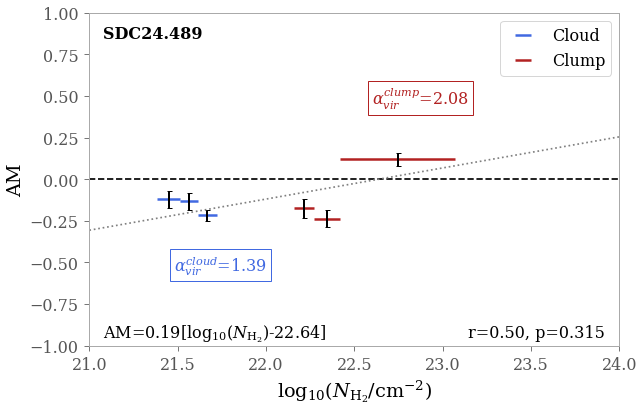}
	\caption{}
	\end{subfigure}
	\begin{subfigure}[H]{0.45\textwidth}
	\includegraphics[width=\columnwidth, height=0.22\textheight]{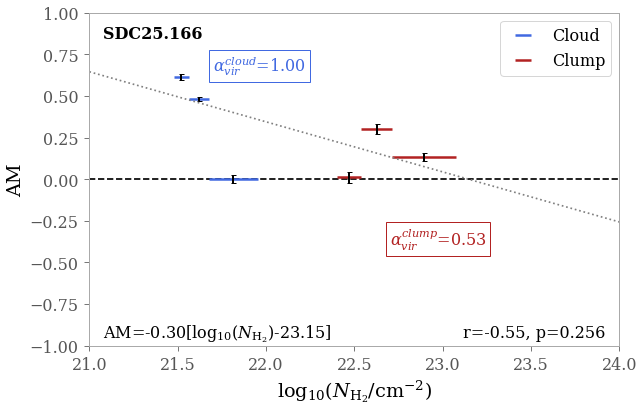}
	\caption{}
	\end{subfigure}
	\begin{subfigure}[H]{0.45\textwidth}
	\includegraphics[width=\columnwidth, height=0.22\textheight]{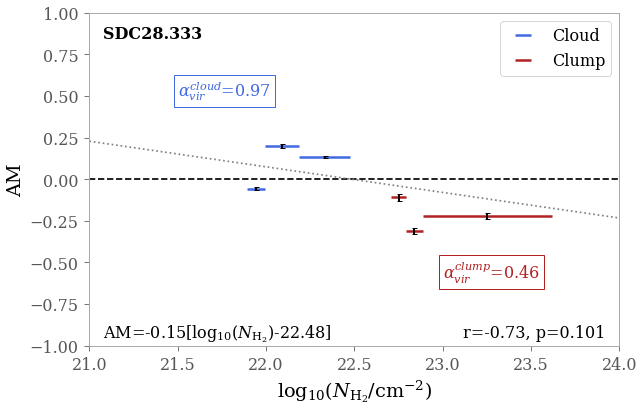}
	\caption{}
	\end{subfigure}
	\begin{subfigure}[H]{0.45\textwidth}
	\includegraphics[width=\columnwidth, height=0.22\textheight]{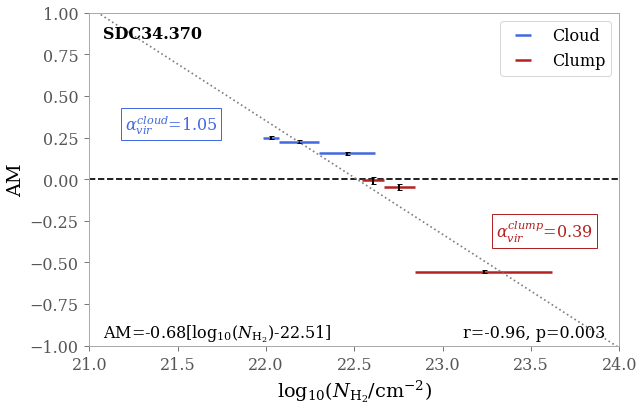}
	\caption{}
	\end{subfigure}
	\begin{subfigure}[H]{0.45\textwidth}
	\includegraphics[width=\columnwidth, height=0.22\textheight]{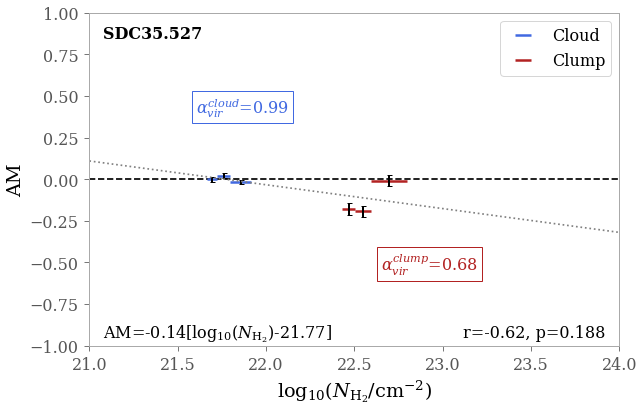}
	\caption{}
	\label{fig:SDC35.527HRO}
	\end{subfigure}
	\begin{subfigure}[H]{0.45\textwidth}
	\includegraphics[width=\columnwidth, height=0.22\textheight]{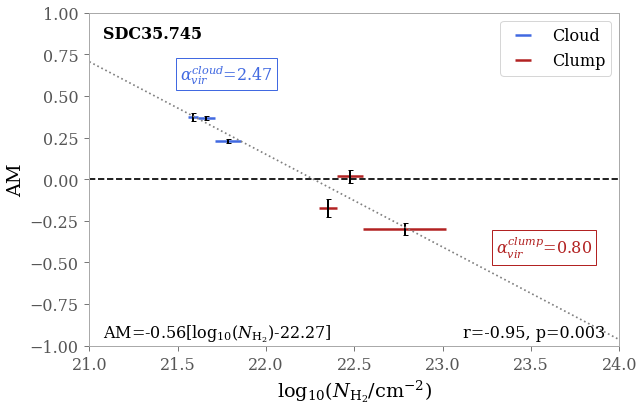}
	\caption{}
	\end{subfigure}
	\begin{subfigure}[H]{0.45\textwidth}
	\includegraphics[width=\columnwidth, height=0.22\textheight]{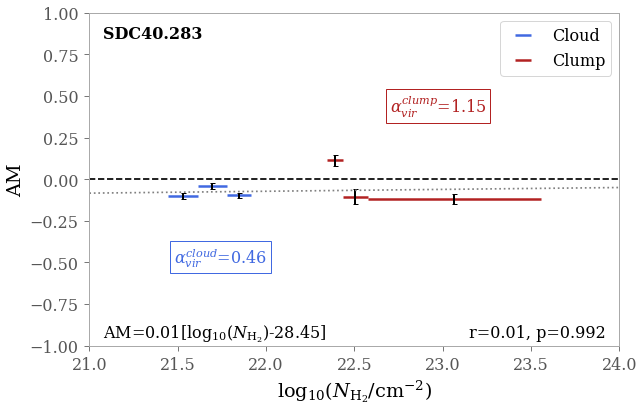}
	\caption{}
	\end{subfigure}
    \caption{HRO alignment measures for the clouds (blue) and clumps (red). Each point spans the range of column densities over which the AM was calculated. Dotted grey lines depict the best-fit line to the points, weighted by their errors. The equation of the best-fit line and the unweighted Pearson correlation coefficient (r) along with its associated p-value are provided in the bottom left and bottom right corners respectively. The virial ratios of each cloud and clump are also shown, with the same colour scheme as the AM points.}
    \label{fig:HRO}
\end{figure*}

Figure \ref{fig:HRO} shows the alignment measures for each cloud-clump pair. Whilst there is no obvious common trend for all of the samples, in general it can be seen that the AM within the clump tends to be lower than the AM in the cloud, displaying a shift in the relative orientation of the $B$-field from more parallel to the density structures to more perpendicular to the density structures with increasing density, as expected for structures whose dynamics is governed by gravity. This is supported by the best-fit lines: six out of the eight sources have a best-fit line with a negative gradient. Three of these - SDC18.624, SDC34.370, and SDC35.745 - have an unweighted Pearson's correlation coefficient r-value of $<$-0.8, with p-value$<$0.05, indicating a strong negative correlation between AM and column density for these sources. We can also see from these sources' best-fit lines that the transition from parallel to perpendicular occurs at around $\log_{10}N_{\rm H_2}\sim 22-22.5$, similarly to what has been found in previous studies. We discuss a possible link between the velocity dispersion ($\delta v$) and virial ratio values ($\alpha_{\rm vir}$) of the clouds/clumps and the AM they display in Section \ref{sec:ourresultsdiscussion}.

\section{Magnetic field strength}
\label{sec:Bfieldstrength}

Estimating the magnetic field strength in star-forming regions is subject to a large number of assumptions which do not always apply to the regions that are being studied. As a result, magnetic field strength measurements from polarised emission observations are highly uncertain and unreliable. Despite those caveats, magnetic field strength measurements from polarised emission are frequently calculated and interpreted to infer the importance of magnetic fields in the star formation process. In this section, we present the magnetic field strength values derived from the most commonly used methods to evaluate whether the picture inferred from those is compatible with the results presented so far.

\subsection{Davis-Chandrasekhar-Fermi (DCF) Method}

The Davis-Chandrasekhar-Fermi (DCF) method \citep{1951PhRv...81..890D,1953ApJ...118..113C} is a widely used method of estimating the mean magnetic field strength in the plane of the sky from dust polarisation measurements. The equation for the mean plane-of-sky (POS) magnetic field strength $B_{\rm{POS}}$ is given by:

\begin{equation}
B_{\rm{POS}} = \sqrt{4 \pi \rho} \frac{\delta v}{\delta \sin(\theta_B)}
\label{eq:dcfsin}
\end{equation}
where $\rho$ is the gas mass density, $\delta v$ is the turbulent line-of-sight velocity dispersion, and $\delta\sin(\theta_B)$ is the dispersion of the $\sin$e of the polarisation angles. Usually, the small angle approximation $\delta \sin(\theta_B) \sim \delta \theta_B$ is made, leading to the equation:

\begin{equation}
B_{\rm{POS}} = \sqrt{4 \pi \rho} \frac{\delta v}{\delta \theta_B}
\label{eq:dcfsa}
\end{equation}

These equations assume incompressible Alfv\'enic waves, along with energy equipartition between the turbulent kinetic energy and the turbulent magnetic energy, a dominant linear ordered magnetic field structure, and isotropic turbulence.

The dispersions, of $\theta_B$ or $\sin(\theta_B)$, are calculated by first fitting a Gaussian to their distributions. From the standard deviation ($\sigma_f$) of the best fit we correct for the mean error in $\theta_B$ and $\sin(\theta_B)$ such that we have: $\delta x = \sqrt{\sigma_f^2 - \bar{\Delta x}^2}$ where $x$ is $\theta_B$ or $\sin(\theta_B)$, and $\Delta x$ their associated errors. We provide the calculated values of $\delta \theta_B$ and $\delta \rm{sin}(\theta_B)$ in Table \ref{tab:angdisps}. In some cases, the mean errors in the measurements are larger than the dispersion, meaning that a corrected dispersion value was not defined. These are marked with a `-' in the table. Estimates for $\rho$ are calculated using the mass and radius estimates in \citet{2023MNRAS.525.2935P} and using the approximation of a uniform density sphere. Values for $\delta v$ are also taken from \citet{2023MNRAS.525.2935P}, using the N$_{2}$H$^{+}$(1-0) values on clump scale and the $^{13}$CO(1-0) values on cloud scale.

\begin{table}
    \centering
\caption{Values of  $\delta \theta_B$ and $\delta \sin(\theta_B)$ for both cloud ({\it Planck}) and clump (POL-2) scales. These values are calculated from the standard deviation of the Gaussian fitted to the distribution.}
\label{tab:angdisps}
    \begin{tabular}{>{\centering\arraybackslash}p{0.125\linewidth}>{\centering\arraybackslash}p{0.125\linewidth}>{\centering\arraybackslash}p{0.125\linewidth}>{\centering\arraybackslash}p{0.175\linewidth}>{\centering\arraybackslash}p{0.175\linewidth}}\toprule
         Source&  $\delta \theta_B^{\rm{cloud}} (^\circ)$&  $\delta \theta_B^{\rm{clump}} (^\circ)$&  $\delta \sin(\theta_B)^{\rm{cloud}}$& $\delta \sin(\theta_B)^{\rm{clump}}$\\\midrule
         SDC18.624
&  5.6
&  47.3&  0.099& -\\
         SDC24.489
&  -&  27.8&  -& 0.096\\
         SDC25.166
&  6.6
&  24.5&  0.130& 0.030\\
         SDC28.333
&  6.8
&  60.4&  0.120& 0.506\\
         SDC34.370
&  13.6
&  91.5&  0.233& 0.340\\
         SDC35.527
&  2.0
&  32.4&  0.038& 0.186\\
         SDC35.745
&  10.1
&  61.4&  0.174& 0.061\\
         SDC40.283&  6.6
&  72.6&  0.116& 0.522\\ \bottomrule
    \end{tabular}

\end{table}

We also calculate the Median Absolute Deviation (MAD) of the $\theta_B$ and $\sin(\theta_B)$ distributions. MAD is a measure of dispersion that is more robust to outliers and does not require a normally distributed dataset - in some cases a Gaussian fit may not be ideal or feasible. In order to enable direct comparison with the standard deviation gained from the Gaussian fit, we use the MAD scaled by a factor of $\sim$1.48 - an established conversion factor for large samples that normalises the MAD to the scale of the standard deviation of a normal distribution \citep{LEYS2013764}. Where the ratio $\frac{1.48 \times \rm{MAD}}{\sigma_f}$ is either $<0.9$ or $>1.1$ we also include the value of the magnetic field strength calculated from the scaled MAD value in Table \ref{tab:uncorrectedBpos}, otherwise we only provide the value calculated from the standard deviation of the Gaussian fit.

The results of applying equations \ref{eq:dcfsin} and \ref{eq:dcfsa} are shown in columns 2 and 3, and 4 and 5, of Table \ref{tab:uncorrectedBpos} respectively. We note that both of these equations should only be used for angular dispersions less than 25$^\circ$ \citep{2001ApJ...546..980O}, which is not the case for most of the clumps in our sample, but we include the results of applying the DCF to these sources anyway for illustration. The DCF equations are also based on the idea of Gaussian turbulent deviations from the ordered magnetic field, and so if the distributions of $\theta_B$ and $\sin(\theta_B)$ are not Gaussian (see e.g. SDC34.370 in Figure~\ref{fig:pseudovectorhists}) it is not expected that the DCF will be suitable. 

A correction factor - conventionally named $Q$, with a value between 0 and 1 - is often applied to the DCF equation to correct for overestimation of the magnetic field strength due to beam integration effects. The exact value of what this correction factor should be is uncertain, but simulation studies \citep[e.g.][]{2001ApJ...546..980O, 2001ApJ...559.1005P, liuCalibratingDavisChandrasekhar2021} have been carried out to determine its value. According to \citet{liuCalibratingDavisChandrasekhar2021} the value of $Q$ also varies depending on the properties of the region being studied. 

On cloud scale, the results of applying equations \ref{eq:dcfsin} and \ref{eq:dcfsa} without any correction factor are mostly consistent, while on clump scale the $B$-field strength gained from equation \ref{eq:dcfsin} is higher than that from equation \ref{eq:dcfsa}, sometimes by a large factor, although, as previously noted, the DCF equations are not expected to be reliable when applied to the clump scale measurements in any case. The field strength on cloud scale is also consistently higher than the strength on clump scale, which is the opposite of what would be expected when considering magnetic flux-freezing and findings from Zeeman measurements that show that the LOS magnetic field strength increases with increasing density in molecular clouds \citep[e.g.][]{2012ARA&A..50...29C}. Values of the magnetic field strength previously found in molecular clouds are on the order of a few to tens of $\mu$G, while in clumps they tend to be in the range of a few tens to hundreds $\mu$G \citep[e.g.][and references therein]{2023ASPC..534..193P}. The clump scale values that we obtain for our sample appear to be consistent with those, but our cloud scale values are much higher. This could be due to the limited number of independent {\it Planck} polarisation measurements in our clouds, which could bias the dispersion to lower values. 

Note that, unlike other studies \citep[e.g.][]{pillaiMAGNETICFIELDSHIGHMASS2015}, we have not subtracted a mean field prior to the estimation of the dispersion. However, this is accounted for in the analysis of the angular dispersion function presented in Section~\ref{sec:adf}.

\subsection{Skalidis-Tassis (ST) Method}

The Skalidis-Tassis (ST) method alters the DCF method in order to also take into account compressible non-Alfv\'enic turbulence and remove the assumption that the turbulence is isotropic \citep{2021A&A...647A.186S}. The POS magnetic field strength is given by: 

\begin{equation}
B_{\rm{POS}} = \sqrt{2 \pi \rho} \frac{\delta v}{\sqrt{\delta \theta_B}}
\label{eq:ST}
\end{equation}

Columns 6 and 7 of Table \ref{tab:uncorrectedBpos} show the results of applying the ST equation to our sample. On clump scale, the $B$-field strengths obtained from the ST method seem to be similar to those obtained from the classical DCF equation within a factor of about 2, but on cloud scale the $B$-field strength is reduced by a factor of 3 to 10 depending on the cloud. The values of $B_{\rm{POS}}$ on clump and cloud scale are now for the most part comparable, although the strength on clump scale still tends to be smaller than the strength on cloud scale.

\subsection{Angular Dispersion Function (ADF)}
\label{sec:adf}

The angular dispersion function (ADF) is a method commonly used to calculate the plane-of-sky magnetic field strength by taking into account the large-scale and turbulent contributions to the magnetic field \citep[][hereafter Hou09]{2009ApJ...706.1504H}. The method works by fitting a structure function of the form:

\begin{equation}
1 - \langle \cos\left[\Delta \theta_B (\mathscr{l})\right]\rangle \simeq \frac{1}{N}\frac{\langle B_t ^2 \rangle}{\langle B_o^2 \rangle}  \left(1 - \exp\left[-\frac{\mathscr{l}^2}{2\left(\delta^2 + 2W^2\right)}\right]\right) + a\mathscr{l}^2 \label{eq:adf}
\end{equation}
where $\Delta \theta_B (\mathscr{l})$ is the difference in angle between two pseudovector angles at a given lag $\mathscr{l}$, $N$ is the number of turbulent cells along the line-of-sight within the beam, $B_t$ is the turbulent magnetic field strength, $B_o$ is the ordered magnetic field strength, $\delta$ is the turbulent correlation length, $W$ is the beam radius ($\rm{FWHM}/2\sqrt{2{\rm ln}2}$, where $\rm{FWHM}$ represents the full width at half maximum of the telescope beam) of the telescope used to take the polarisation measurements, and $a$ is a constant that quantifies the amount of curvature in the large scale magnetic field. At the distances of the sources in our sample, it is unlikely that the beam will be able to resolve the turbulent correlation length, and so values of $\delta$ gained from these fits will not be accurate estimates. We fit the ADF in two ways - for the first method we fit all three parameters $\frac{1}{N}\frac{\langle B_t ^2 \rangle}{\langle B_o^2 \rangle}$, $\delta$, and $a$, and for the second method we keep $\delta$ fixed at 0 (i.e. we consider it negligible compared to the beam size $W$) and only fit the other two parameters. We fit the ADF using reduced-$\chi^2$ minimisation. Plots of the ADF for our sources are given in Appendix \ref{sec:appendixADF}, with the best-fitting curves for the case where all three parameters $\frac{1}{N}\frac{\langle B_t ^2 \rangle}{\langle B_o^2 \rangle}$, $\delta$, and $a$ were fit.

\begin{table*}
\centering
\caption{ADF fitting results: the reduced chi-squared value of the fit, the turbulent-to-ordered magnetic field strength ratio divided by the number of turbulent cells along the LOS, the turbulent correlation length, and the curvature of the $B$-field are given for both clump (POL-2) and cloud ({\it Planck}) scales. Rows where $\delta$ is marked as '-' are where $\delta$ was set to 0.
}
\label{tab:ADF}
\begin{tabular}{ccccccccc}
\toprule
                        & \multicolumn{4}{c}{POL-2}                                                                                                                  & \multicolumn{4}{c}{{\it Planck}}                                                                                                                        \\ \hline
                        Source& $\chi^2$ & $\frac{1}{N}\frac{\langle B_t ^2 \rangle}{\langle B_o^2 \rangle}$ & $\delta$ (arcsec)              & $a$ (10$^{-6}$arcsec$^{-2}$)& $\chi^2$ & $\frac{1}{N}\frac{\langle B_t ^2 \rangle}{\langle B_o^2 \rangle}$ & $\delta$ (arcmin) & \multicolumn{1}{c}{$a$ (10$^{-6}$arcmin$^{-2}$)} \\
\multirow{2}{*}{SDC18.624} & 6.92     & 0.2162 ($\pm$0.0035)& 0.48 ($\pm$7.41) & 27.6 ($\pm$1.6)& 2.94& 0.0217 ($\pm$0.0003)& 0.00 (-)& 17.6 ($\pm$2.1)\\
                        & 1.87     & 0.2268 ($\pm$0.0072)& -                              & 2.6 ($\pm$10.7)& 2.45& 0.0217 ($\pm$0.0003)& -& 17.6 ($\pm$2.1)\\
\multirow{2}{*}{SDC24.489} & 1.09     & 0.1732 ($\pm$0.0047)& 0.00 (-)                        & 24.8 ($\pm$3.4)& 0.77& 0.0039 ($\pm$0.0025)& 0.00 (-)& 58.5 ($\pm$41.2)\\
                        & 1.06     & 0.1766 ($\pm$0.0041)& -                              & 21.3 ($\pm$2.4)& 1.26& 0.0027 ($\pm$0.0024)& -& 81.8 ($\pm$38.9)\\
\multirow{2}{*}{SDC25.166} & 3.35     & 0.2014 ($\pm$0.0017)& 0.00 (-)                        & 6.1 ($\pm$0.4)& 1.59& 0.0166 ($\pm$0.0093)& 2.28 ($\pm$2.47)& 170.0 ($\pm$72.0)\\
                        & 3.02     & 0.1989 ($\pm$0.0020)& -                              & 7.2 ($\pm$0.5)& 1.11& 0.0130 ($\pm$0.0018)& -& 198.7 ($\pm$22.1)\\
\multirow{2}{*}{SDC28.333} & 5.17     & 0.1827 ($\pm$0.0043)& 1.85 ($\pm$1.89)                & 45.0 ($\pm$3.2)& 1.35& 0.0179 ($\pm$0.0004)& 0.00 (-)& 10.6 ($\pm$4.7)\\
                        & 3.51     & 0.1810 ($\pm$0.0024)& -                              & 46.2 ($\pm$2.1)& 0.67& 0.0179 ($\pm$0.0004)& -& 10.6 ($\pm$4.7)\\
\multirow{2}{*}{SDC34.370} & 2.92& 0.1525 ($\pm$0.0052)&  1.70 ($\pm$2.39)&  86.3 ($\pm$4.8)& 0.17& 0.0243 ($\pm$0.0016)& 2.59 ($\pm$0.30)& 148.6 ($\pm$9.9)\\
                        & 1.24&  0.1474 ($\pm$0.0035)& -                              &  93.8 ($\pm$5.2)& 2.24& 0.0156 ($\pm$0.0008)& -& 238.3 ($\pm$12.7)\\
\multirow{2}{*}{SDC35.527} & 4.20     & 0.2154 ($\pm$0.0026)& 0.00 (-)                        & 11.4 ($\pm$0.7)& 32.62& 0.0838 ($\pm$0.0395)& 4.18 ($\pm$1.23)& 0.0 ($\pm$241.9)\\
                        & 3.78     & 0.2154 ($\pm$0.0026)& -                              & 11.4 ($\pm$0.7)& 7.21& 0.0211 ($\pm$0.0020)& -& 607.1 ($\pm$32.7)\\
\multirow{2}{*}{SDC35.745} & 3.51     & 0.3419 ($\pm$0.0036)& 0.00 (-)                        & 3.2 ($\pm$0.8)& 5.49& 0.0235 ($\pm$0.0008)& 0.00 (-)& 77.9 ($\pm$8.7)\\
                        & 3.22     & 0.3419 ($\pm$0.0036)& -                              & 3.2 ($\pm$0.8)& 0.69& 0.0278 ($\pm$0.0017)& -& 0.0 ($\pm$27.9)\\
\multirow{2}{*}{SDC40.283} & 2.63     & 0.2600 ($\pm$0.0059)& 0.00 (-)                        & 40.4 ($\pm$4.5)& 1.01& 0.0191 ($\pm$0.0060)& 2.66 ($\pm$1.07)& 0.0 ($\pm$48.3)\\
                        & 1.97     & 0.2600 ($\pm$0.0059)& -                              & 40.4 ($\pm$4.5)& 0.41& 0.0115 ($\pm$0.0009)& -& 81.7 ($\pm$14.6)\\ \bottomrule
\end{tabular}
\end{table*}

Table \ref{tab:ADF} shows the results obtained from fitting the ADF for our sources on both clump and cloud scales. As expected, we can see that the values of $\delta$ that are fitted are either 0 or too large to be realistic. When fitting the ADF keeping $\delta$ fixed at 0, the values obtained for $\frac{1}{N}\frac{\langle B_t ^2 \rangle}{\langle B_o^2 \rangle}$ and $a$ usually stay similar to when $\delta$ was also fitted. The decrease in $\chi^2$ that is seen when keeping $\delta$ fixed at 0 compared to when $\delta$ is fitted is due to the use of the number of degrees of freedom, $DoF = n-m$,  in the denominator, where $n$ is the number of fitted points and $m$ is the number of parameters being fitted. When fitting $\frac{\langle B_t ^2 \rangle}{\langle B_o^2 \rangle}$, $\delta$, and $a$, $m$=3, but when $\delta$ is fixed at 0, $m$=2.

There is also an ambiguity as to how exactly the fit is carried out, and the results will strongly depend on factors such as for which range of lags the fit is carried out, binning of the lags, initial restrictions on fitted parameters, etc. This means that even when using the same data, it is likely that different studies may not obtain the same values for the fitted parameters.

The reduced-$\chi^2$ values in columns 2 and 6 of Table \ref{tab:ADF} show that the fits tend to be poor, especially the fit for {\it Planck} for SDC35.527. This exemplifies how difficult it is to gain an accurate estimate of the magnetic field strengths in these regions. While we refrain from reading too much into the results from the ADF fitting due to the uncertainties in their reliability, we can note that $\frac{1}{N}\frac{\langle B_t ^2 \rangle}{\langle B_o^2 \rangle}$ on average decreases by a factor of $\sim$10 from clump to cloud scales, with the ratios being on the order of 0.2 on clump scale and 0.02 on cloud scale. Since this quantity reflects the dispersion of magnetic field pseudovector angles, this is expected from what we can see in the distributions in Figure \ref{fig:pseudovectorhists} and our observations that the {\it Planck} $B$-field tends to align strongly with the Galactic plane, while the POL-2 $B$-field varies much more.

The turbulent-to-total magnetic field strength ratio $\frac{\langle B_t ^2 \rangle}{\langle B_{\rm{POS}}^2 \rangle}$ can be calculated using:

\begin{equation}
\frac{\langle B_t ^2 \rangle}{\langle B_{\rm{POS}}^2 \rangle} = \frac{1}{1+\left(\frac{\langle B_t ^2 \rangle}{\langle B_o^2 \rangle}\right)^{-1}}
\end{equation}

According to \citet{2009ApJ...706.1504H}, the number of turbulent cells along the LOS, $N$, should be calculated using the equation 

\begin{equation}
N = \frac{\left(\delta^2+2W^2\right)\Delta'}{\sqrt{2\pi}\delta^3}
\label{eq:adfN}
\end{equation}
where $\Delta'$ is the LOS depth of the cloud. Another method to calculate the number of turbulent cells along the LOS, $N_{\rm{LOS}}$, was presented by \citet{choTECHNIQUECONSTRAININGDRIVING2016} - this method estimates $N_{\rm{LOS}} = (\frac{\delta v}{\delta V_c})^2$, where $\delta v$ is the turbulent line-of-sight velocity dispersion and $\delta V_c$ is the dispersion of centroid velocities within the cloud. Using the \citet{choTECHNIQUECONSTRAININGDRIVING2016} method - which we refer to hereafter as CY16 - we obtain the results presented in columns 2 and 3 of Table \ref{tab:Nvals}. We use the velocity dispersions from \citet{2023MNRAS.525.2935P} for $\delta v$, and the standard deviations of the Gaussian fits to the distribution of velocity centroids gained from spectral fits to the molecular line data for $\delta V_c$. For the $^{13}$CO(1-0) data we first smooth the data cube to {\it Planck} resolution before carrying out the spectral fits. The N$_{2}$H$^{+}$(1-0) data already has a spatial resolution greater than that of the JCMT POL-2 data; in this case $\delta V_c$ may be smaller than it would be at POL-2 resolution, and so the estimates of $N_{\rm{LOS}}$ can be seen as upper limits. This means that the corresponding values of $B_{\rm{POS}}$ on clump scale are lower limits. We do also note that the CY16 method only accounts for the number of turbulent cells along the LOS, and not for the number of cells smoothed in the POS. We also calculate lower limits on $N$ using the Hou09 method, assuming that the turbulent correlation length is not resolved by the polarisation data i.e. $\delta < 2\sqrt{2\rm{ln}(2)}W$, by setting $\delta$ equal to the resolution of the data and calculating $N$ using Equation (\ref{eq:adfN}). We set the LOS depth of the cloud $\Delta'$ to the radius estimates from \citet{2023MNRAS.525.2935P}, where we use the radius rather than twice this number due to Hou09 defining $\Delta'$ as the distance corresponding to the half-maximum of the autocorrelation of the polarised flux. Since the \citet{2023MNRAS.525.2935P} radius values are calculated assuming a spherical geometry, it is possible that they still overestimate $\Delta'$, especially for more elongated structures, but they are suitable as a rough estimate. 

A number of studies have used the ADF to estimate $\delta$ in star-forming regions. The original application of the ADF method in \citet{2009ApJ...706.1504H} led to a value for $\delta$ of 16mpc in OMC-1. Studies after that have found values ranging from 10s to 100s of mpc \citep[e.g.][]{2022ApJ...925...30L}. However, most of these studies were carried out on regions of core scale and below. \citet{2022FrASS...9.3556L} find that the turbulent correlation length derived from the ADF tends to have a strong correlation with the size of the telescope beam. They propose that this means that the magnetic field is correlated with turbulence at multiple scales, and the measured values of $\delta$ gained from these fits are limited by factors such as the beam resolution, the maximum spatial scale recovered by the observations, and the size of the cloud. They also note that this correlation could be due to higher densities reducing the mean free path, and therefore the turbulent correlation length. In practice, the turbulent correlation length is expected to have a lower limit set by the ambipolar diffusion scale \citep{2007ARA&A..45..565M}, which also scales with density, but empirical values for the ambipolar diffusion scale are also in contention \citep{2022FrASS...9.3556L}. We therefore do not have a robust idea of how exactly $\delta$ may vary from IRDC to GMC scale. 

The results of the Hou09 method for the calculation of $N$ are shown in columns 4 and 5 of Table \ref{tab:Nvals}. We can see that these values are almost all lower than those from the CY16 method as expected, with the exception of $N_{\rm clump}$ for SDC34.370. $N_{\rm cloud}$ from the Hou09 method is also consistently lower than the corresponding $N_{\rm clump}$ due to using $\delta$ and $W$ on cloud scale that is $\sim20$ times the $\delta$ and $W$ we use on clump scale, while the factor between $\Delta'$ on these two scales is considerably less. Due to the many uncertainties involved in the estimation of $N$ using the Hou09 method we choose to use the CY16 correction to calculate $B_{\rm POS}$.

\begin{table}
\centering
\caption{Values of $N$ obtained from the CY16 method (columns 2 and 3), and lower limit values of $N$ obtained from the Hou09 method (columns 4 and 5).}
\label{tab:Nvals}
\begin{tabular}{@{}c>{\centering\arraybackslash}p{0.15\linewidth}>{\centering\arraybackslash}p{0.15\linewidth}@{}>{\centering\arraybackslash}p{0.15\linewidth}>{\centering\arraybackslash}p{0.15\linewidth}}\toprule
 & \multicolumn{2}{c}{CY16} & \multicolumn{2}{c}{Hou09}\\\midrule

       Source& $N_{\rm{cloud}}$ & $N_{\rm{clump}}$  & $N_{\rm{cloud}}$ &$N_{\rm{clump}}$ \\  
SDC18.624 & 33.8& 5.1
 & 1.3&3.6
\\
SDC24.489 & 108.8& 7.0
 & 1.0&2.6
\\
SDC25.166 & 5.5& 5.7
 & 0.8&3.3
\\
SDC28.333 & 9.3& 5.5
 & 2.5&4.5
\\
SDC34.370  & 47.4& 1.9
 & 1.2&4.6
\\
SDC35.527 & 3.4& 5.0
 & 1.0&3.5
\\
SDC35.745 & 5.7& 6.2
 & 0.8&3.0
\\
SDC40.283 & 1.5& 3.5 & 0.8&2.8
\\ \bottomrule 
\end{tabular}
\end{table}

\begin{table*}
    \centering
\caption{$B$-field strengths from the classical DCF equation (equation \ref{eq:dcfsa}), the original DCF equation not assuming the small angle approximation (equation \ref{eq:dcfsin}), the ST equation (equation \ref{eq:ST}), and the ADF (equation \ref{eq:adfbpos}). Where two values are displayed, both the standard deviation and MAD have been used (see text). The bottom row gives the average values of the $B$-field strengths on cloud and clump scale (calculated from the standard deviation measures of angle dispersion). No correction factors have been applied for any of the methods here.}
\label{tab:uncorrectedBpos}
    \begin{tabular}{ccccccccc}\toprule
         &  \multicolumn{2}{c}{Classical DCF}&  \multicolumn{2}{c}{sin-DCF}&  \multicolumn{2}{c}{ST}&  \multicolumn{2}{c}{ADF}\\\midrule
         Source&  $B_{\rm{POS}}^{\rm{cloud}}$ ($\mu$G) &  $B_{\rm{POS}}^{\rm{clump}}$ ($\mu$G) &  $B_{\rm{POS}}^{\rm{cloud}}$ ($\mu$G) &  $B_{\rm{POS}}^{\rm{clump}}$ ($\mu$G) &  $B_{\rm{POS}}^{\rm{cloud}}$ ($\mu$G) &  $B_{\rm{POS}}^{\rm{clump}}$ ($\mu$G) &  $B_{\rm{POS}}^{\rm{cloud}}$ ($\mu$G) &  $B_{\rm{POS}}^{\rm{clump}}$ ($\mu$G) \\
         SDC18.624&  709, 620&  45, 52&  700, 618&  -, 160&  157, 147&  29, 31&  477&  89
\\
         SDC24.489&  -, -&  81&  -, -&  406, 247&  -, -&  40&  201&  102
\\
         SDC25.166&  155&  88&  137&  1271, 406&  37&  41&  140&  92
\\
         SDC28.333&  377&  52, 61&  373&  109, 127&  92&  38, 41&  336&  141
\\
         SDC34.370&  254&  42, 64&  259&  195&  87&  37, 46&  391&  182\\
         SDC35.527&  602, 217&  47&  565, 216&  144, 125&  80, 48&  25&  77&  64
\\
         SDC35.745&  211&  29, 33&  215&  514, 122&  63&  22, 23&  247&  63
\\
         SDC40.283&  100&  43, 62&  99&  105, 119&  24&  34, 41&  83&  120
\\ 
 \textbf{Average}& \textbf{344}& \textbf{54}& \textbf{335}& \textbf{392}& \textbf{77}& \textbf{33}& \textbf{244}& \textbf{107}\\ \bottomrule
    \end{tabular}
\end{table*}

\begin{table}
\centering
\caption{$B$-field strengths from the ADF, corrected for line-of-sight integration with the estimates of $N$ calculated using the CY16 method.}
\label{tab:ADFcorr}
\begin{tabular}{@{}ccc@{}}
\toprule
       Source& $B_{\rm{POS}}^{\rm{cloud}}$ ($\mu$G) & $B_{\rm{POS}}^{\rm{clump}}$ ($\mu$G) \\ \midrule
SDC18.624& 107& 52
\\
SDC24.489& 23& 53
\\
SDC25.166& 62& 52
\\
SDC28.333& 118& 78
\\
SDC34.370& 82& 139\\
SDC35.527& 45& 37
\\
SDC35.745& 108& 38
\\
SDC40.283& 69& 79\\ 
 \textbf{Average}& \textbf{77}&\textbf{66}\\ \bottomrule
\end{tabular}
\end{table}

The POS magnetic field strength can be calculated from the ADF fitted parameters by replacing $\delta \theta_B$ in the original DCF equation by $\left(\frac{\langle B_t ^2 \rangle}{\langle B_{\rm{POS}}^2 \rangle}\right)^{\frac{1}{2}}$:

\begin{equation}
B_{\rm{POS}} =  \sqrt{4 \pi \rho}\delta v \left({\frac{\langle B_t ^2 \rangle}{\langle B_{\rm{POS}}^2 \rangle}}\right)^{-\frac{1}{2}}
\label{eq:adfbpos}
\end{equation}

In columns 8 and 9 of Table \ref{tab:uncorrectedBpos} we present the $B_{\rm{POS}}$ values derived from Equation~(\ref{eq:adfbpos}) using $N=1$. Here, the $B_{\rm{POS}}$ on clump scale is consistently higher than those obtained from the original DCF and ST methods, but tends to be on par with, or lower than, those from the sin-DCF method. The results on cloud scale are more mixed; the $B_{\rm{POS}}$ on cloud scale using the ADF method significantly varies between clouds, but is also consistently higher than those derived from the ST method. The $B$-field strengths presented in Table \ref{tab:ADFcorr}, now corrected for line-of-sight integration using the CY16 $N_{\rm{LOS}}$ values, also show a broad range of values on cloud scale, but all of the values are now lower than those from the original DCF and sin-DCF methods and tend to be similar to those from the ST method. On clump scale though $B_{\rm{POS}}$ values derived from the classical DCF, ST, and corrected ADF methods  generally agree within a factor of $\sim$2.

\subsection{Reliability of $B$-field strength estimates}

\begin{figure}
	\includegraphics[width=\columnwidth]{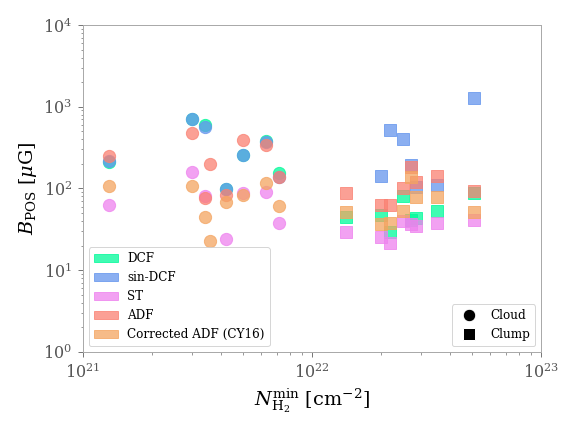}
	\caption{$B_{\rm POS}$ values calculated for our sample using the classical DCF, sin-DCF, ST, ADF, and corrected ADF (with the CY16 $N_{\rm LOS}$ values) methods, against the H$_2$ column density of the bounding contour of each clump or cloud.}
    \label{fig:B-N}
\end{figure}

As we have seen, each method requires choices to be made - often dictated by observational limitations - on how certain parameters are best estimated. The impact of those choices can be as large as an order of magnitude on the estimated $B$-field strength. For instance, the estimation of the cloud/clump volume density, one of parameters of the DCF equation, is often poorly constrained, requiring multiple approximations and assumptions as to the 3D structure of the object being investigated based on POS observations; these have the potential to be extremely inaccurate. All of these methods also result in an `average' field strength across the region to which they are applied, and usually require the concurrent use of quantities that may be resolved to varying degrees, adding an unknown ambiguity into the estimation of the magnetic field strength. In addition, several of the assumptions that are made in the derivation of the $B$-field strength equations do not apply to the clouds and scales we are investigating. More importantly, across all methods we trialled the magnetic field strength on clump scale is consistently lower than the $B$-field strength on cloud scale, unlike what one expects in the case of magnetic flux conservation. Figure \ref{fig:B-N} shows that in general, the relationship between our calculated $B_{\rm POS}$ values and H$_2$ column density is largely flat. The alignment of the magnetic field with respect to the LOS may be affecting these estimates, especially if this changes greatly between cloud and clump scales, but once again without any knowledge of the LOS component of the magnetic field we cannot quantify to what extent this plays a role. Multiple studies \citep[e.g.][]{2018MNRAS.474.5122K, 2025A&A...700A.256P} have investigated how projection angle changes the quantities that we are able to observe using ideal-MHD simulations, with a common finding being that as the angle between the LOS and the mean magnetic field direction decreases, the observed angular dispersion increases, sometimes by up to an order of magnitude. This would directly translate into a potential order of magnitude variation in observed $B_{\rm POS}$ values simply due to the relative angle between the LOS and magnetic field, not even taking into account the other factors that may affect the inferred values of $B_{\rm POS}$ from observations.

While all of the above caveats likely play a significant role in shifting the calculated $B$-field strengths from their `true' values, we must also consider the errors in the quantities used to calculate these strengths - namely $M$, $R$, and $\delta v$. We summarise the approximate errors in these values in Table \ref{tab:errors}, which are taken from \citet{2023MNRAS.525.2935P}. These are upper-bound values, to ensure that we do not underestimate the error. We calculate the 68\% confidence interval of $B_{\rm POS}$ values that would be obtained by incorporating these uncertainties along with the uncertainties in $\delta \theta_B$, $\delta \sin (\theta_B)$, or $\frac{\langle B_t ^2 \rangle}{\langle B_{\rm{POS}}^2 \rangle}$ as appropriate from the fittings. The lower and upper limits of the 68\% confidence interval of $B_{\rm POS}$ after taking all of these errors into account are shown in Table \ref{tab:Bposranges}. It can be seen that even after factoring in the errors in the intrinsic quantities, this does not solve the issue of the $B$-field strength in the clouds tending to be greater than those in their corresponding clumps. This strongly suggests that the derived $B$-field strengths are just not reliable. We discuss this further in Section \ref{sec:energybalance}.

\begin{table}
    \centering
\caption{Approximate errors in the intrinsic quantities used to calculate $B$-field strengths.}
\label{tab:errors}
    \begin{tabular}{c>{\centering\arraybackslash}p{0.25\linewidth}cc}\toprule
         &  $M$&  $R$& $\delta v$\\\midrule
         Cloud&  -25\%, +50\%&  $\pm$20\%& $\pm$30\%\\
         Clump&  $\sim \pm$ 10 to 25\% \tablefootnote{See Table \ref{tab:SamplePropertiesTable}}&  $\pm$20\%& $\pm$10\%\\ \bottomrule
    \end{tabular}

\end{table}

\begin{table*}
    \centering
        \caption{Lower (first number) and upper (second number) bounds of the values of $B_{\rm POS}$ presented in Tables \ref{tab:uncorrectedBpos} (columns 2-9) and \ref{tab:ADFcorr} (columns 10 and 11) after taking into account the errors in the quantities used to calculate $B_{\rm POS}$. These bounds represent the 64\% confidence interval around the applicable value. For the classical DCF, sin-DCF, and ST methods, these ranges are only applicable to the $B_{\rm POS}$ values calculated from the standard deviation measure of angle dispersion.}
    \begin{tabular}{>{\centering\arraybackslash}p{0.065\linewidth}>{\centering\arraybackslash}p{0.065\linewidth}>{\centering\arraybackslash}p{0.065\linewidth}>{\centering\arraybackslash}p{0.065\linewidth}>{\centering\arraybackslash}p{0.065\linewidth}>{\centering\arraybackslash}p{0.065\linewidth}>{\centering\arraybackslash}p{0.065\linewidth}>{\centering\arraybackslash}p{0.065\linewidth}>{\centering\arraybackslash}p{0.065\linewidth}>{\centering\arraybackslash}p{0.065\linewidth}>{\centering\arraybackslash}p{0.065\linewidth}}\toprule
         &  \multicolumn{2}{c}{Classical DCF}&  \multicolumn{2}{c}{sin-DCF}& \multicolumn{2}{c}{ST}& \multicolumn{2}{c}{ADF}&\multicolumn{2}{c}{Corrected ADF (CY16)}\\\midrule
         Source&  $B_{\rm{POS}}^{\rm{cloud}}$ ($\mu$G) &  $B_{\rm{POS}}^{\rm{clump}}$ ($\mu$G)&  $B_{\rm{POS}}^{\rm{cloud}}$ ($\mu$G) &  $B_{\rm{POS}}^{\rm{clump}}$ ($\mu$G)& $B_{\rm{POS}}^{\rm{cloud}}$ ($\mu$G) & $B_{\rm{POS}}^{\rm{clump}}$ ($\mu$G)& $B_{\rm{POS}}^{\rm{cloud}}$ ($\mu$G) & $B_{\rm{POS}}^{\rm{clump}}$ ($\mu$G)&$B_{\rm{POS}}^{\rm{cloud}}$ ($\mu$G) & $B_{\rm{POS}}^{\rm{clump}}$ ($\mu$G)\\
         SDC18.624&  409, 1081&  33, 67&  404, 1072&  -, -& 91, 242& 22, 43& 282, 745& 66, 130&70, 150& 39, 76\\
         SDC24.489&  -, -&  58, 117&  -, -&  288, 612& -, -& 29, 57& 99, 356& 75, 149&12, 35& 39, 76\\
         SDC25.166&  90, 238&  64, 126&  78, 212&  843, 2010& 21, 57& 30, 58& 70, 238& 69, 131&37, 95& 39, 72\\
         SDC28.333&  215, 569&  38, 75&  220, 571&  74, 165& 52, 140& 28, 54& 196, 516& 105, 200&79, 162& 58, 109\\
         SDC34.370&  146, 396&  29, 64&  147, 404&  142, 279& 51, 134& 27, 54& 230, 600& 137, 268&53, 116& 106, 196\\
         SDC35.527&  348, 948&  34, 72&  314, 893&  103, 221& 47, 123& 18, 37& 41, 126& 47, 97&27, 66& 28, 57\\
         SDC35.745&  120, 325&  20, 42&  121, 334&  328, 803& 37, 98& 15, 30& 140, 375& 46, 87&74, 150& 28, 52\\ 
 SDC40.283
& 59, 155& 28, 67& 56, 150& 73, 153& 13, 36& 24, 49& 45, 130& 89, 170& 43, 98&60, 112\\
 \textbf{Average}& \textbf{198, 530}& \textbf{38, 79}& \textbf{191, 519}& \textbf{264, 606}& \textbf{45, 119}& \textbf{24, 48}& \textbf{138, 386}& \textbf{79, 154}& \textbf{49, 109}&\textbf{50, 94}\\ \bottomrule
    \end{tabular}
    \label{tab:Bposranges}
\end{table*}

\section{Energy balance}
\label{sec:energybalance}

Quantifying the energy balance between gravity, turbulence, and magnetic fields is central to star formation studies so that their respective importance can be assessed. As discussed in the previous section, the estimates of the magnetic field strength for our sample - as for any $B$-field strength derived from polarised emission - are uncertain, given the multiple challenges and caveats. However, because similar measurements are routinely  presented in star formation literature, their analysis can give us an important point of comparison with published work, along with providing us with a different perspective on the role of the magnetic field from that obtained from the morphological analysis presented in Section~\ref{sec:magfielddirection}. The magnetic field strength values that we use in this section are those in Table \ref{tab:ADFcorr}. In the following, we present the picture of the relative importance of gravity, turbulence, and magnetic field that we would infer from an analysis using these derived magnetic field strength values.

The quantities that we compute are the gravitational energy density $u_G$, the magnetic pressure $P_B$, and the turbulent pressure $P_T$, as well as the magnetic virial parameter $\alpha_{\rm{mag}}=\frac{2E_{\rm{K}}+E_{\rm{B}}}{|E_{\rm{G}}|}$, where $E_{\rm{K}}$, $E_{\rm{B}}$, and $E_{\rm{G}}$ are the kinetic, magnetic, and gravitational energies respectively, which quantifies the relative importance of magnetic and kinetic forces compared to gravitational forces.

$u_G$ is calculated using the equation:
\begin{equation}
u_G = \frac{9}{20 \pi} G \frac{M^2}{R^4}
\label{eq:u_G}
\end{equation}
where, following \citet{2023MNRAS.525.2935P}, we make the assumption of spherical geometry and uniform density. 

$P_B$ is calculated as:
\begin{equation}
P_B = \frac{1}{8\pi} B_{\rm tot}^2
\end{equation}
where $B_{\rm tot}$, the total magnetic field strength, is calculated as $B_{\rm tot}=\frac{4}{\pi}B_{\rm POS}$ using the statistical correction factor from \citet{2004ApJ...600..279C}.

The equation for $P_T$ is:
\begin{equation}
P_T = \frac{3}{2} \rho \delta v^2
\label{eq:P_T}
\end{equation}
assuming isotropic turbulence.

$\alpha_{\rm{mag}}$ is calculated using:

\begin{equation}
\alpha_{\rm{mag}} = \alpha_{\rm vir} + \frac{5R}{GM} \frac{v_A^2}{6}
\label{eq:alpha_mag}
\end{equation}
where \begin{math}v_A = \frac{B_{\rm tot}}{\sqrt{4 \pi \rho}}\end{math} is the Alfv\'en velocity.  

\begin{figure*}
	\includegraphics[width=\textwidth]{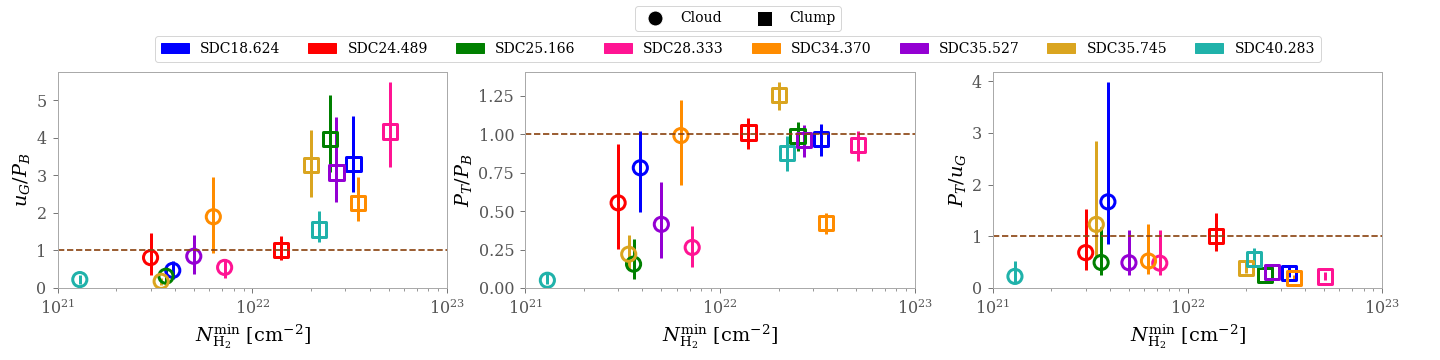}
	\caption{Ratios between the gravitational, turbulent, and magnetic energy densities in the clumps and clouds. Horizontal dashed brown lines mark where each of the quantities is equal to 1. The $x$-axis of each plot is the $\rm{H_2}$ column density of the bounding contour of each clump or cloud. Errorbars show the 68\% confidence interval based on the uncertainties in the intrinsic values used to calculate each ratio.}
    \label{fig:energybalance}
\end{figure*}

In Figure \ref{fig:energybalance} we show the ratios between $u_G$, $P_B$, and $P_T$ as a function of the minimum, bounding column density of  our clouds and clumps. Errorbars are calculated using the uncertainty estimates in $M$, $R$, and $\delta v$ in Table \ref{tab:errors}, in conjunction with the best fit uncertainties on $\frac{1}{N}\frac{\langle B_t ^2 \rangle}{\langle B_o^2 \rangle}$ presented in Table \ref{tab:ADF}.

$\frac{u_G}{P_B}$ (Figure~\ref{fig:energybalance} left panel) shows that the clouds are usually magnetically sub- to trans- critical, with the exception of the parent molecular cloud of SDC34.370 which is trans- to supercritical given the measurement uncertainties. The importance of gravity increases with column density, with almost all of the clumps being magnetically super-critical. Only SDC24.489 is trans-critical according to $\frac{u_G}{P_B}$. 

$\frac{P_T}{P_B}$ (Figure~\ref{fig:energybalance} middle panel) shows that the clumps are mainly trans-Alfv\'enic, with almost all of the clouds being sub-Alfv\'enic. While the observed decreased importance of the magnetic field on clump scale is consistent with the difference in $B$-field morphologies between the clouds and their clumps (see Figure~\ref{fig:smoothedresults}), it is likely that this trend is exacerbated as a result of systematic overestimation of $B$-field strength on cloud scale - due to the often low number of independent measurements of the magnetic field in our cloud sample - and underestimation on clump scale. \citet{2026A&A...706A..60T} have carried out non-ideal MHD simulations of a collapsing cloud and calculated `observed’ values of $B_{\rm POS}$, which they compared to the `true' values from their simulations. They show that even when the POS $B$-field is optimally observed - i.e. the viewing LOS is perpendicular to the initial $B$-field direction - the $B$-$N$ relation is almost flat, not displaying the expected increase present in their simulation. This is a direct artefact of projection effects and mismatch between quantities that are measured along the LOS (i.e. the column density) and the POS (i.e. the $B$-field strength). These results from \citet{2026A&A...706A..60T} are in agreement with the nearly flat $B$-field strength trend we observe between our sample of clumps and clouds (see Figure \ref{fig:B-N}). 

$\frac{P_T}{u_G}$ (Figure~\ref{fig:energybalance} right panel) does not involve the $B$-field strength and is therefore not affected by $B-$field measurements. This ratio is analogous to the virial ratio, except that it measures specifically the ratio of turbulent to gravitational energies rather than overall kinetic energy to gravitational energy. It can be seen that a similar picture is drawn from $\frac{P_T}{u_G}$ and the virial ratios. 

\begin{figure}
	\includegraphics[width=\columnwidth]{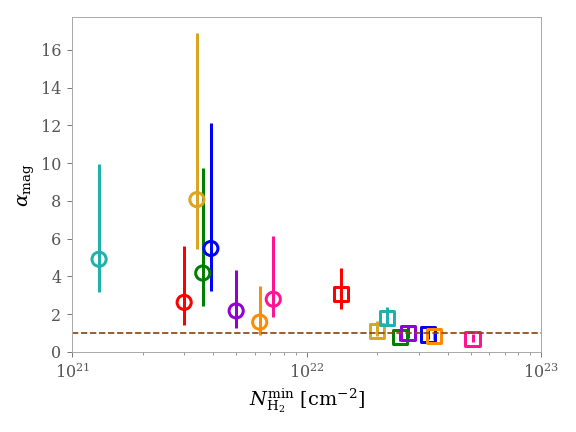}
	\caption{The magnetic virial parameters of the clumps and clouds. A horizontal dashed brown line demarcates $a_{\rm mag} = 1$. The $x$-axis of each plot is the $\rm{H_2}$ column density of the bounding contour of each clump or cloud. The colour scheme is the same as that for Figure \ref{fig:energybalance}.}
    \label{fig:a_mag}
\end{figure}

The $\alpha_{\rm{mag}}$ ratio adds the contribution of the magnetic field support to the virial parameters and thus includes contributions from all three energy sources. Figure \ref{fig:a_mag} shows that all of the clouds and some of the clumps are now in the stability regime, but with the majority of the clumps still being close to, or slightly below, the boundary for stability. If we were to take all of these results at face value, the overwhelming picture would appear to be that molecular clouds are supported against collapse by magnetic fields, but on clump scales gravity and turbulence start to take over, pushing the clumps towards collapse.

\begin{table}
\centering
\caption{Values of $r_B^{\rm{low}}$,  $r_B^{\rm{obs}}$, and $F$ (see text) for each source.}
\label{tab:Bfactor}
\begin{tabular}{@{}ccc@{}>{\centering\arraybackslash}p{0.2\linewidth}}
\toprule
       Source& $r_B^{\rm{low}}$& $r_B^{\rm{obs}}$ &$F$\\ \midrule
SDC18.624& 0.78          & 2.06&$2.1-2.6$
\\
SDC24.489& 0.39& 0.44&$1.1, 2.3$
\\
SDC25.166& 0.34          & 1.20&$1.2-3.5$
\\
SDC28.333& 0.55          & 1.51&$1.5-2.7$
\\
SDC34.370& 0.54          & 0.59&$1.1, 1.7$
\\
SDC35.527& 0.64          & 1.22&$1.2-1.9$
\\
SDC35.745& 0.66          & 2.83&$2.8-4.3$
\\
SDC40.283& 0.33& 0.87&$2.6, 1.1$\\ \bottomrule
\end{tabular}
\end{table}

Despite large uncertainties on the magnetic field strength estimates, the observed trend $\left(\frac{u_G}{P_B}\right)_{\rm cloud} < \left(\frac{u_G}{P_B}\right)_{\rm clump}$ seen in  Figure~\ref{fig:energybalance} is compatible with the idea of a dynamical decoupling of the clumps triggered by collapse. This condition sets a lower limit $r_B^{\rm{low}}$ on the $\frac{B_{\rm{cloud}}}{B_{\rm{clump}}}$ ratio, where $B_{\rm cloud}$ is the magnetic field strength on cloud scale and $B_{\rm clump}$ is the magnetic field strength on clump scale, given by:

\begin{equation}
\left(\frac{B_{\rm{cloud}}}{B_{\rm{clump}}}\right) > r_B^{\rm{low}} = \frac{M_{\rm{cloud}}}{M_{\rm{clump}}}\left(\frac{R_{\rm{clump}}}{R_{\rm{cloud}}}\right)^2
\label{rlow}
\end{equation}

However, we also know that, due to magnetic flux conservation, we must have:
\begin{equation}
\left(\frac{B_{\rm{cloud}}}{B_{\rm{clump}}}\right)<1
\label{rup}
\end{equation}

By comparing the observed ratio $r_B^{\rm{obs}}=\left(\frac{B_{\rm{cloud}}}{B_{\rm{clump}}}\right)_{\rm{obs}}$ - based on the values from the corrected ADF method in Table \ref{tab:ADFcorr} -  to the range provided by the two conditions in Equations (\ref{rlow}) and (\ref{rup}), we can quantify the range of systematic error made in the estimate of this ratio.   Table \ref{tab:Bfactor} shows the values of $r_B^{\rm{low}}$, $r_B^{\rm{obs}}$, and $F =\left[r_B^{\rm{obs}},\frac{r_B^{\rm{obs}}}{r_B^{\rm{low}}}\right]$, i.e. the over-estimation factor range. 

For sources that already have a $r_B^{\rm{obs}}$ between $r_B^{\rm{low}}$ and 1 (SDC24.489, SDC34.370, and SDC40.283), the first number listed for $F$ in Table \ref{tab:Bfactor} is the maximum potential overestimation factor based on $r_B^{\rm{low}}$, and the second number is the largest factor by which $r_B^{\rm{obs}}$ is potentially underestimated based on the upper limit of 1. For all other sources the values given are the range of factors by which $r_B^{\rm{obs}}$ is potentially overestimated given the above assumptions. The values in column 4 of Table \ref{tab:Bfactor} show that $\frac{B_{\rm{cloud}}}{B_{\rm{clump}}}$ appears to be erroneous by a factor on the order of $\sim2$, with the maximum overestimation factor being 4.3 for SDC35.745. Of course, if the assumptions that we have made do not hold, the values of $F$ in Table \ref{tab:Bfactor} are not valid.

\section{Discussion}
\label{sec:discussion}

This research presents an investigation into the relationship between the magnetic field morphology on cloud ($\sim$10~pc) and clump ($\sim$1~pc) scales in molecular clouds in the Milky Way, applying a consistent analysis across multiple independent case studies with differing properties. It follows a large amount of single-case study research into the properties of the magnetic field on different scales in star-forming regions within our Galaxy.  

\subsection{Comparison with previous literature}

\subsubsection{Other studies on the sources in our sample}

Some of the sources in our sample have already been analysed in previous studies. SDC18.624 was recently the subject of a paper by \citet{2025A&A...696A.163L}, whose JCMT POL-2 data we also use here. At clump scale, they find that gravity dominates over turbulence but both are negligible compared to the magnetic field. Since they do not study the large-scale magnetic field of the cloud we cannot compare the results on this scale. \citet{2025A&A...696A.163L} derive a POS magnetic field strength of 110$\mu G$ in the clump with the ADF method, which is two times higher than the 52$\mu G$ that we derive using the same method. There are however multiple differences in the way that we apply the method. 
Firstly, the \citet{2025A&A...696A.163L} study uses a value of $N$=0.48 from the Hou09 method to correct their $\frac{1}{N}\frac{\langle B_t ^2 \rangle}{\langle B_o^2 \rangle}$ obtained from the ADF fit, while we use $N_{\rm{LOS}}$=5.1 from the CY16 method. This means that their uncorrected ADF value is $\sim$76$\mu$G, while ours is 89$\mu$G, which are not largely different values. There are also differences in the way that we fit the ADF, the main one being that while they have not mentioned what range of lags they fit the ADF over, it appears that they may have fit the ADF across the whole range of lags. Meanwhile, we fit only above the beam size and use the reduced-chi squared value to decide how many points above this are included in the fit. This would account for the difference in fitted parameters that we obtain: 0.22, 0.48’’, and 27.6$\times$10$^{-6}$ arcsec$^{-2}$ compared to the 0.33, 24.0'', and 0.269$\times$10$^{-6}$ arcsec$^{-2}$ that they obtain for $\frac{1}{N}\frac{\langle B_t ^2 \rangle}{\langle B_o^2 \rangle}$, $\delta$, and $a$ respectively. In the final calculation of the magnetic field strength using the DCF equation, we also use different turbulent velocities and likely different densities. \citet{2025A&A...696A.163L} directly use the turbulent-to-ordered field strength in calculating $B_{\rm{POS}}$, while we convert this to a turbulent-to-total field strength fraction before substituting it into the equation. Taking into account all of these differences, it is not surprising that we obtain largely differing values for $B_{\rm{POS}}$, with the main difference stemming from the correction factor we use to account for beam smoothing.

\citet{2024ApJ...966..120L} recently conducted a multi-scale analysis of SDC28.333 using polarisation data from {\it Planck}, JCMT, and ALMA for cloud, clump, and core scale respectively. They average the {\it Planck} polarisation data within a circle of 10pc radius towards the IRDC, and the POL-2 polarisation data in circles of 1pc radius centred on high density regions in the clump, and recalculate a single angle for each of these regions. When they calculate the difference between the averaged cloud-scale angle and the angle in each of the clump-scale regions, they find that half of the clump scale regions’ averaged magnetic fields lie within 30$^\circ$ of the averaged cloud scale magnetic field, and the other half’s have a difference of 60$^\circ$-90$^\circ$. 
This is similar to what we see in Figure \ref{fig:SDC28.333vecthistsmoothed}, where the {\it Planck} histogram distribution peaks close to 0$^\circ$ and the smoothed POL-2 distribution displays two peaks - one close to -30$^\circ$ and one close to -60$^\circ$. Since we only consider the overall average direction within the clump, our average direction of the magnetic field within SDC28.333 is between these two regimes. They also find that from cloud to clump scale, the alignment of the magnetic field with the density structures transitions from preferentially parallel to more perpendicular, compatible with what we find in our study.
 \citet{2024ApJ...966..120L} also calculate magnetic field strengths from the {\it{Planck}} and POL-2 polarisation data, however they do so with {\it Planck} observations in a circle of 15pc radius, and their clump-scale observations are based on JCMT POL-2 measurements in four circular regions of radii 1pc centred on a selection of the high-density regions in the IRDC. For the cloud-scale magnetic field strength, \citet{2024ApJ...966..120L} use the original DCF method with two correction factors – the CY16 correction factor to account for turbulent cells along the LOS, and a factor of 0.5 following \citet{2001ApJ...546..980O}. From the former they obtain a $B_{\rm{POS}}$ of 74$\mu$G, and from the latter they obtain 270$\mu G$. Due to differences in the values of $\delta v$ and $\delta V_c$ between our study and theirs, our estimate of the number of turbulent cells along the line of sight from the CY16 method differs significantly from theirs – we calculated $N$=9.3 from the CY16 method, while they have $N$=53. If we disregard any correction factors, their value of $B_{\rm{POS}}$ would be $\sim$540$\mu$G, which is larger than the 377$\mu$G we obtain from the DCF method with no correction factors. This is perhaps due to \citet{2024ApJ...966..120L} finding a $\delta \theta_B$ of 3.9$^\circ$, which is smaller than our $\delta \theta_B$ = 6.8$^\circ$ - likely since we probe a much larger area with our {\it{Planck}} measurements – as well as differences in $\delta v$ and the density we substitute into the equation. 
For the clump-scale magnetic field, \citet{2024ApJ...966..120L} apply the ADF to their four regions, and once again use two correction factors – CY16 and a factor of 0.21 from \citet{liuCalibratingDavisChandrasekhar2021}. With the CY16 correction the $B$-field strengths obtained are in the range 39-81$\mu$G, while with the correction factor of 0.21 they obtain 75-154$\mu$G. Extrapolating back from this, their $B_{\rm{POS}}$ from the ADF without any correction factors would be in the range $\sim$335-735$\mu$G. These are all much higher than the 141$\mu$G we calculate. 
We obtain $\frac{\langle B_t ^2 \rangle}{\langle B_o^2 \rangle}$=0.18 (ignoring the $N$ correction factor) within the IRDC from our ADF fit, leading to $\frac{\langle B_t ^2 \rangle}{\langle B_{\rm POS}^2 \rangle}$=0.15. \citet{2024ApJ...966..120L} derive values in the range 0.05-0.10. This accounts for a factor difference of about 1.5-3 between our uncorrected ADF estimate and theirs. 
The much larger $B_{\rm{POS}}$ estimates in \citet{2024ApJ...966..120L} likely in this case also stem from the density estimates– their densities are more than a factor of two larger than ours – as well as the velocity dispersions– which are 1.3-1.9 times larger than the velocity dispersion of 1.42km s$^{-1}$ that we use. We note that \citet{2024ApJ...966..120L} have - similarly to our study - measured $B$-field strength on both cloud and clump scale in SDC28.333 and find the same trend that we find of similar magnetic field strengths on cloud and clump scale.
Overall, \citet{2024ApJ...966..120L} propose that the diffuse gas is magnetically subcritical, sub-Alfv\'enic, and super-virial, while at clump scales gravity dominates over the magnetic field, but magnetic fields and turbulence together are likely near equilibrium with gravity.

\citet{ 2025ApJ...985..222H} have also recently studied the magnetic fields of SDC28.333 with data from {\it Planck}, JCMT, and ALMA, although they concentrate mostly on a region in the most high-density part of the IRDC – the P2 clump. Rather than calculating averaged quantities across the whole region, they observe how parameters vary across the studied area. They estimate the magnetic field strengths across the P2 clump with POL-2 polarisation data using the DCF method with a correction factor of 0.5, which leads to $B_{\rm{POS}}$ in the range 96-772$\mu$G (mean=330$\mu$G). Doubling their mean $B_{\rm{POS}}$ to obtain their uncorrected DCF estimate, they obtain $\sim$660$\mu$G. This is more than ten times larger than the estimate that we obtain in the clump – 52$\mu$G. This is once again majorly due to the much larger variation in angles that we sample due to considering the IRDC as a whole; the average $\delta \theta_B$ that \citet{ 2025ApJ...985..222H} use in their estimation is 11$^\circ$, while ours is 60.4$^\circ$. As with all of these estimates, differences in the values estimated for $\rho$ and $\delta v$ will also play a role in obtaining differing values. Using $B_{\rm{POS}}$ estimates now corrected by a factor of 0.28 following \citet{liuCalibratingDavisChandrasekhar2021}, \citet{ 2025ApJ...985..222H} find that the P2 clump is overall magnetically supercritical, with gravity dominating over both the turbulence and magnetic field together. Turbulence also dominates over the magnetic field.

SDC34.370 has been the subject of numerous magnetic field studies. \citet{2019ApJ...883...95S} looks at the magnetic field in the clump using JCMT POL-2 data. They divide the clump into north, central, and south regions, noting that the field direction in the north part of the clump is almost parallel to the elongated clump, while in the central and south regions it is dominantly perpendicular. They also comment on the fact that the large-scale {\it Planck} field is parallel to the filament, similar to the magnetic field in the northern region of the clump. These findings are consistent with those found in \citet{2019ApJ...878...10T}, whose magnetic field morphology in the clump was derived from 350$\mu$m SHARP dust polarisation data. Our magnetic field morphology for SDC34.370 also agrees with these studies. 
\citet{2019ApJ...883...95S} find magnetic field strengths of 470$\mu G$, 100$\mu G$, and 60$\mu G$ in the central, northern and southern parts of the clump respectively using the DCF method corrected by a factor of 0.5. Comparing twice these values to our estimate of $B_{\rm{POS}}$ from the uncorrected DCF method, 42$\mu$G, we see that our value is much smaller. This would be due to the much larger $\delta \theta_B \sim 90^\circ$ that we use due to sampling across the whole clump - \citet{2019ApJ...883...95S} have $\delta \theta_B = 10.8^\circ$ in the central region of the IRDC. Our estimation of $\delta \theta_B$ is of course inappropriate in this case. \citet{2019ApJ...883...95S} also obtain a $B_{\rm{POS}}$ estimate of 90$\mu$G from the ADF method, where they use $N$=5.5. Their value not including correction for beam integration is 200$\mu$G. Our values from the same method - 139$\mu$G and 182$\mu$G respectively - are comparable, with the fact that we use $N$=1.9 to correct for beam integration effects accounting for the larger difference between our corrected ADF value and theirs. 
\citet{2019ApJ...878...10T} obtain 190$\mu G$, 340$\mu G$, and 60$\mu G$ in the 2pc-scale regions around the MM1, MM2, and MM3 cores in SDC34.370 respectively from the DCF method corrected with a factor of 0.5, where they estimate their $\delta \theta_B$ based on the fit to the polarisation dispersion functions at 10'', which allows separation of the large-scale and small-scale $B$-field. Their $\delta \theta_B$ values are 17.3$^\circ$, 9.0$^\circ$, and 20$^\circ$ for the MM1, MM2 and MM3 regions respectively. Similarly to \citet{2019ApJ...883...95S}, their smaller angular dispersions result in larger estimates of $B_{\rm{POS}}$ compared to ours.

\citet{ 2018ApJ...859..151L} studied SDC35.527 using JCMT POL-2 data - we use the same data here as well. They observe that in the less dense regions of the IRDC, the magnetic field is parallel to density structures, while in the denser central regions it is perpendicular. This is perhaps akin to what is seen in SDC34.370 from our observations as well. Their estimation of the POS magnetic field strength in the clump leads to a value of 50$\mu$G using the DCF method with a correction factor of 0.5. We obtain a value of 47$\mu$G without the use of a correction factor – our value is under two times less than theirs. This difference likely mainly stems from the fact that they use the dispersion in angles only from the central part of the clump where the field is ordered. This means that they obtain $\delta \theta_B \sim $15$^\circ$, while our $\delta \theta_B$ is 32.4$^\circ$. 
\citet{ 2018ApJ...859..151L} do note that their estimate should be taken as an upper limit, since the $B$-field varies more in the other regions of the clump that they did not include when calculating their angular dispersion, and also the resolution of the NH$_3$ (1, 1) molecular line data that they use to calculate their $\delta v$ is larger than that of the JCMT POL-2 data, which is also the case for our N$_2$H$^+$ data. In addition, they use an empirical relation from \citet{2010ApJ...725..466C} that was based on Zeeman observations to calculate the maximum total field strength, from which they obtain 64$\mu$G – similar to the 65$\mu$G they get from the DCF with a conversion factor of 1.3 to convert from POS to total magnetic field strength. According to their findings, SDC35.527 is likely to be unstable against gravity. Their upper limit value of 50$\mu$G could still be reconcilable with the 37$\mu$G we gain from the corrected ADF method, since ours is theoretically a lower limit. 

Overall, while the magnetic field morphology that we observe is unsurprisingly consistent with those presented in the studies mentioned here, the magnetic field strengths vary largely between our measurements and others. This is consistent with the results of our Section~\ref{sec:Bfieldstrength} that highlight how parameter and method dependent magnetic field strength measurements are. 

\subsubsection{Magnetic fields in star formation in a wider context}

Generally in observational studies that look at the magnetic field in star-forming regions, it is concluded that the importance of the magnetic field does vary between scales. As previously mentioned, where exactly this transition occurs is however debated. Multiple observational studies have found a change in the alignment of the magnetic field relative to density structures, which is thought to demarcate the change from a magnetically dominated regime to a gravitationally dominated one \citep[e.g.][]{2016A&A...586A.135P, 2016A&A...586A.138P, 2019A&A...629A..96S, 2019MNRAS.484.3604L, 2020MNRAS.494.1971C, 2021ApJ...918...39L, 2023ApJ...945..160L, 2023ApJ...945...34B, 2023sf2a.conf..313O, 2024A&A...686A.202J, 2024MNRAS.533.1938C}. This is generally observed to occur around $N_{\rm{H_2}} \sim 10^{22}$~cm$^{-2}$, although this value varies between studies and it has also been suggested that it may vary between regions. 

Other attempts to quantify the relative importance of observed magnetic fields take the form of calculating: the energy balance between gravitational, turbulent, and magnetic energy densities, the mass-to-flux ratio, the virial ratio, or the Alfv\'enic Mach number. As previously discussed, the conclusions of studies such as these on cloud and clump scales are mixed \citep[e.g.][]{2022ApJ...925...30L,2023ASPC..534..193P}, but tend to show similar trends to what we have found here, despite our magnetic field strength estimates being unreliable as clearly evidenced by the uniform or decreasing computed $B$-field strengths between cloud and clump scales. 
Note that measuring the $B$-field strength on both cloud and clump scale for a sample of Galactic plane sources has not been done before. Therefore this unreliability of $B$-field strength measurements, while often mentioned in the literature, is clearly revealed here for the first time. It is most likely that the same systematics are present in any $B$-field strength measurements from polarised dust emission. 

In light of all this, the analysis relating to the morphology of the magnetic field is likely to be more robust than that involving quantitative calculations with the use of the magnetic field strength estimates (with the caveat that the magnetic field morphology is also subject to certain drawbacks which we discuss in the next section). However, we do note that both analyses seem to result in the same conclusion: the magnetic field plays a larger role on cloud scales than it does on clump scales. It is possible that the trend we see from our energy balance analysis is not simply a spurious result, but does in fact reflect the true conditions in our sources, just scaled up or down depending on which scale we are looking at. As we mentioned in Sections \ref{sec:Bfieldstrength} and \ref{sec:energybalance}, we expect that the magnetic field strength estimates on cloud scale are overestimated, while on clump scale they are underestimated, but we do not know to what extent. Table \ref{tab:Bfactor} shows that in most cases, even with a magnetic field strength that is smaller in the cloud than in the clump – which is what would usually be expected – it is possible for the magnetic field to be more energetically important in the cloud compared to the clump. 

To achieve a decreased importance of the magnetic field in more dense, smaller regions, a mechanism that diffuses the magnetic field from the gas must be at play. Ambipolar diffusion can make this happen. While the timescales for the complete removal of the magnetic field through ambipolar diffusion are on the order of 10 times the free-fall timescale, molecular clouds are most likely to be magnetically critical as opposed to heavily sub-critical (see Figure~\ref{fig:energybalance}) and so only a partial removal of the field would trigger collapse on timescales that are much shorter. An increase by a factor of $\frac{\rho}{\rho_0}\sim2$ of the gas density through ambipolar diffusion should be enough to trigger collapse. If we assume that the formation of the observed clumps is the consequence of such an increase of density by ambipolar diffusion, then we can derive the timescale $t_{\rm{fc}}$ of that process to occur via:

\begin{equation}
t_{\rm{fc}} = \frac{(l_{\rm{ic}}-l_{\rm{cd}})}{v_{\rm{drift}}}
\label{tfc}
\end{equation}

\noindent where $v_{\rm{drift}}$ is the ion-neutral drift velocity, $l_{\rm{ic}}$ is the initial clump radius before contraction starts, and $l_{\rm{cd}}$ is the clump radius after a contraction leading to an increase of $\sim2$ in density and can be thus expressed as a function of the initial clump size as:

\begin{equation}
l_{\rm{cd}}=\frac{l_{\rm{ic}}}{2^{1/3}}.
\end{equation}

\noindent In cgs units, the drift velocity is expressed by \citep[e.g.][]{2010ApJ...720.1612M}:

\begin{equation}
v_{\rm{drift}}= \frac{B^2}{4\pi\gamma_{\rm{AD}} x_i \frac{m_i}{m_n}\rho^2 l_{\rm{ic}}}
\label{vdrift}
\end{equation}

\noindent where $x_i\simeq10^{-7}-10^{-6}$ is the ionisation fraction, $\gamma_{\rm{AD}}=3.7\times10^{13}$cm$^3$g$^{-1}$s$^{-1}$ is the ion-neutral coupling coefficient, $\frac{m_i}{m_n}\simeq12.5$ is the mass ratio of ions-to-neutrals, and $\rho\sim1\times10^{-21}$g cm$^{-3}$ is the gas mass density of the molecular cloud. Our clump samples are typically 10 times denser than their parent clouds, which means that $l_{\rm{ic}}$ is $10^{1/3}$ larger than the average clump radius of $\sim 2$~pc, leading to $l_{\rm{ic}}\sim 4$~pc. Assuming a cloud-scale $B-$field strength of $30 \mu$G leads to $v_{\rm{drift}}\simeq 0.1 -1$~km/s, and $t_{\rm{fc}} = 0.8-8$~Myr, for the range of ionisation fractions considered. If we consider instead a magnetic field strength of $\sim10\,\mu$G, then the clump formation time would increase by nearly a factor of 10, to $t_{\rm{fc}} = 7-70$~Myr. What these calculations are showing is that a marginally sub-critical cloud, with low ionisation fraction and a relatively strong $B-$field, could become, as a result of ambipolar diffusion, super-critical and collapse on a timescale that is shorter than the cloud free-fall time of $\sim2$~Myr. But even in such a favourable situation a question remains: Why $l_{\rm{ic}}=4$~pc?

A potentially relevant scale could be the minimum wavelength over which Alfv\'en waves can propagate, which is given by \citep{1969ApJ...156..445K, 2010ApJ...720.1612M}:

\begin{equation}
\lambda_{\rm{min}}=\pi v_{\rm{A}} \tau_{\rm{ni}}
\end{equation}

\noindent where $v_{\rm{A}} =\frac{B}{\sqrt{\mu_0\rho}}$ is the Alfv\'en wave speed, and $\tau_{ni}=\frac{1}{\gamma_{\rm{AD}}\frac{m_i}{m_n}x_i\rho}$ is the ion-neutral collision time. Below this scale, ion-neutral friction is too strong for Alfv\'en waves to develop, effectively removing kinetic support to the region. Using the same parameters as we used previously, we estimate $\lambda_{\rm{min}}=0.5-5$~pc for ionisation fractions between $10^{-6}-10^{-7}$ and a $B-$field strength of $30\,\mu$G. For a $B-$field strength of $10\,\mu$G, the minimum wavelength range drops to $\lambda_{\rm{min}}=0.2-1.5$~pc. Still, these calculations show that for low ionisation fractions, and relatively high $B-$field strengths, clumps can form under the combined effect of ambipolar contraction of the cloud and the dissipation of Alfv\'en waves on timescales and spatial scales that are compatible with observations. 

Note that the formation of parsec-scale clumps might be facilitated by a sharp decrease of the ionisation fraction at some density. In their model, \citet{2012ApJ...761...67B} propose that it is the transition between UV-dominated and cosmic ray-dominated ionisation regimes that sets a specific length scale for clump formation in trans-critical clouds on timescales of $\sim2$ to $\sim20$ cloud free-fall times ($\sim6$ to $\sim53$ Myr), depending on the initial conditions of their non-ideal MHD simulations \citep{2014ApJ...780...40B}. These studies show that the shape of the ionisation fraction profile of molecular clouds, which is observationally poorly constrained, can play a significant role in setting the stage for clump collapse and cluster formation. However, as \citet{2014MNRAS.443..230H} have shown, if large-scale flows drive mass accretion along field lines onto clouds on timescales that are much shorter than the ambipolar timescale, then ambipolar diffusion becomes irrelevant.

\subsection{The evolution of the magnetic field during star formation}
\label{sec:ourresultsdiscussion}

The principal result emerging from  our investigation is that the mean direction of the magnetic field projected onto the POS is consistently different on cloud and clump scale for all of the sources except one, with a difference in the average directions of $>$30$^\circ$. This difference implies a change in the physics governing the morphology of the magnetic field between these two scales. We note that our JCMT data filters out scales above $5'$, while the minimum resolvable scale of the {\it Planck} data is also $5'$ - the spatial scales of the two datasets do not overlap.
The difference in the magnetic field morphology from cloud to clump scale implies that at clump scales, the magnetic field is sufficiently weak that it can be distorted by gravity, turbulence, or inflows.

\begin{figure}
	\includegraphics[width=\columnwidth]{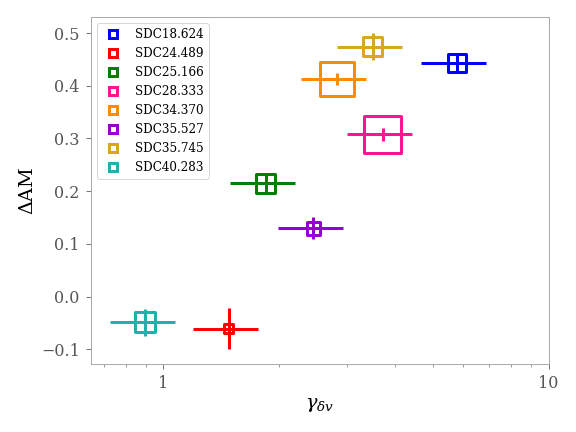}
	\caption{The difference in the average AMs from the HRO analysis within each clump and its parent molecular cloud versus the factor between the cloud's and clump's velocity dispersions. Marker size is correlated with the mass of the clump for each source.}
    \label{fig:AMvsdv}
\end{figure}

\begin{figure}
	\includegraphics[width=\columnwidth]{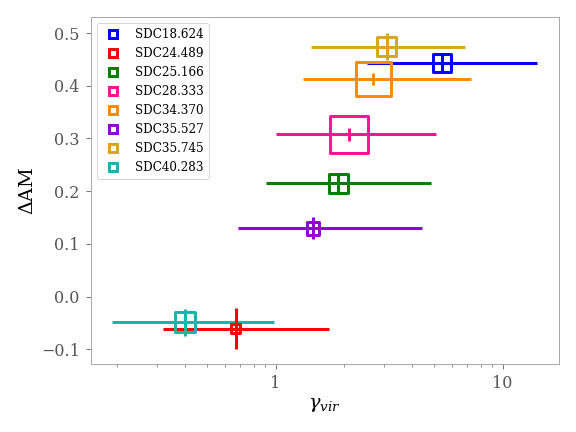}
	\caption{The difference in the average AMs from the HRO analysis within each clump and its parent molecular cloud versus the factor between the cloud's and clump's virial ratios. Marker size is correlated with the mass of the clump for each source.}
    \label{fig:AMvsVR}
\end{figure}

Figure \ref{fig:AMvsdv} shows the relationship between $\gamma_{\delta v} = \delta v_{\rm{cloud}}/\delta v_{\rm{clump}}$ and the difference in the average AMs from the HRO analysis, $\Delta$AM = AM$_{\rm{cloud}}$ - AM$_{\rm{clump}}$, within the whole cloud or clump. A positive correlation is evident, where the difference between the cloud's and clump's average AMs increases - corresponding to a more perpendicular relative alignment of the $B$-field and density structure in the clump compared to the cloud - with the factor between the velocity dispersions of the cloud and clump. In Figure \ref{fig:AMvsVR} we show the equivalent plot to Figure \ref{fig:AMvsdv} but this time plotting $\gamma_{\rm vir} = \alpha^{\rm{cloud}}_{\rm vir}/\alpha^{\rm{clump}}_{\rm vir}$ on the x-axis. Since $\delta v$ is involved in the estimation of $\alpha_{\rm vir}$ the two plots are not independent, but we can now see that the linear correlation still holds. In order to quantify this correlation we calculate the Pearson correlation coefficients for these two sets of data. The relationship shown in Figure \ref{fig:AMvsdv} has a Pearson’s correlation coefficient of 0.80 with a p-value of 0.02, while the relationship in Figure \ref{fig:AMvsVR} has a value of 0.85 with a p-value of 0.01, indicating that there is a strong positive correlation between $\Delta{\rm AM}$ and both $\gamma_{\delta v}$ and $\gamma_{\rm vir}$. The low p-values suggest that it is extremely unlikely that these linear relationships are random. 

From Figures \ref{fig:AMvsdv} and \ref{fig:AMvsVR}, we can see that as $\gamma_{\delta v}$ and $\gamma_{\rm vir}$ get closer to 1, the clouds and their clumps show similar relative alignment between  magnetic field and density structures, i.e. $\Delta {\rm AM} \sim 0$. For clouds the closest to $\gamma_{\rm vir}=1$ (i.e. SDC24.489, SDC40.283, and SDC35.527), $\Delta {\rm AM} \sim 0$ occurs at ${\rm AM}\sim0$, which means that, on average, there is no preferential alignment between $B$-field and density gradients. In those cases, clumps just seem to sample the global properties of the clouds in which they are embedded, and $B$-field is unimportant at all scales. Note, however, that this does not necessarily mean that the mean direction of the magnetic field is similar on the two scales - in fact, according to our analysis in Section \ref{sec:magfielddirection}, they are notably different for all sources except SDC40.283. 

Sources with the most extreme $\gamma_{\delta v}$ and $\gamma_{\rm vir}$ values (i.e. SDC28.333, SDC34.370, SDC18.624, and SDC35.745) also have the largest  $\Delta$AM values. This suggests that for those clouds the large difference in the dynamical state of the clouds is matched by a clear difference in the relative orientation of density structures and $B$-field. Contrary to sources located at the lower corners of Figures~\ref{fig:AMvsdv} and \ref{fig:AMvsVR}, gas properties for clouds and clumps located at the top corner are markedly different. It is also interesting to note that one of the most extreme clouds in Figures~\ref{fig:AMvsdv} and \ref{fig:AMvsVR}, SDC18.624, has been proposed to be an example of cloud-cloud collision \citep{2018ApJ...861...19D}. 

The clear positive correlations seen in Figures~\ref{fig:AMvsdv} and \ref{fig:AMvsVR} can be interpreted in two ways - either as an evolutionary track, or as differences in the initial conditions of cloud formation. Whether the arrangement of clouds in Figures~\ref{fig:AMvsdv} and \ref{fig:AMvsVR} represents an evolutionary sequence or different initial conditions is key to our understanding of cluster formation. 

If one considers the $\Delta\rm{AM}-\gamma$ correlation as an evolutionary sequence, then evolution must start towards the bottom left corner. There, gas located within the clump and its parent clump have similar properties, i.e. the clump is not yet decoupled from its cloud. The dynamical state of the cloud/clump is likely to be one of early collapse. As the cloud and clumps within become larger and more massive as a result of the continuous accretion flows occurring across all scales \citep{2019MNRAS.490.3061V}, massive star formation proceeds, ionising and injecting energy preferentially in the lower density gas \citep{2019A&A...628A..21W}. This might have the effect of increasing the velocity dispersion of the gas on the largest scales, and also producing a tighter coupling between magnetic fields and gas at lower densities. The impact of such feedback might thus lead to larger values of both $\Delta\rm{AM}$ and $\gamma$.

If we now consider that the $\Delta\rm{AM}-\gamma$ correlation is the result of different initial conditions, then Figure~\ref{fig:AMvsdv} might be seen as a correlation between converging flow strength and magnetic field strength. With that interpretation, clouds at the top right corner of the plot are those for which magnetic fields have the largest impact on cloud structures and for which the flow velocities (via the measurement of the velocity dispersion) are also the largest. For those clouds, the flows must be directed along the field lines efficiently accumulating matter within clumps at the centre which then may collapse as their mass to flux ratio increases. The recent study by \citet{2026arXiv260512604P} seems to suggest that the DR21 high-mass star-forming region follows such a scenario. On the other hand, clouds at the bottom left corner of the plots might have converging flows that are mostly driven by gravity, resulting in slower and less focussed inflows. 

We note that SDC40.283 has $\gamma_{\delta v}<1$, implying an approximately flat/increasing velocity dispersion profile from cloud to clump scale, thus inconsistent with the clump being dynamically decoupled. A minority of cloud-clump pairs from \citet{2023MNRAS.525.2935P}'s sample fall in this category. For those, it is the gas velocity dispersion on cloud scale that is surprisingly low. They might represent cases that are at the earliest stages of molecular cloud evolution and for which the high velocity dispersion typically observed at tens of parsec scale has not yet developed.

Recent studies with ALMA have suggested that the high-mass end of the core mass function (CMF) evolves in time \citep[e.g.][]{2023A&A...674A..75N, 2024ApJ...966..171M, 2025A&A...696A.151C}, with high-mass cores gaining mass as the result of clump collapse \citep[e.g.][]{2013A&A...555A.112P, 2021MNRAS.508.2964A, 2024MNRAS.528.1172R}. Mapping the CMF of the clumps studied here could thus provide indirect evidence for a potential evolutionary sequence of the cloud-clump pairs in the $\Delta$AM - $\gamma$ plane. The cores in some of the more well-studied sources in our sample  - e.g. SDC28.333, SDC34.370, and SDC35.527 - have in fact already been mapped, with studies even fitting the slope of the CMF of SDC28.333 and finding that it is much shallower than the slope of the Salpeter initial mass function \citep{2018ApJ...862..105L, 2019ApJ...873...31K}. With greater coverage across samples more robust comparisons can be drawn.
Also, by mapping and connecting gas flows across scales for our sample of cloud-clump pairs one could investigate the initial condition hypothesis. This is in fact a key aspect of the new {\it Panta Rei} ALMA large programme (PIs: N. Peretto, A. Traficante, S. Clarke, M. Merello) which, by mapping the velocity field of a sample of 286 clumps already observed with ALMAGAL \citep{2025A&A...696A.149M} and SEDIGISM \citep{2017A&A...601A.124S}, will be able to track and characterise star cluster-forming flows.

\subsection{Limitations}

Our study, like any other dedicated to polarisation observations, is limited by our ability to infer the 3-dimensional properties of the cloud density structure and $B-$field from their observed projections onto the plane-of-sky (POS). For instance, when looking at a region of the sky where the mean magnetic field is mainly oriented along the line-of-sight (LOS), the observed POS magnetic field direction tends to be dominated by the turbulent component of the 3D field \citep{2018MNRAS.474.5122K, liMagnetizedInterstellarMolecular2019}.  It therefore can be difficult to distinguish between a region with a weak magnetic field/high turbulence, or a region with a $B$-field preferentially oriented along the LOS. Additionally, the projection of the relative orientation of two distinct 3D vector fields (such as $B-$fields and density gradients for instance) onto the POS will systematically tend to decrease the orientation differences and thus make the projected vector fields appear more parallel than what they truly are in 3D  \citep{2016A&A...586A.138P, 2019A&A...629A..96S, 2020MNRAS.497.4196S} - consequently, observed parallel alignment in the POS may be a projection artefact.
The impact of projection on the HRO analysis has also been investigated by a few authors \citep[see e.g.][]{2021MNRAS.503.5425B, 2023MNRAS.521.3830M}. In those studies, it has been shown that absolute AM values are directly impacted by the orientation of $B-$field with respect to the LOS, in particular in extreme cases (i.e. perpendicular or parallel). However, the variations of AM measurements from cloud to clump, $\Delta$AM, for a sample of clouds with unknown orientations is the best way to mitigate uncertainties linked to projection effects. The trends observed in Figures~\ref{fig:AMvsdv} and \ref{fig:AMvsVR} are thus likely to be robust and contain key information regarding the magneto-dynamics of molecular clouds and the formation of star clusters.

Another matter of contention is the reliability of the magnetic field morphology towards our sample of clouds, which are all large structures that lie in the Galactic plane and so are likely to suffer from LOS confusion. We use two methods to probe the magnetic field direction within the clouds - {\it Planck} polarisation data and the Velocity Gradient Theory. Each of these have their own drawbacks. We have already discussed the correction that was made to the {\it Planck} polarisation observations in Section \ref{sec:dustpolobs}, where the assumption that the magnetic field is expected to be aligned with the Galactic plane influences the method used to correct for bandpass mismatch leakage. We use the VGT to compare to the {\it Planck} observations and also correct for LOS contamination, but this is in itself a new and developing method which may not be fully established as of yet. It should be noted that the VGT papers often mention a change in the application of the VGT in self-gravitating regions, with the magnetic field pseudovectors calculated from the VGT being parallel to the thin channel velocity gradients in these regions rather than perpendicular. We have not applied this correction to any of the sources due to the exact region to which this correction should be applied being unknown. This does not seem to make a large difference to the results of the VGT - this is perhaps because the distance of the clouds means that the scales at which regions become self-gravitating are not sufficiently resolved in the telescope beam to result in a total rotation of the pseudovector direction. Since both {\it Planck} and the VGT appear to show magnetic field pseudovectors that preferentially lie parallel to the Galactic plane, which is what is expected from Galactic dynamo effects \citep[e.g.][]{2020A&A...641A.165N, 2023ARA&A..61..561B}, we assume that the pseudovectors obtained from {\it Planck} are reasonable to trace the magnetic field in our clouds. We also note that at 353GHz - the observing frequency of {\it Planck} (and JCMT) - the effect of Faraday rotation is negligible and so should not affect the results. 

Finally, it should also be noted that the study presented here is based on relatively small sample statistics, even though a sample of eight cloud-clump pairs is larger than most magnetic field studies published in literature. Increasing statistics would be a natural step forward to test the robustness of the trends presented in Figures~\ref{fig:AMvsdv} and \ref{fig:AMvsVR}.

\section{Summary}
\label{sec:summary}

In this study we analyse the magnetic fields of eight IRDCs and their parent molecular clouds to observe whether the dynamical changes between these two scales found in recent studies is also replicated in the magnetic field. {\it Planck} dust polarisation observations, as well as the VGT, are used to derive the magnetic fields of the larger scale ($\sim$10~pc) more diffuse clouds, with JCMT POL-2 polarisation data utilised for the dense small-scale ($\sim$1~pc) clumps. We find evidence that there is indeed an imprint of this dynamical change between cloud and clump scales in the magnetic field – the majority of samples display a marked change in magnetic field direction from cloud to clump scale, showing that memory of the magnetic field is lost.

An investigation into the correlation of the magnetic field with density structures revealed that in general, magnetic field pseudovectors tend to be more aligned to the diffuse density structures but rotate to become more perpendicular as density increases, in agreement with previous observations. This result was not consistent across all sources however, and it was observed that the greater the difference in behaviour between {the cloud- and clump-scale magnetic fields in terms of alignment to density structure, the larger the difference in kinematic properties (velocity dispersion and virial ratio) between the cloud and clump. We propose that this positive correlation between magnetic field-density alignment and kinematic properties could signify either the evolution of a molecular cloud as it advances through the star formation process, or a difference in the initial conditions.

We also trial multiple methods of calculating magnetic field strength, finding that results vary widely across methods. The methods also tend to provide a larger magnetic field strength in the clouds compared to the clumps, against the expectations of magnetic flux conservation. We conclude that these methods are not suitable for use on our sources due to the multiple assumptions involved. Despite all of these caveats, analysing the energy balance between gravity, turbulence, and magnetic fields using the magnetic field strength values obtained from one of these methods leads to conclusions in agreement with previous studies – clouds are supported against collapse by magnetic fields, while gravity and turbulence begin to take over in clumps. 

\section*{Acknowledgements}

We thank the referee for their insightful comments that helped to significantly increase the quality of this paper.
The authors thank Alex Lazarian and Yue Hu for their helpful provision of information regarding the VGT.
RR is supported by the Science and Technology Facilities Council
(STFC). NP acknowledges the support of STFC consolidated grant number ST/N00033X/1 and STFC Small Award number APP30146.
GAF acknowledges funding by the Deutsche Forschungsgemeinschaft (DFG, German Research Foundation) under Germany’s Excellence Strategy – EXC 3037 – 533607693, the DFG through SFB 1601 ``Habitats of massive stars across cosmic time'' (sub-project B1), and from the University of Cologne and its Global Faculty programme.
PMK acknowledges support from the National Science and Technology Council (NSTC) in Taiwan through grants NSTC 112-2112-M-001-049-, NSTC 113-2112-M-001-016-, and NSTC 114-2112-M-001-007-.

%%%%%%%%%%%%%%%%%%%%%%%%%%%%%%%%%%%%%%%%%%%%%%%%%%
\section*{Data Availability}

The JCMT POL-2 polarisation data is publicly available at the Canadian Astronomy Data Centre (CADC) website. These observations were obtained by the James Clerk Maxwell Telescope, operated by the East Asian Observatory on behalf of The National Astronomical Observatory of Japan; Academia Sinica Institute of Astronomy and Astrophysics; the Korea Astronomy and Space Science Institute; the National Astronomical Research Institute of Thailand; Center for Astronomical Mega-Science (as well as the National Key R\&D Program of China with No. 2017YFA0402700). Additional funding support is provided by the Science and Technology Facilities Council of the United Kingdom and participating universities and organizations in the United Kingdom and Canada.

The {\it Planck} polarisation data is publicly available at the {\it Planck} Legacy Archive. Data are based on observations obtained with {\it Planck} (http://www.esa.int/Planck), an ESA science mission with instruments and contributions directly funded by ESA Member States, NASA, and Canada.

%%%%%%%%%%%%%%%%%%%% REFERENCES %%%%%%%%%%%%%%%%%%

% The best way to enter references is to use BibTeX:

\bibliographystyle{mnras}
\bibliography{References} % if your bibtex file is called example.bib

% Alternatively you could enter them by hand, like this:
% This method is tedious and prone to error if you have lots of references
%\begin{thebibliography}{99}
%\bibitem[\protect\citeauthoryear{Author}{2012}]{Author2012}
%Author A.~N., 2013, Journal of Improbable Astronomy, 1, 1
%\bibitem[\protect\citeauthoryear{Others}{2013}]{Others2013}
%Others S., 2012, Journal of Interesting Stuff, 17, 198
%\end{thebibliography}

%%%%%%%%%%%%%%%%%%%%%%%%%%%%%%%%%%%%%%%%%%%%%%%%%%

%%%%%%%%%%%%%%%%% APPENDICES %%%%%%%%%%%%%%%%%%%%%

\appendix

\section{Effect of increasing the polarised intensity signal-to-noise ratio}
\label{sec:appendixsnr>3}
In this paper, we selected polarisation pseudovectors that have $P/\delta P>2$ and $I/dI>10$. Here we investigate how our results may vary if we were to use $P/\delta P>3$ instead.

Tables \ref{tab:BfieldstrengthsSNR>3} and \ref{tab:ADFcorrSNR>3} show the values of $B_{\rm POS}$ that would be obtained if we were to only use polarisation pseudovectors fulfilling $P/\delta P>3$, with everything else in our calculations being the same as in our analysis in Section \ref{sec:Bfieldstrength}. We only calculate $B_{\rm POS}$ where the angular dispersion has been estimated from the standard deviation of the Gaussian fitted to the distribution of angles (not the MAD values). It can be seen that there is for the most part not a vast difference between the outcomes of using $P/\delta P>3$ compared to $P/\delta P>2$, with the substantial difference in the average clump-scale $B$-field strength for the sin-DCF method mostly being due to the loss of the large $B_{\rm POS}$ value for SDC25.166 when applying $P/\delta P>3$. The corrected ADF $B_{\rm POS}$ values using $P/\delta P>3$ shown in Table \ref{tab:ADFcorrSNR>3} are similar to those in Table \ref{tab:ADFcorr}, showing that our results are not severely affected by the choice of constraint.

Table \ref{tab:HROAM} lists the HRO AM values when applying $P/\delta P>2$ in columns 2 and 3, and when applying $P/\delta P>3$ in columns 4 and 5. As expected, on cloud scale there is not a significant difference between the AM values. On clump scale the values do also tend to be similar for the two situations except for SDC18.624 and SDC24.489, where the AM decreases for SDC18.624 and increases for SDC24.489 by $\sim0.1$ when applying $P/\delta P>3$ instead of $P/\delta P>2$. SDC34.370 also displays a moderate decrease of 0.09. Despite these differences, this does not affect any of our conclusions, as the correlations in Figures \ref{fig:AMvsdv} and \ref{fig:AMvsVR} still hold.
\begin{table*}
    \centering
\caption{Same as Table \ref{tab:uncorrectedBpos}, but filtering polarisation vectors by $P/\delta P>3$ rather than $P/\delta P>2$.}
\label{tab:BfieldstrengthsSNR>3}
    \begin{tabular}{ccccccccc}\toprule
         &  \multicolumn{2}{c}{Classical DCF}&  \multicolumn{2}{c}{sin-DCF}&  \multicolumn{2}{c}{ST}&  \multicolumn{2}{c}{ADF}\\\midrule
 Source& $B_{\rm{POS}}^{\rm{cloud}}$ ($\mu$G)& $B_{\rm{POS}}^{\rm{clump}}$ ($\mu$G)& $B_{\rm{POS}}^{\rm{cloud}}$ ($\mu$G)& $B_{\rm{POS}}^{\rm{clump}}$ ($\mu$G)& $B_{\rm{POS}}^{\rm{cloud}}$ ($\mu$G)& $B_{\rm{POS}}^{\rm{clump}}$ ($\mu$G)& $B_{\rm{POS}}^{\rm{cloud}}$ ($\mu$G)&$B_{\rm{POS}}^{\rm{clump}}$ ($\mu$G)\\
 SDC18.624& 718& 69& 744& -& 158& 36& 487&98
\\
         SDC24.489&  -&  98&  -&  338&  -&  44&  207& 120
\\
         SDC25.166&  136&  88&  136&  -&  35&  41&  101& 106
\\
         SDC28.333&  385&  60&  375&  114&  93&  41&  359& 153
\\
         SDC34.370&  254&  52&  259&  217&  87&  42&  391& 204
\\
         SDC35.527&  503&  56&  491&  114&  73&  27&  76& 73
\\
         SDC35.745&  211&  32&  215&  383&  63&  22&  247& 63
\\
         SDC40.283&  100&  46&  99&  93&  24&  36&  83& 109
\\
         \textbf{Average}&  \textbf{330}&  \textbf{63}&  \textbf{331}&  \textbf{210}&  \textbf{76}&  \textbf{36}&  \textbf{244}& \textbf{116}\\
 & & & & & & & &\\ \bottomrule
    \end{tabular}

\end{table*}

\begin{table}
    \centering
\caption{Same as Table \ref{tab:ADFcorr}, but filtering polarisation vectors by $P/\delta P>3$ rather than $P/\delta P>2$.}
\label{tab:ADFcorrSNR>3}
    \begin{tabular}{ccc}\toprule
         Source&  $B_{\rm{POS}}^{\rm{cloud}}$ ($\mu$G)& $B_{\rm{POS}}^{\rm{clump}}$ ($\mu$G)\\\midrule
         SDC18.624
&  108& 55
\\
         SDC24.489
&  23& 58
\\
         SDC25.166
&  46& 56
\\
         SDC28.333
&  125& 82
\\
         SDC34.370
&  82& 154
\\
         SDC35.527
&  45& 40
\\
         SDC35.745
&  108& 38
\\
         SDC40.283
&  69& 75
\\
         \textbf{Average}&  \textbf{76}& \textbf{70}\\ \bottomrule
    \end{tabular}

\end{table}

\begin{table}
    \centering
\caption{Values of the average HRO AMs within the clouds and clumps, for both the case where polarisation vectors fulfilling $P/\delta P>2$ are considered (columns 2 and 3) and the case where only vectors fulfilling $P/\delta P>3$ are considered (columns 4 and 5).}
\label{tab:HROAM}
    \begin{tabular}{ccccc}\toprule
         &  \multicolumn{2}{c}{$P/\delta P>2$}&  \multicolumn{2}{c}{$P/\delta P>3$}\\\midrule
         Source&  Cloud&  Clump
&  Cloud& Clump
\\
         SDC18.624&  0.12&  -0.32
&  0.12& -0.43
\\
         SDC24.489&  -0.16&  -0.10
&  -0.19& 0.04
\\
         SDC25.166&  0.36&  0.15
&  0.37& 0.14
\\
         SDC28.333&  0.09&  -0.21
&  0.09& -0.25
\\
         SDC34.370&  0.21&  -0.20
&  0.21& -0.29
\\
         SDC35.527&  0.00&  -0.13
&  0.01& -0.11
\\
         SDC35.745&  0.32&  -0.15
&  0.32& -0.16
\\
         SDC40.283&  -0.08&  -0.04
&  -0.08& -0.02
\\ \bottomrule
    \end{tabular}

\end{table}
\section{Magnetic Field pseudovector Images}
\label{sec:appendixmagfieldpseudovectorimages}

The images depicting the magnetic field pseudovectors within the clouds and clumps, similarly to that of SDC34.370 in Figure \ref{fig:magfieldpseudovectors}, of all of the remaining investigated sample are shown here.

\begin{figure*}
	\begin{subfigure}[H]{0.462\textwidth}
	\includegraphics[width=\textwidth]{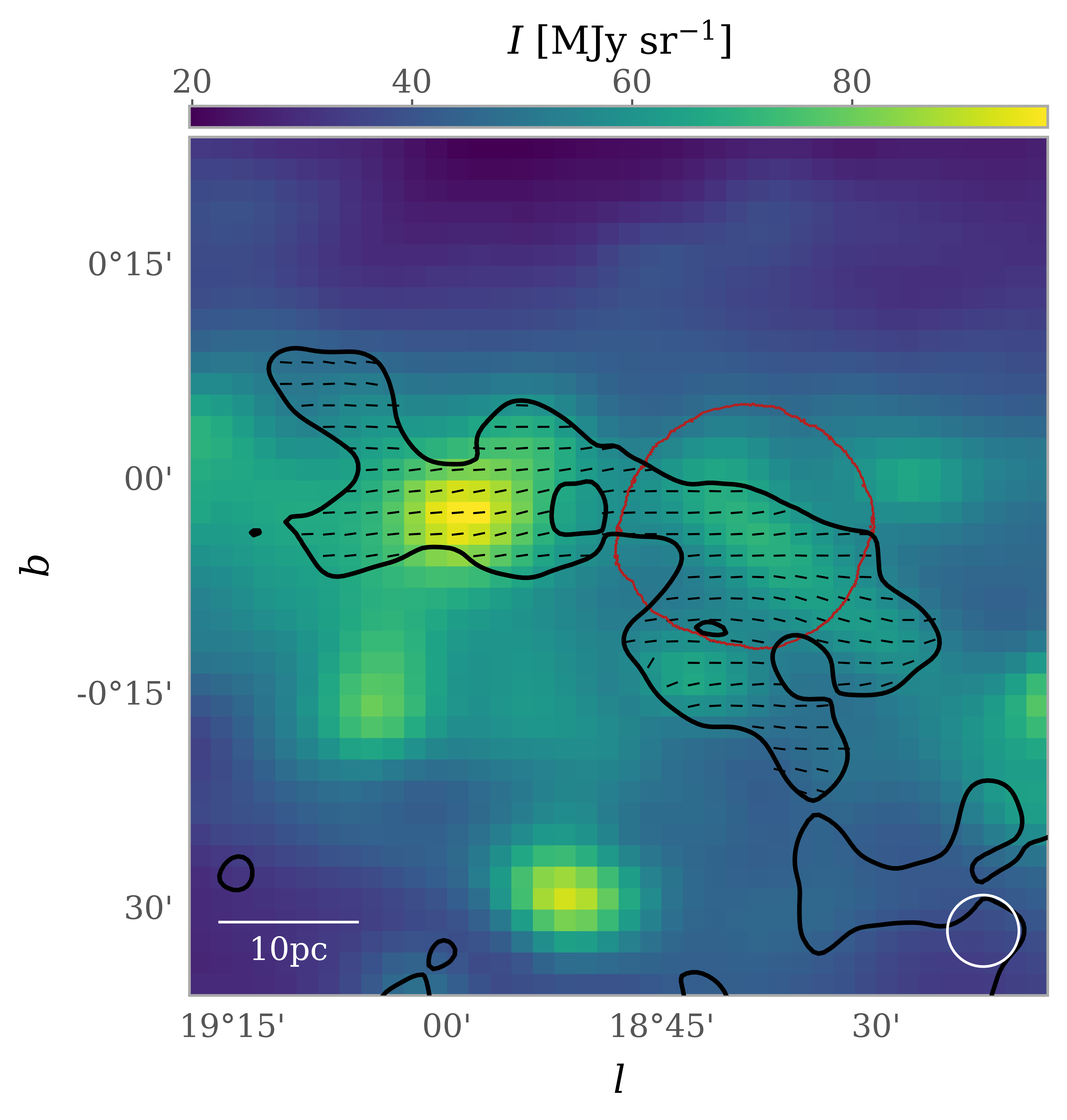}
	\caption{}
	\label{fig:SDC18p624_Planck}
	\end{subfigure}
	\begin{subfigure}[H]{0.528\textwidth}
	\includegraphics[width=\textwidth]{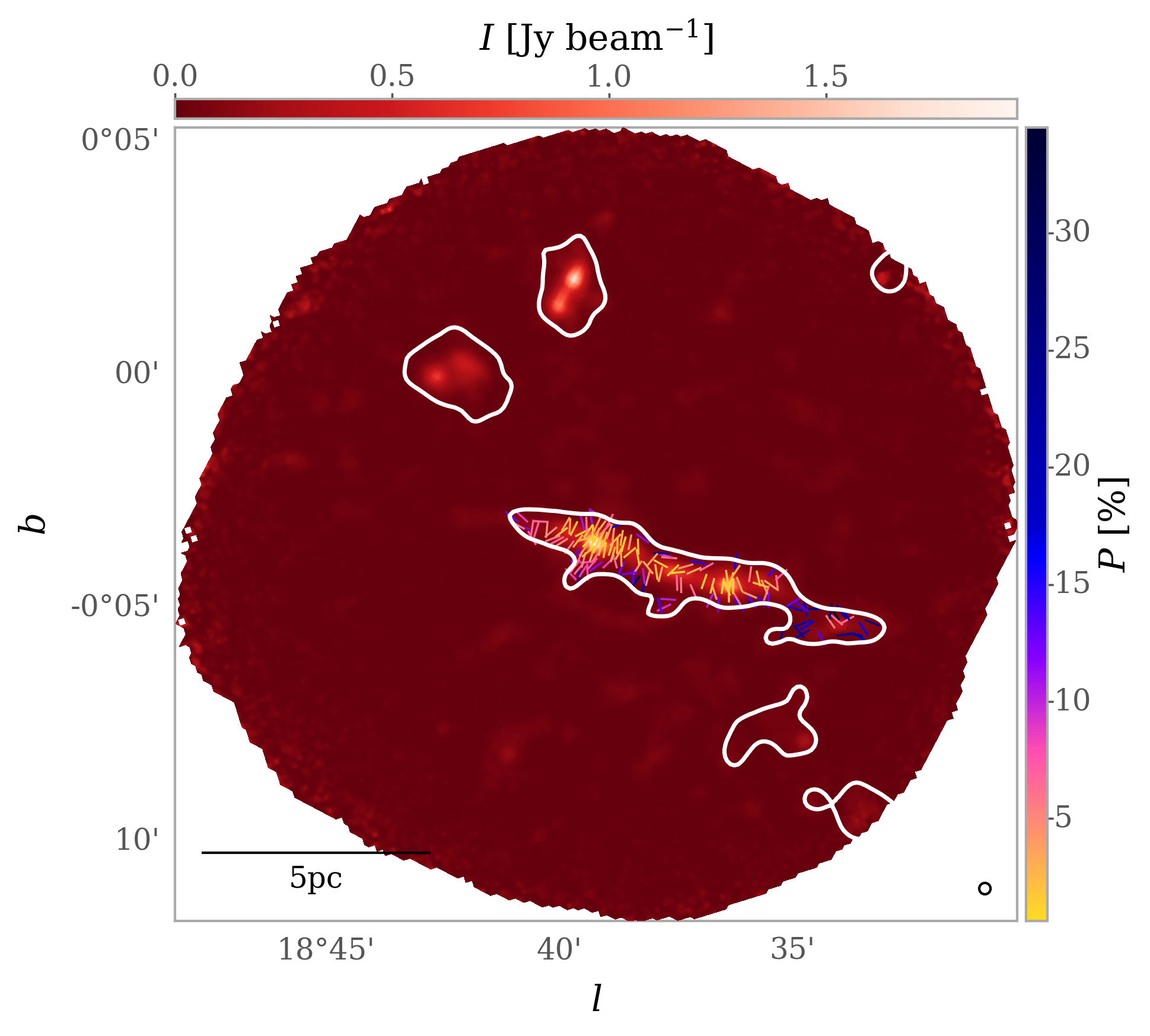}
	\caption{}
	\label{fig:SDC18p624_POL-2}
	\end{subfigure}
	\begin{subfigure}[H]{0.99\textwidth}
    \centering
	\includegraphics[width=\textwidth]{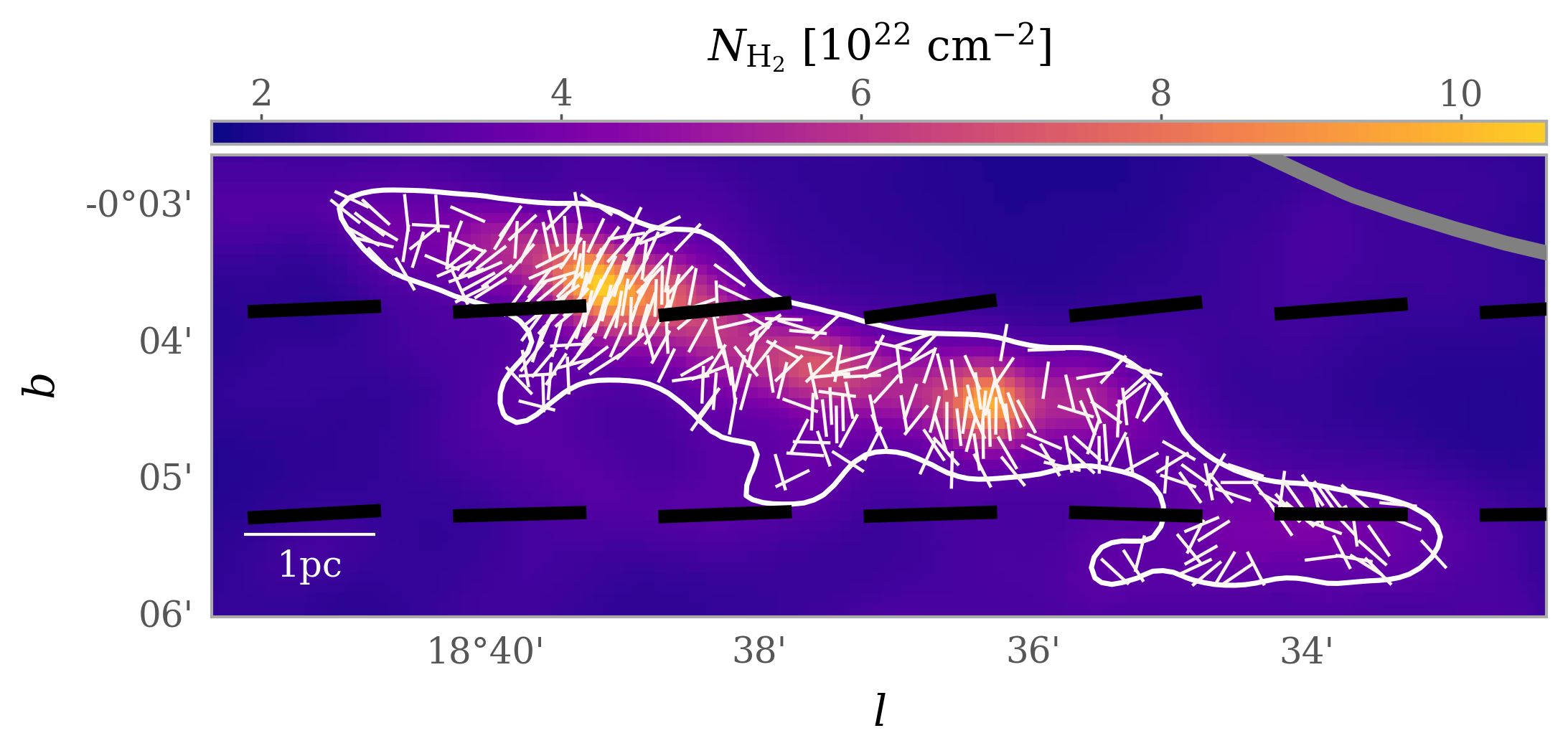}
	\caption{}
	\label{fig:SDC18p624_zoomed}
	\end{subfigure}
    \caption{Same as Figure \ref{fig:magfieldpseudovectors} for SDC18.624.}
    \label{fig:SDC18p624magfieldpseudovectors}
\end{figure*}

\begin{figure*}
	\begin{subfigure}[H]{0.462\textwidth}
	\includegraphics[width=\textwidth]{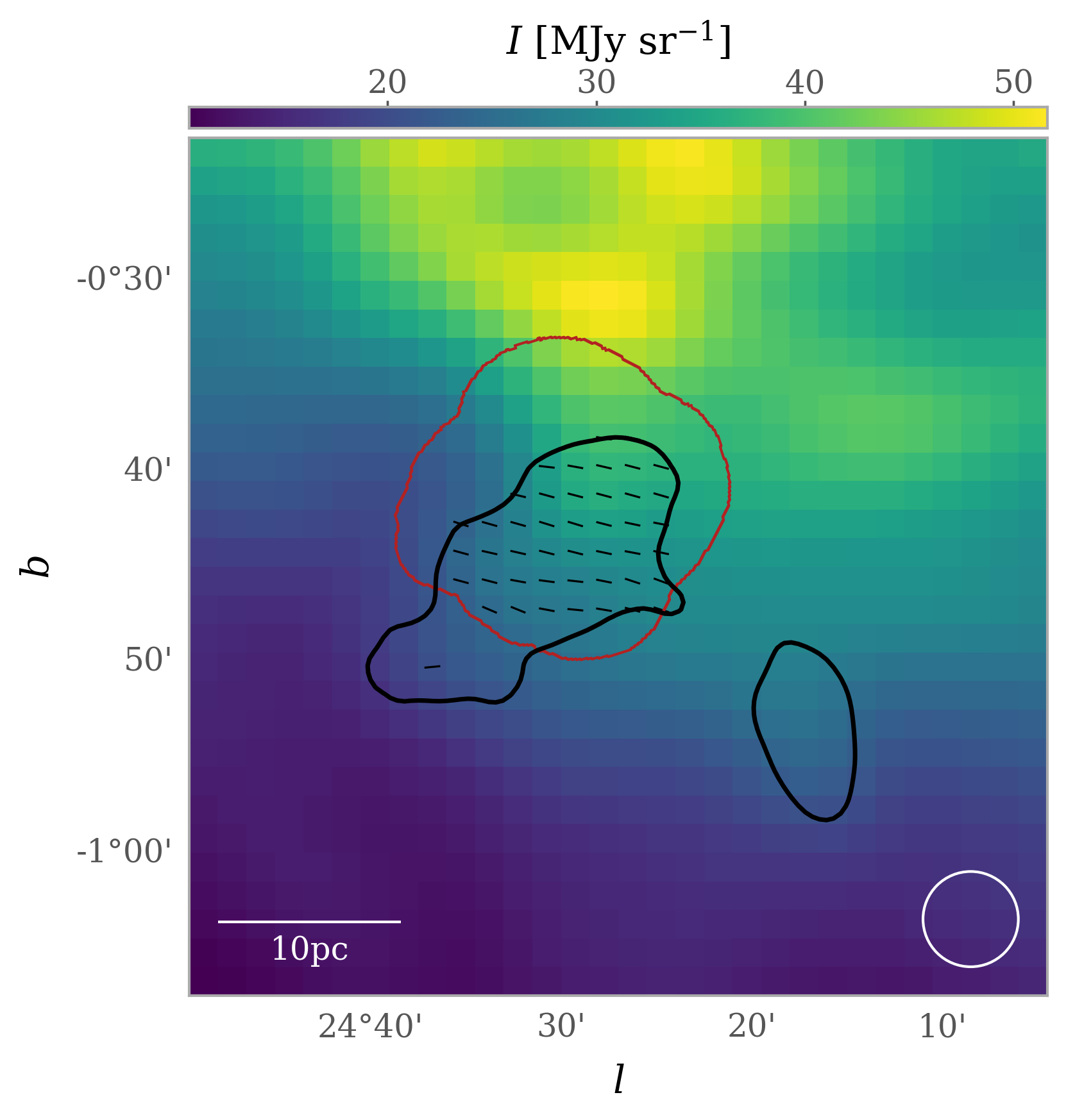}
	\caption{}
	\label{fig:SDC24p489_Planck}
	\end{subfigure}
	\begin{subfigure}[H]{0.528\textwidth}
	\includegraphics[width=\textwidth]{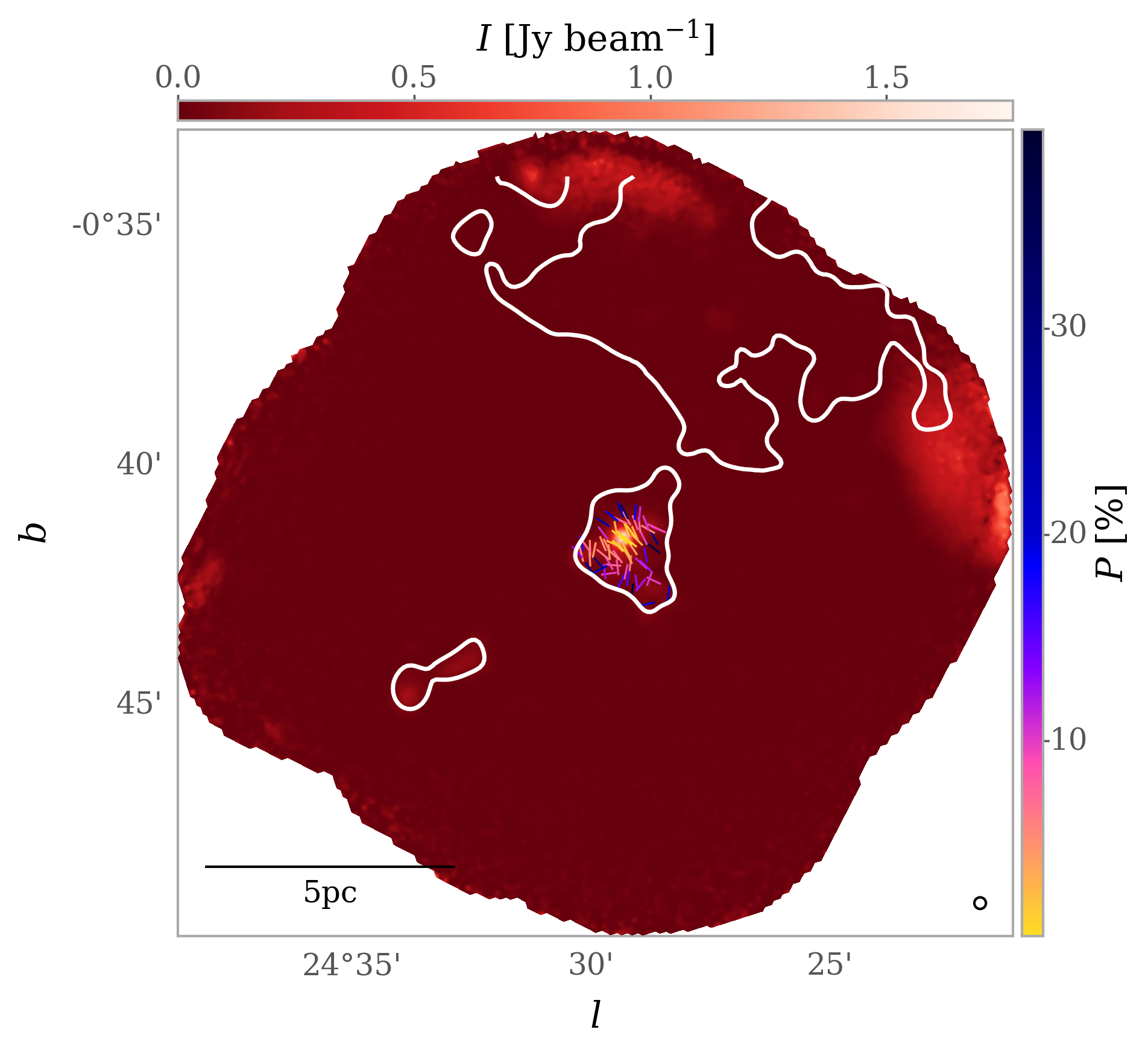}
	\caption{}
	\label{fig:SDC24p489_POL-2}
	\end{subfigure}
	\begin{subfigure}[H]{0.9\textwidth}
    \centering
	\includegraphics[width=0.6\textwidth]{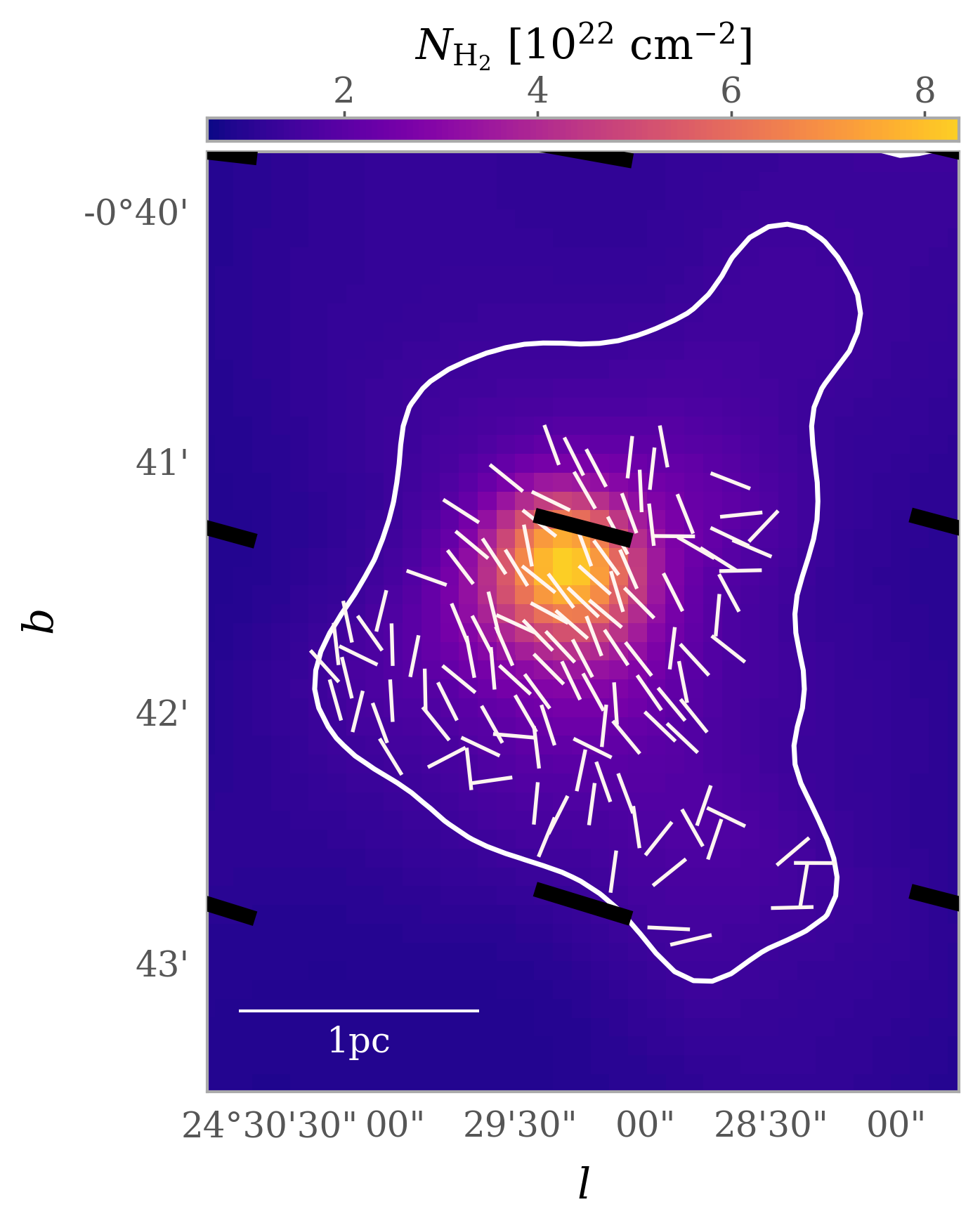}
	\caption{}
	\label{fig:SDC24p489_zoomed}
	\end{subfigure}
    \caption{Same as Figure \ref{fig:magfieldpseudovectors} for SDC24.489.}
    \label{fig:SDC24p489magfieldpseudovectors}
\end{figure*}

\begin{figure*}
	\begin{subfigure}[H]{0.462\textwidth}
	\includegraphics[width=\textwidth]{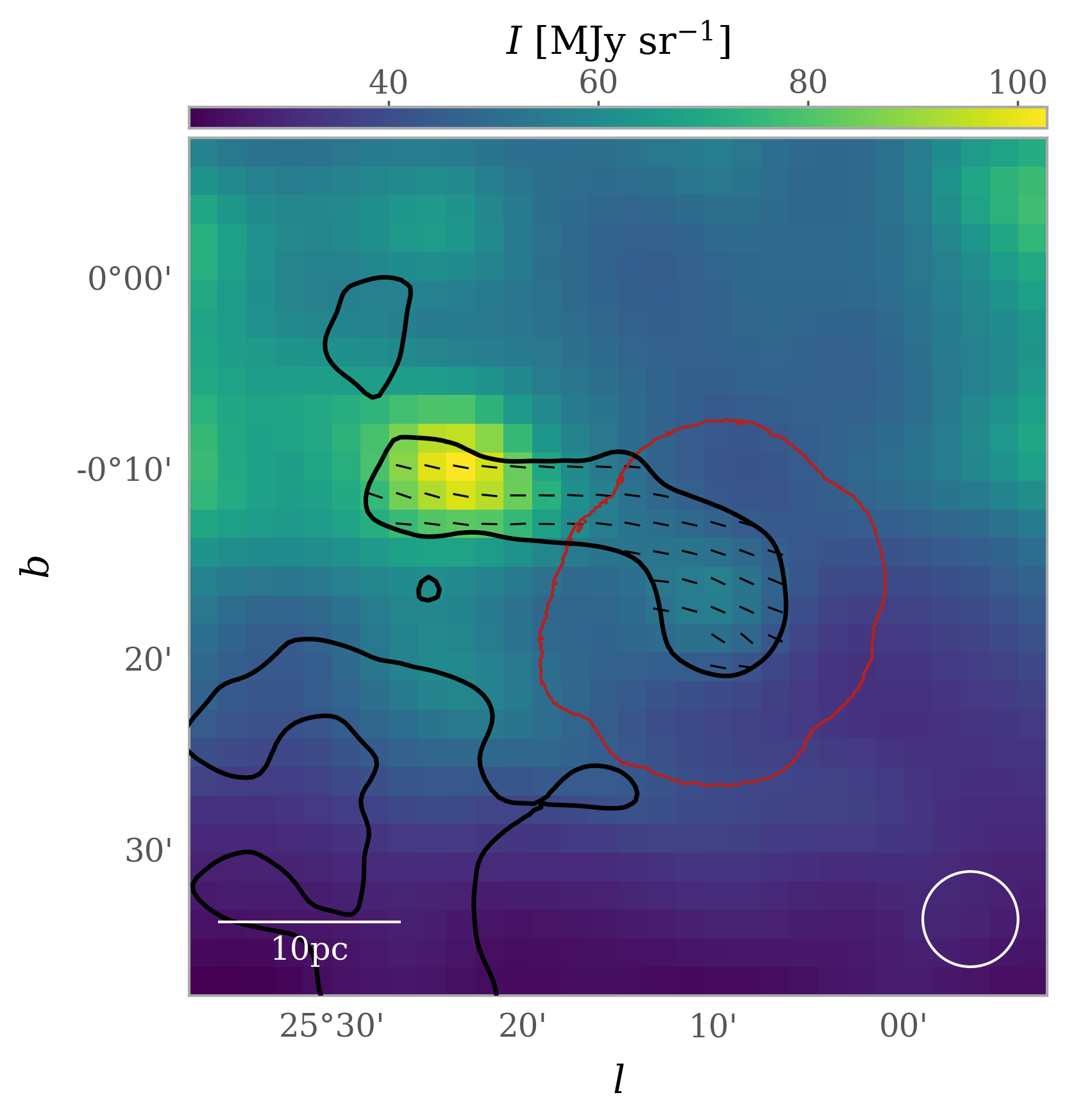}
	\caption{}
	\label{fig:SDC25p166_Planck}
	\end{subfigure}
	\begin{subfigure}[H]{0.5\textwidth}
	\includegraphics[width=\textwidth]{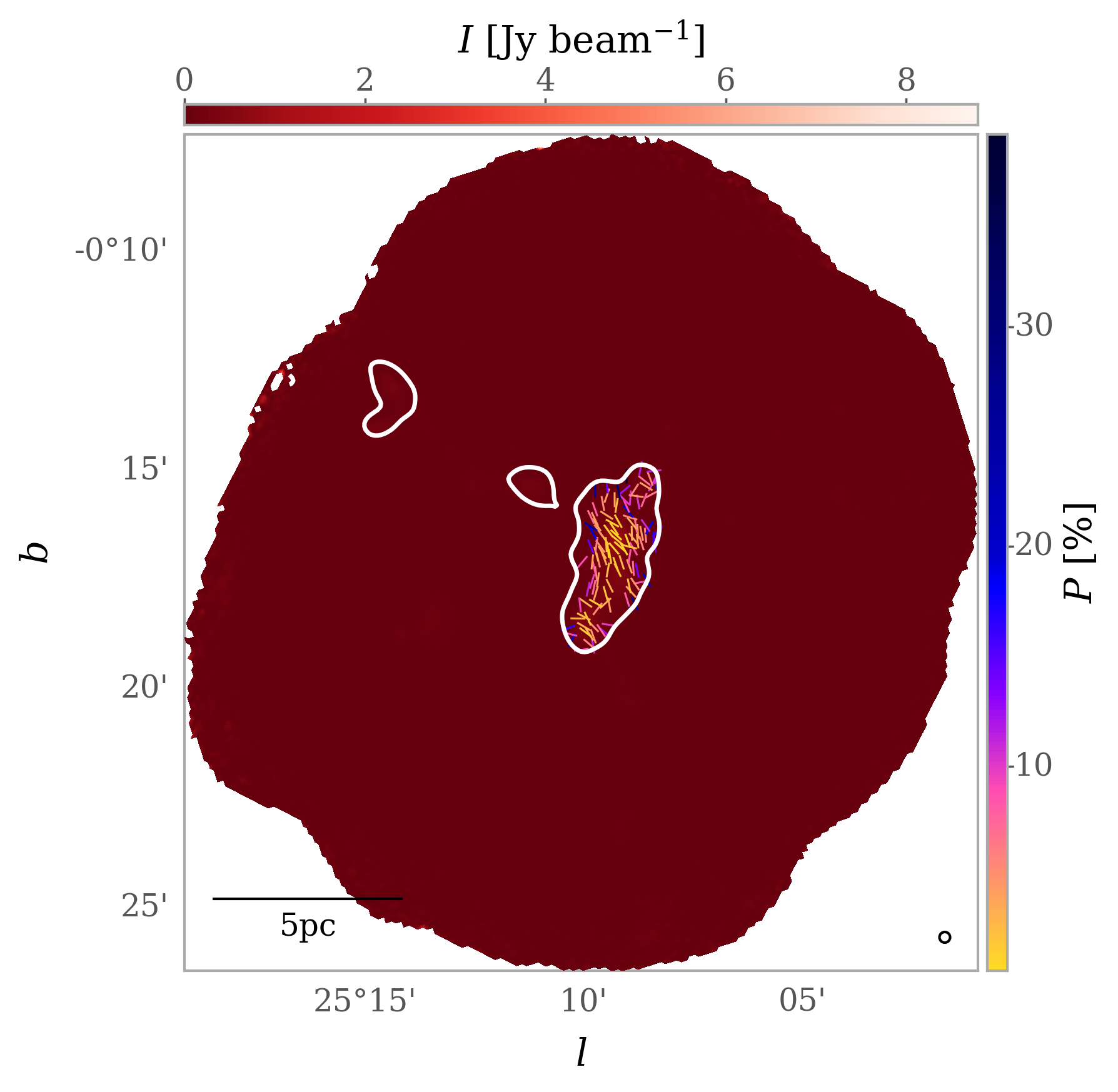}
	\caption{}
	\label{fig:SDC25p166_POL-2}
	\end{subfigure}
	\begin{subfigure}[H]{0.9\textwidth}
    \centering
	\includegraphics[width=0.6\textwidth]{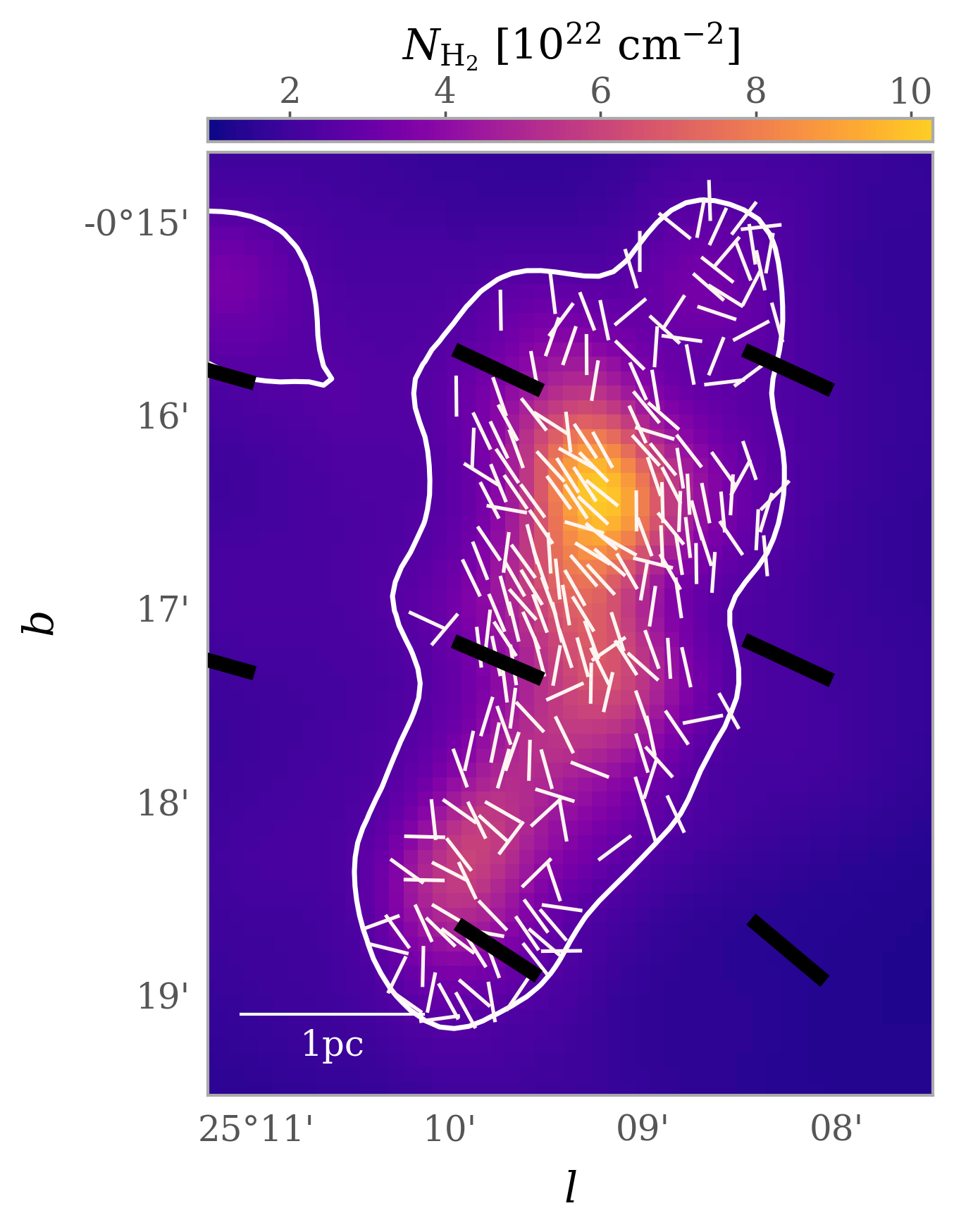}
	\caption{}
	\label{fig:SDC25p166_zoomed}
	\end{subfigure}
    \caption{Same as Figure \ref{fig:magfieldpseudovectors} for SDC25.166.}
    \label{fig:SDC25p166magfieldpseudovectors}
\end{figure*}

\begin{figure*}
	\begin{subfigure}[H]{0.462\textwidth}
	\includegraphics[width=\textwidth]{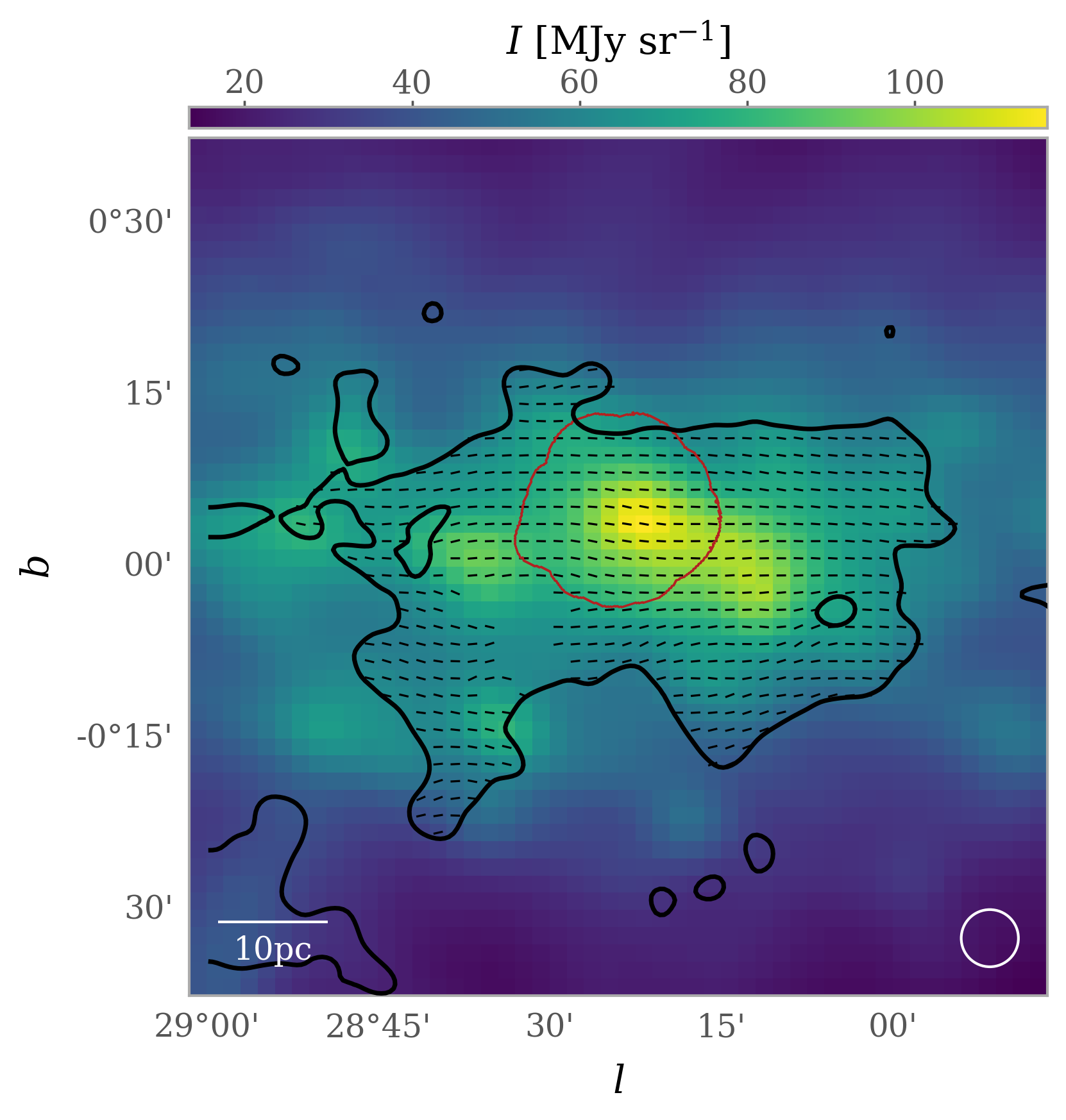}
	\caption{}
	\label{fig:SDC28p333_Planck}
	\end{subfigure}
	\begin{subfigure}[H]{0.528\textwidth}
	\includegraphics[width=\textwidth]{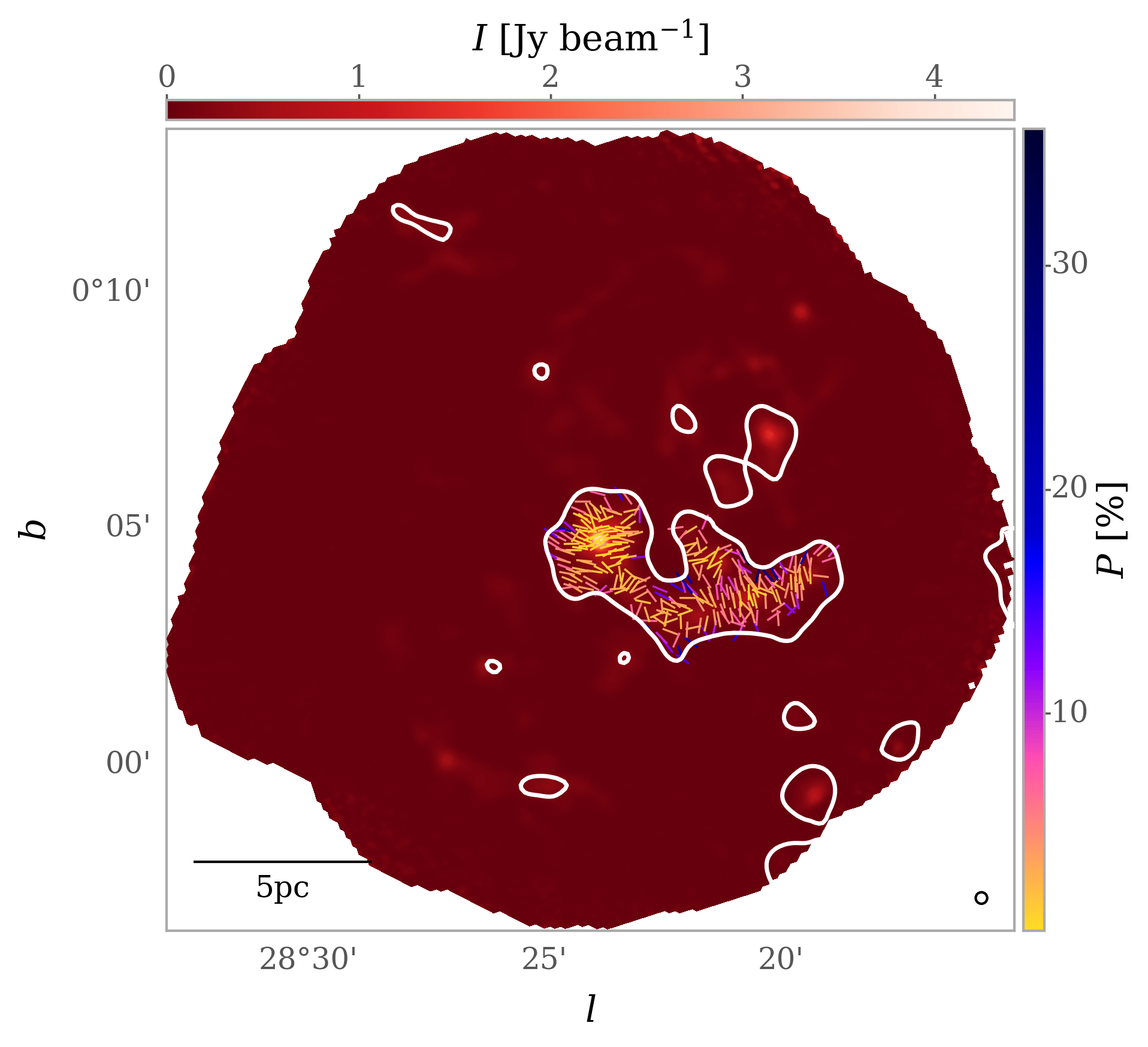}
	\caption{}
	\label{fig:SDC28p333_POL-2}
	\end{subfigure}
	\begin{subfigure}[H]{0.99\textwidth}
    \centering
	\includegraphics[height=0.4\textheight]{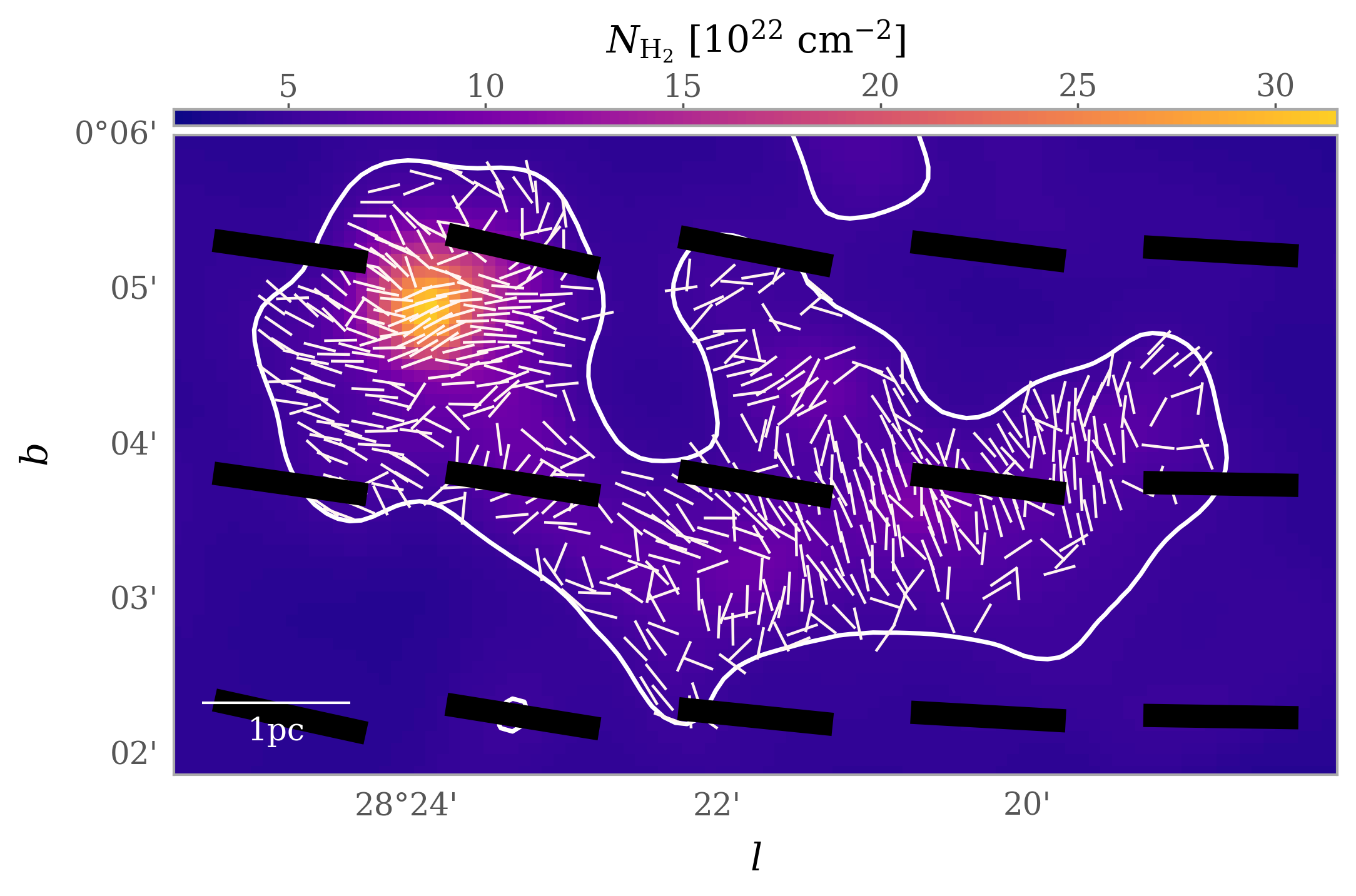}
	\caption{}
	\label{fig:SDC28p333_zoomed}
	\end{subfigure}
    \caption{Same as Figure \ref{fig:magfieldpseudovectors} for SDC28.333.}
    \label{fig:SDC28p333magfieldpseudovectors}
\end{figure*}

\begin{figure*}
	\begin{subfigure}[H]{0.462\textwidth}
	\includegraphics[width=\textwidth]{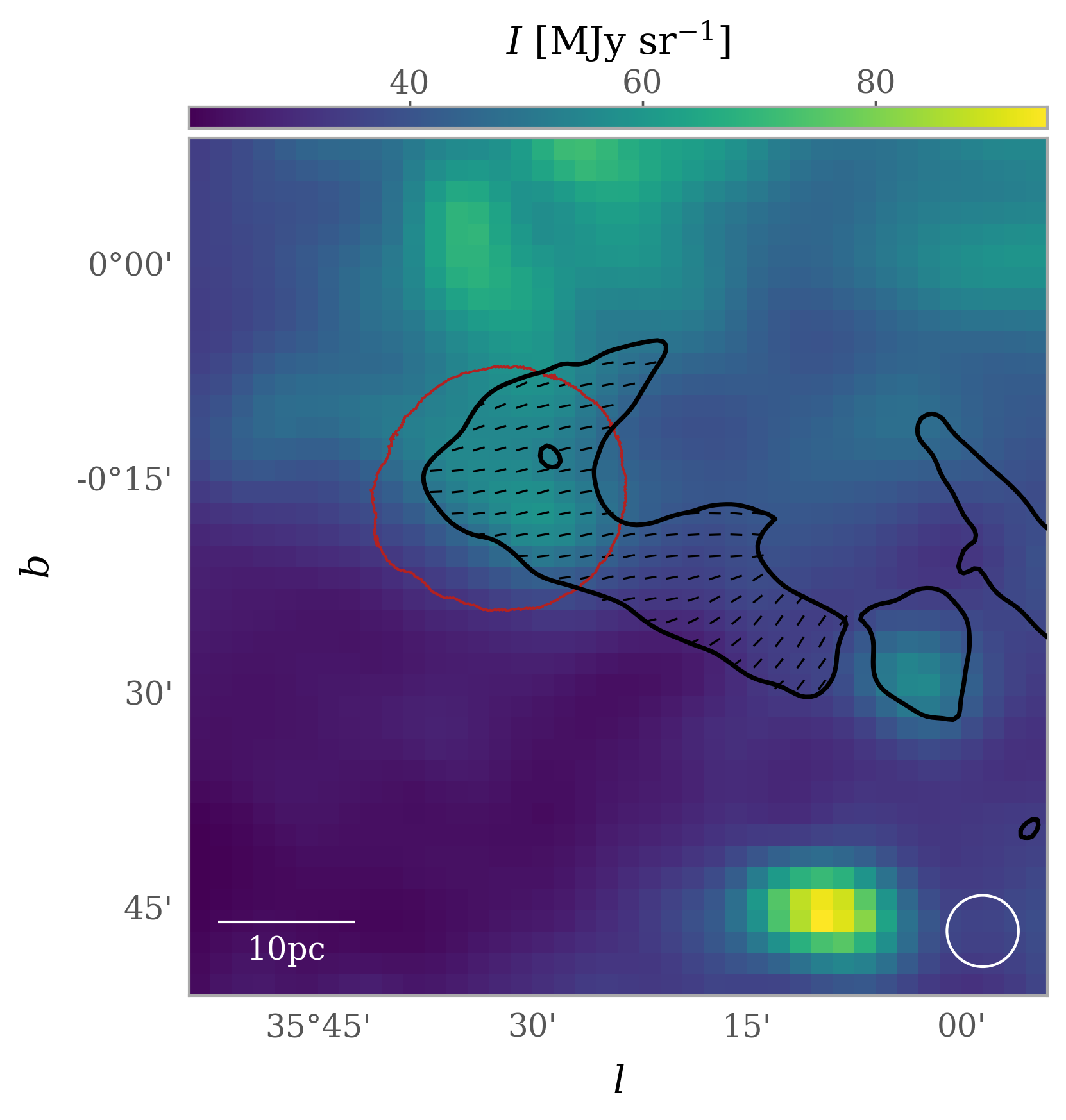}
	\caption{}
	\label{fig:SDC35p527_Planck}
	\end{subfigure}
	\begin{subfigure}[H]{0.528\textwidth}
	\includegraphics[width=\textwidth]{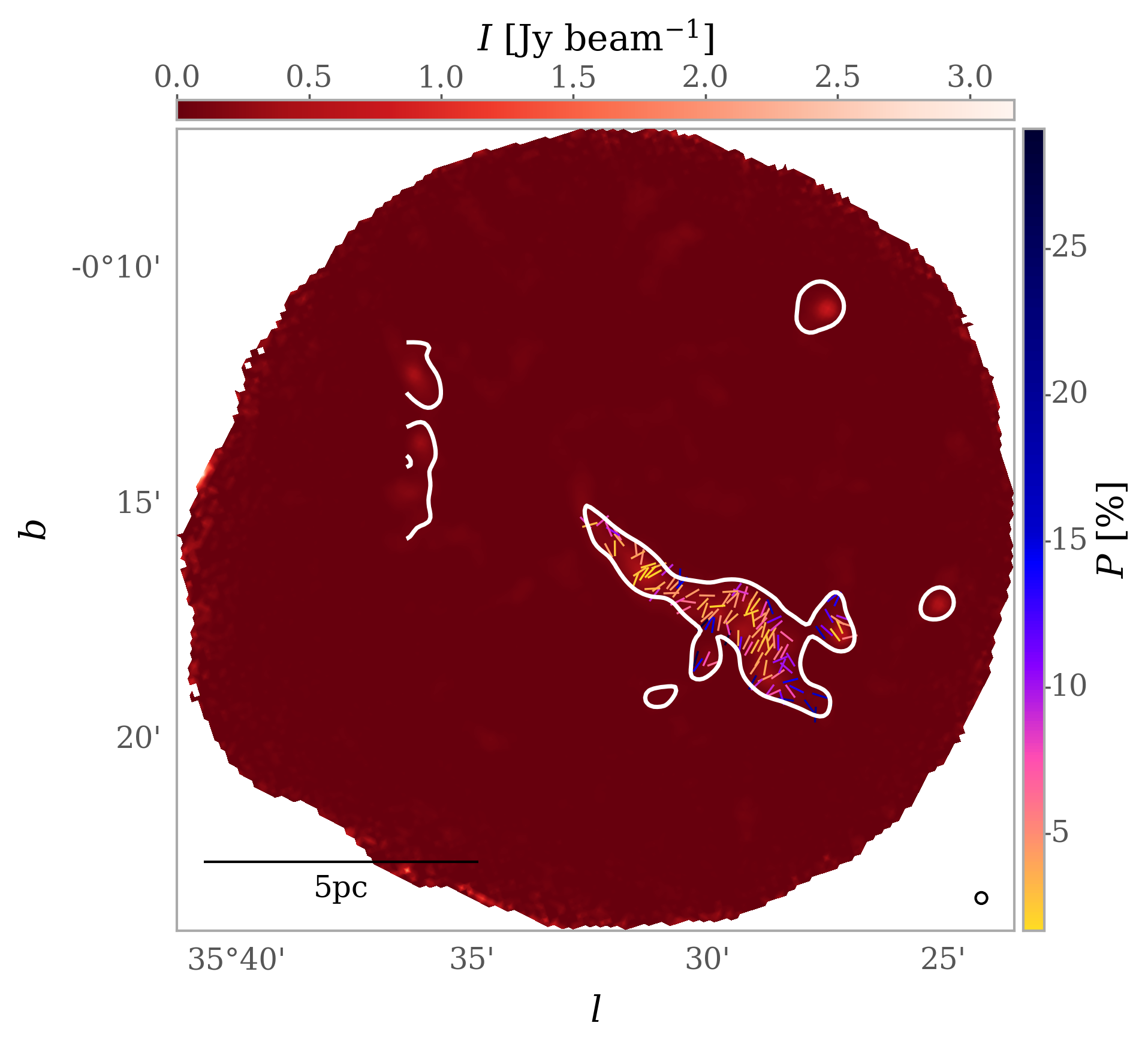}
	\caption{}
	\label{fig:SDC35p527_POL-2}
	\end{subfigure}
	\begin{subfigure}[H]{0.9\textwidth}
    \centering
	\includegraphics[height=0.5\textheight]{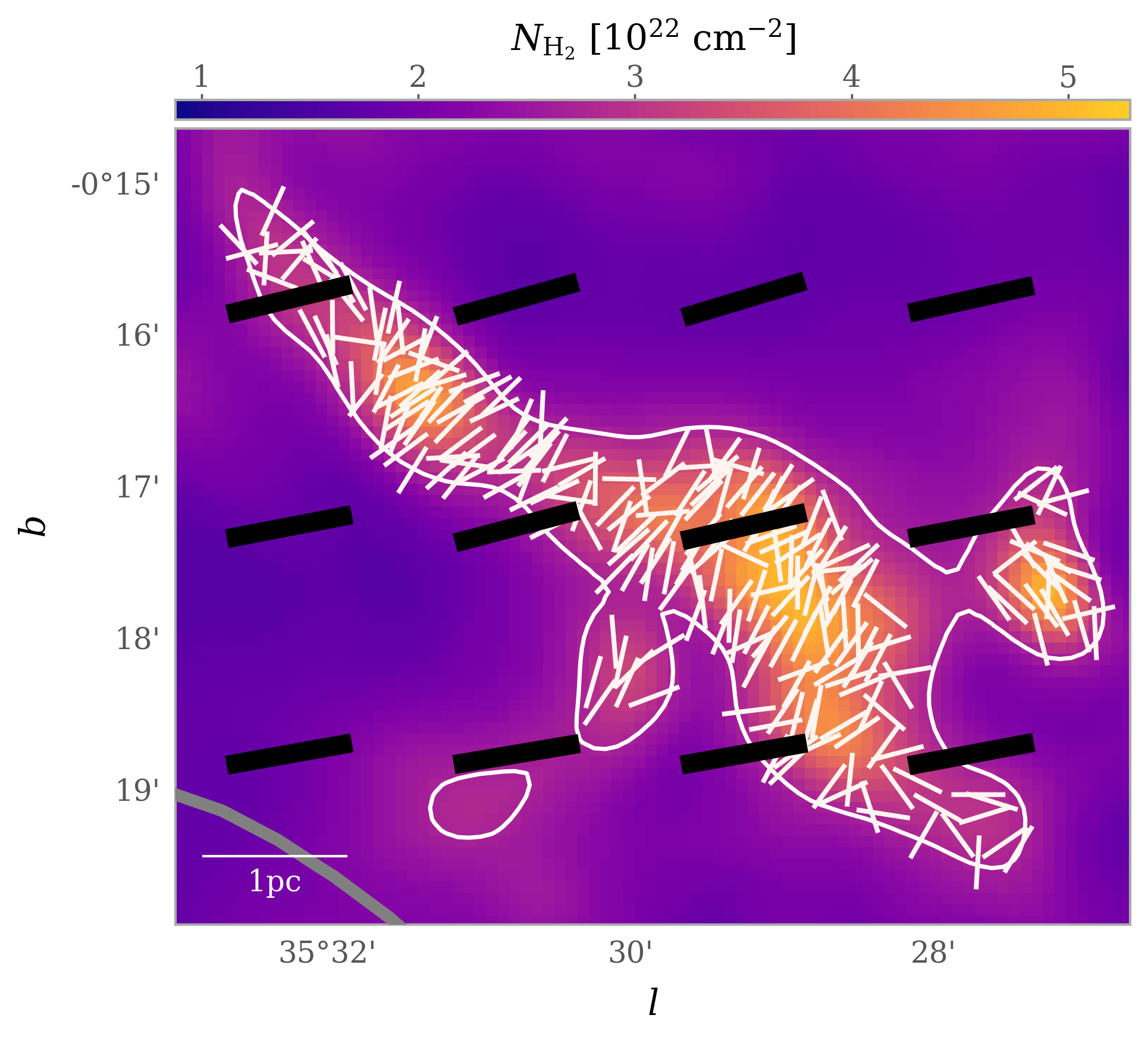}
	\caption{}
	\label{fig:SDC35p527_zoomed}
	\end{subfigure}
    \caption{Same as Figure \ref{fig:magfieldpseudovectors} for SDC35.527.}
    \label{fig:SDC35p527magfieldpseudovectors}
\end{figure*}

\begin{figure*}
	\begin{subfigure}[H]{0.462\textwidth}
	\includegraphics[width=\textwidth]{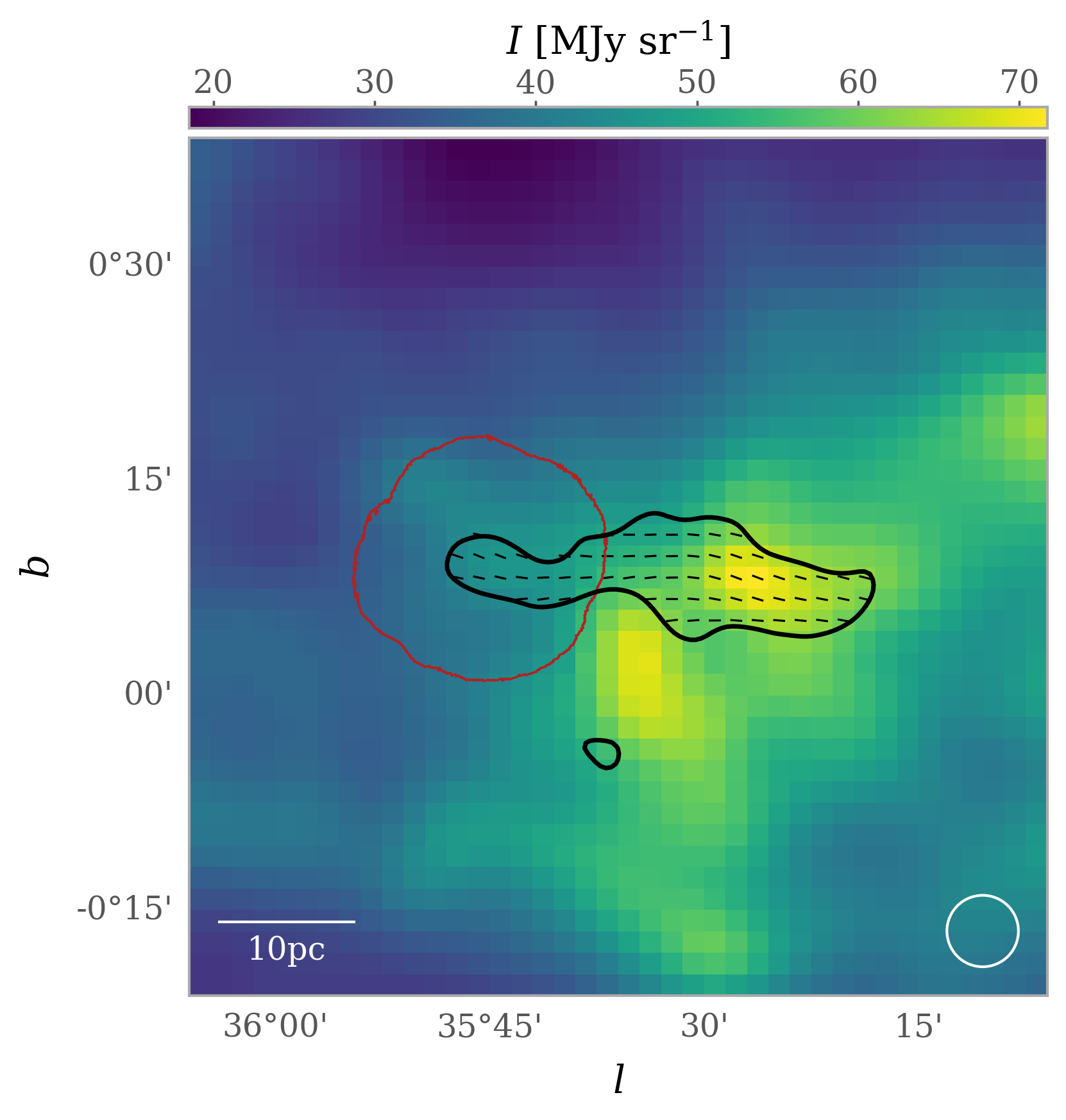}
	\caption{}
	\label{fig:SDC35p745_Planck}
	\end{subfigure}
	\begin{subfigure}[H]{0.528\textwidth}
	\includegraphics[width=\textwidth]{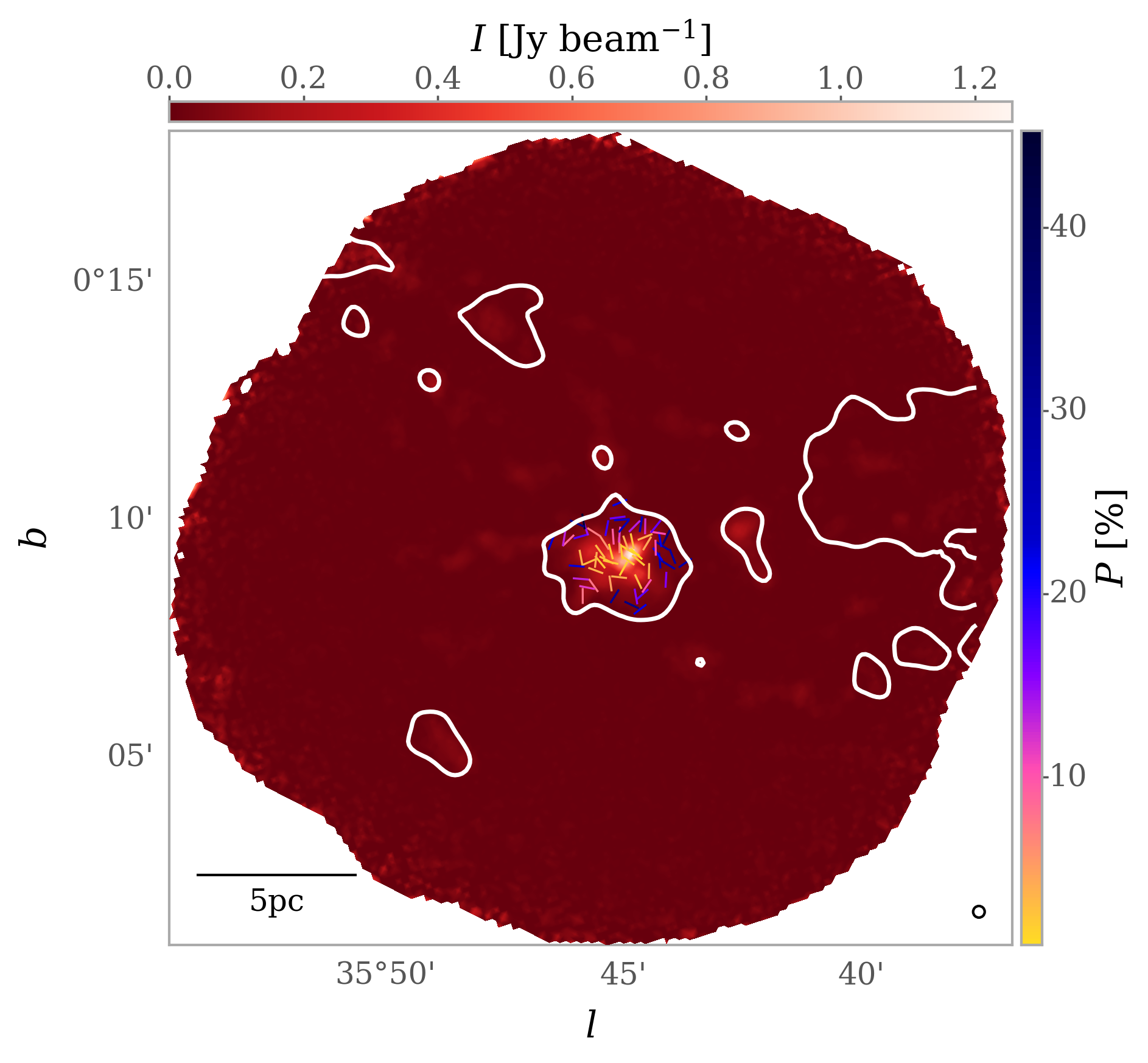}
	\caption{}
	\label{fig:SDC35p745_POL-2}
	\end{subfigure}
	\begin{subfigure}[H]{0.9\textwidth}
    \centering
	\includegraphics[height=0.5\textheight]{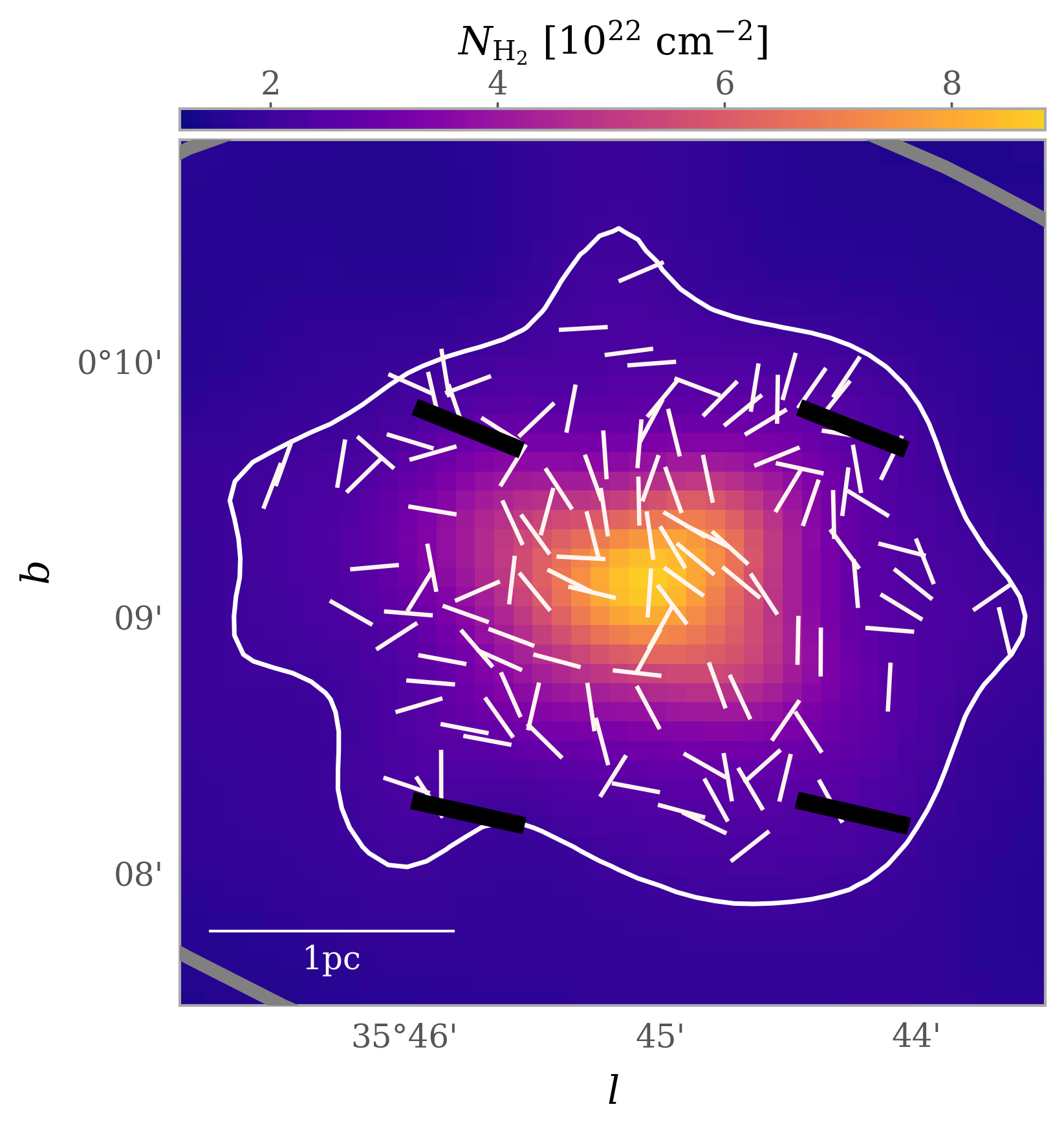}
	\caption{}
	\label{fig:SDC35p745_zoomed}
	\end{subfigure}
    \caption{Same as Figure \ref{fig:magfieldpseudovectors} for SDC35.745.}
    \label{fig:SDC35p745magfieldpseudovectors}
\end{figure*}

\begin{figure*}
	\begin{subfigure}[H]{0.462\textwidth}
	\includegraphics[width=\textwidth]{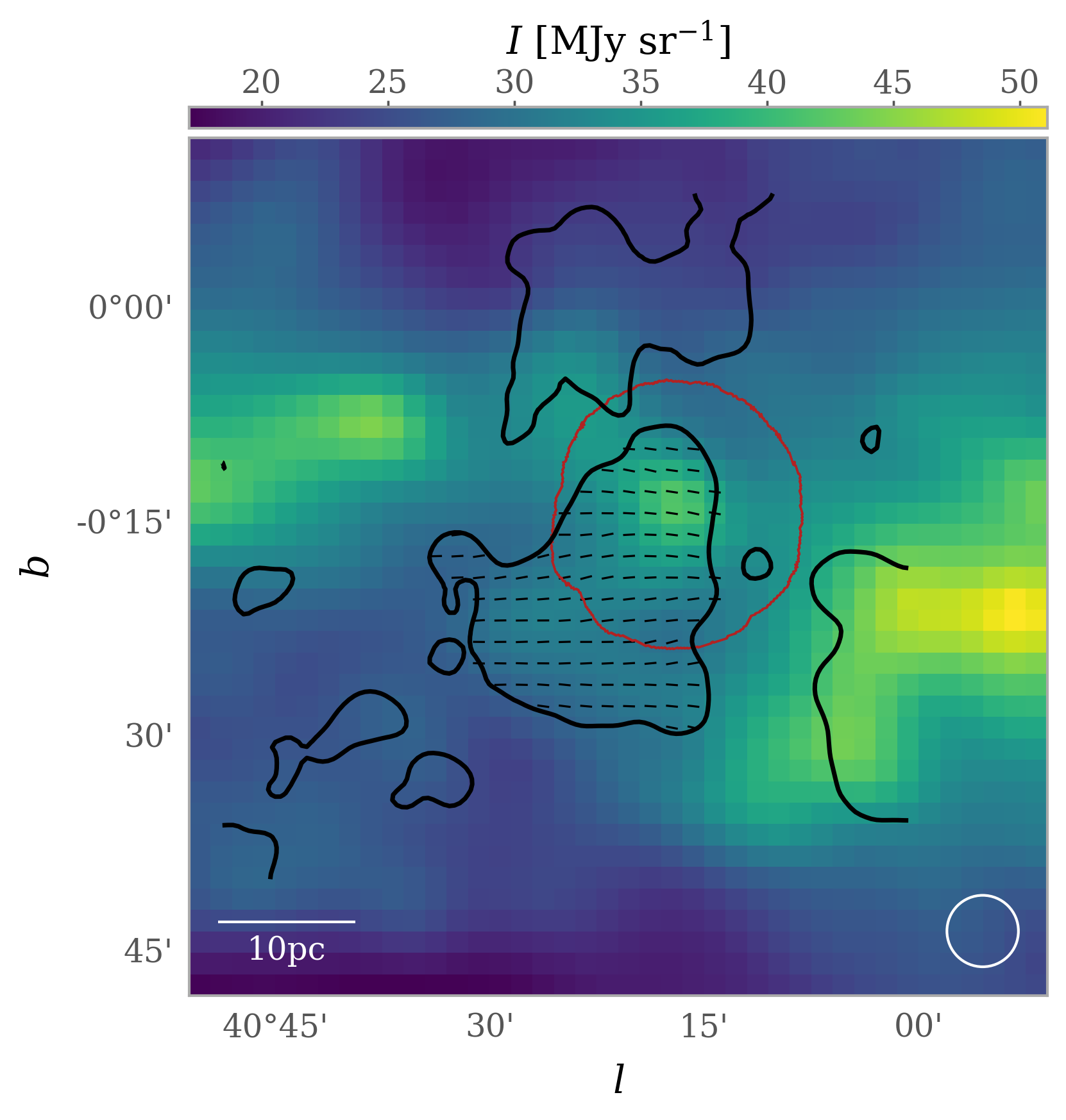}
	\caption{}
	\label{fig:SDC40p283_Planck}
	\end{subfigure}
	\begin{subfigure}[H]{0.49\textwidth}
	\includegraphics[width=\textwidth]{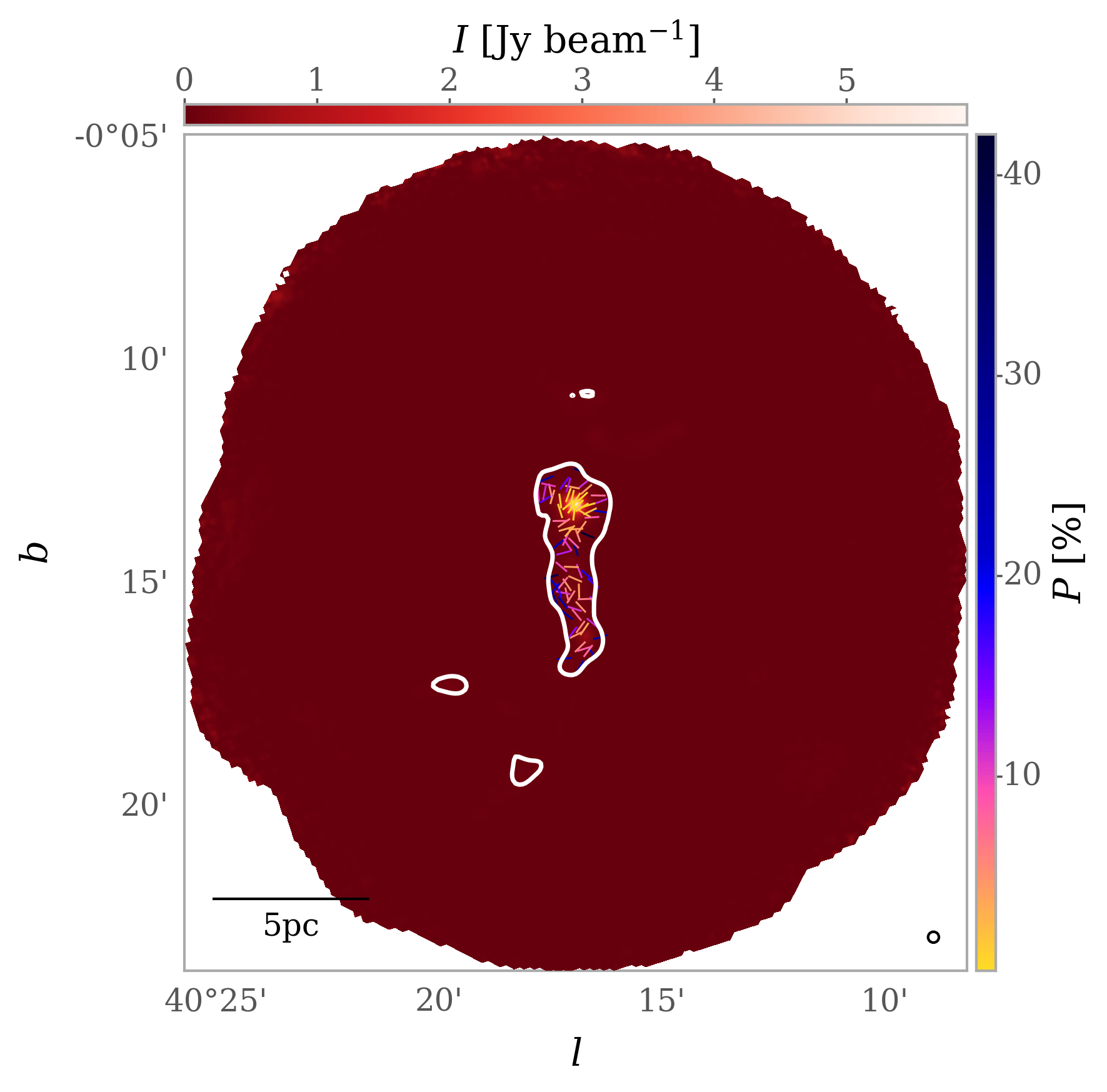}
	\caption{}
	\label{fig:SDC40p283_POL-2}
	\end{subfigure}
	\begin{subfigure}[H]{0.9\textwidth}
    \centering
	\includegraphics[height=0.5\textheight]{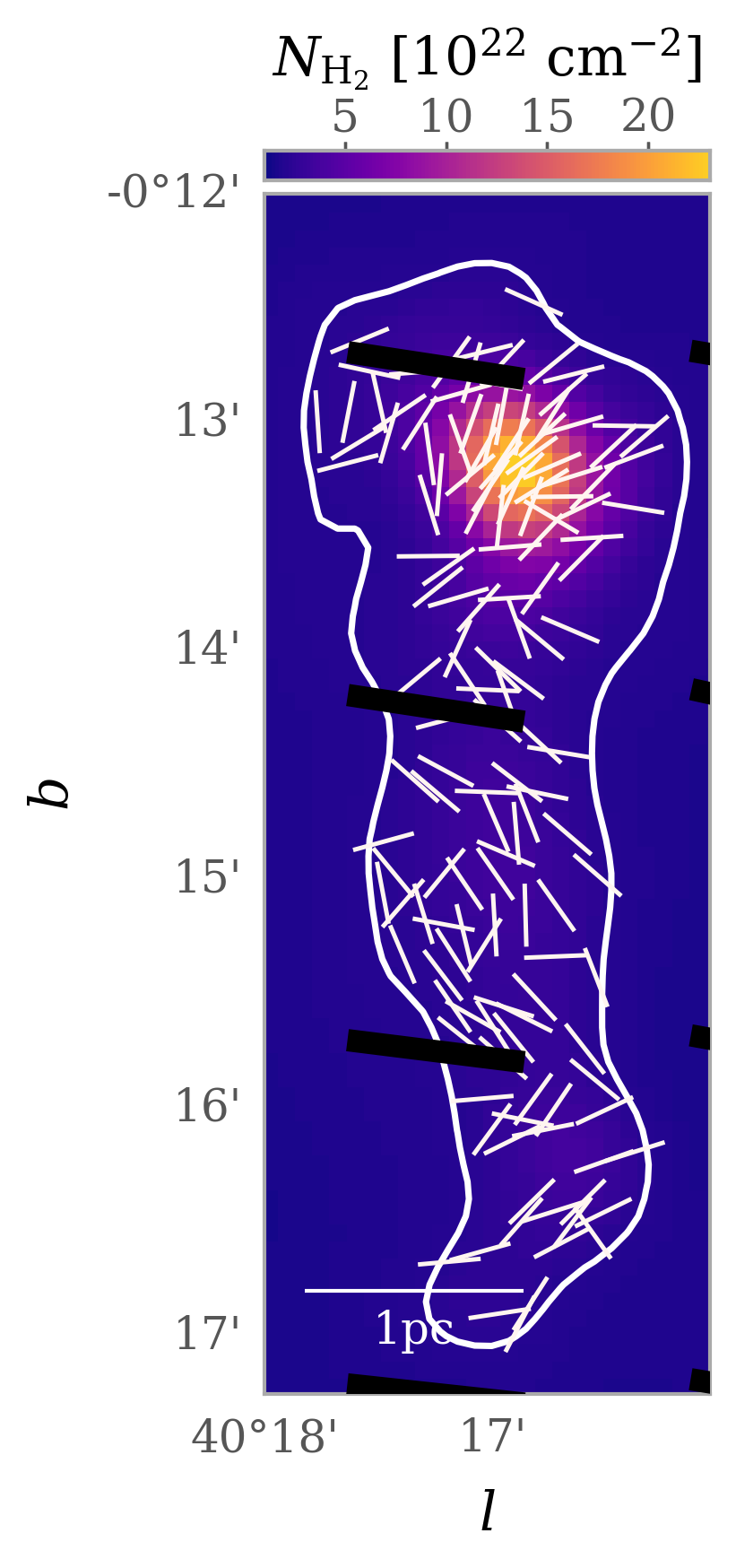}
	\caption{}
	\label{fig:SDC40p283_zoomed}
	\end{subfigure}
    \caption{Same as Figure \ref{fig:magfieldpseudovectors} for SDC40.283.}
    \label{fig:SDC40p283magfieldpseudovectors}
\end{figure*}

\section{Smoothed Histograms}
\label{sec:smoothedhists}

The equivalent histograms to those in Figure \ref{fig:pseudovectorhists} after smoothing the JCMT POL-2 data to {\it Planck} resolution are provided here. The smoothed POL-2 data has been filtered according to $P/\delta P>2$ and $I/\delta I>10$, where $P$, $\delta P$, $I$, and $\delta I$ are also at $5'$ resolution.

\begin{figure*}
	\begin{subfigure}[H]{0.45\textwidth}
	\includegraphics[width=\columnwidth]{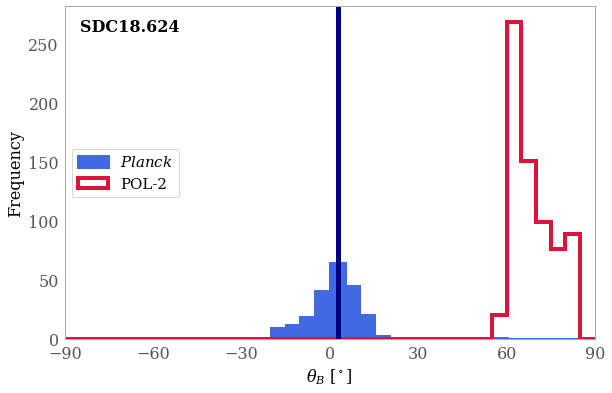}
	\caption{}
	\end{subfigure}
	\begin{subfigure}[H]{0.45\textwidth}
	\includegraphics[width=\columnwidth]{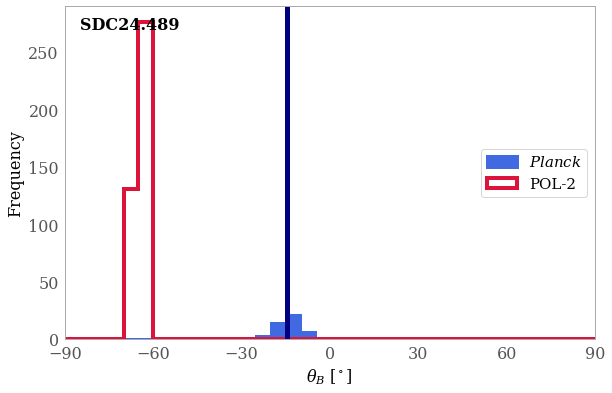}
	\caption{}
	\end{subfigure}
	\begin{subfigure}[H]{0.45\textwidth}
	\includegraphics[width=\columnwidth]{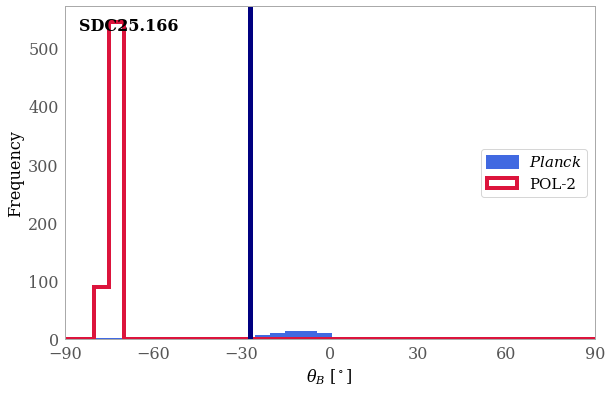}
	\caption{}
	\end{subfigure}
	\begin{subfigure}[H]{0.45\textwidth}
	\includegraphics[width=\columnwidth]{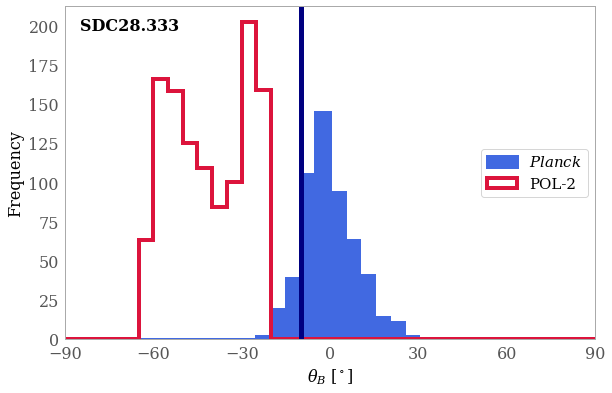}
	\caption{}
	\label{fig:SDC28.333vecthistsmoothed}
	\end{subfigure}
	\begin{subfigure}[H]{0.45\textwidth}
	\includegraphics[width=\columnwidth]{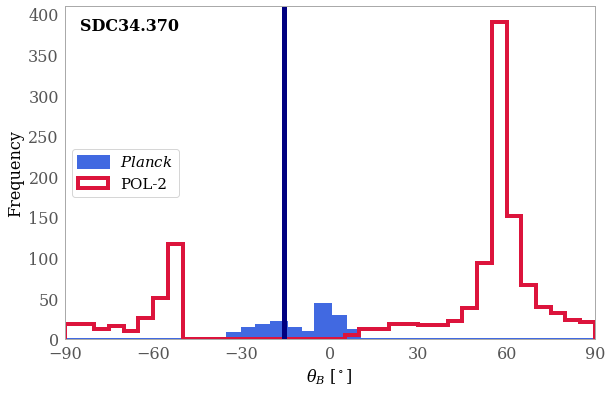}
	\caption{}
	\end{subfigure}
	\begin{subfigure}[H]{0.45\textwidth}
	\includegraphics[width=\columnwidth]{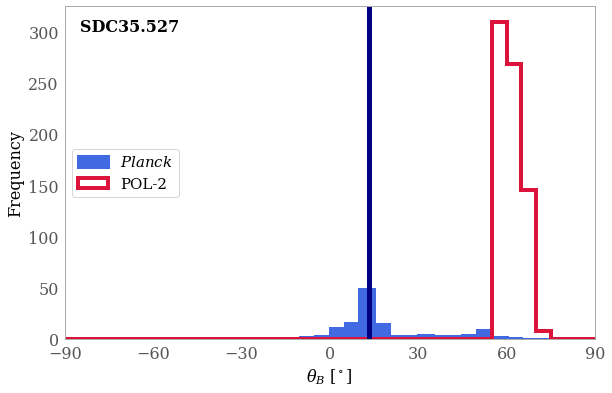}
	\caption{}
	\label{fig:SDC35.527vecthistsmoothed}
	\end{subfigure}
	\begin{subfigure}[H]{0.45\textwidth}
	\includegraphics[width=\columnwidth]{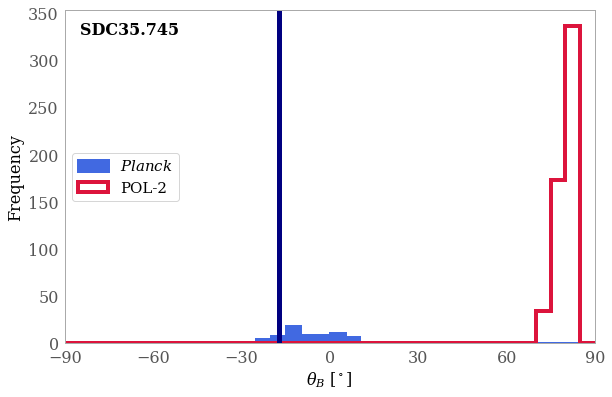}
	\caption{}
	\end{subfigure}
	\begin{subfigure}[H]{0.45\textwidth}
	\includegraphics[width=\columnwidth]{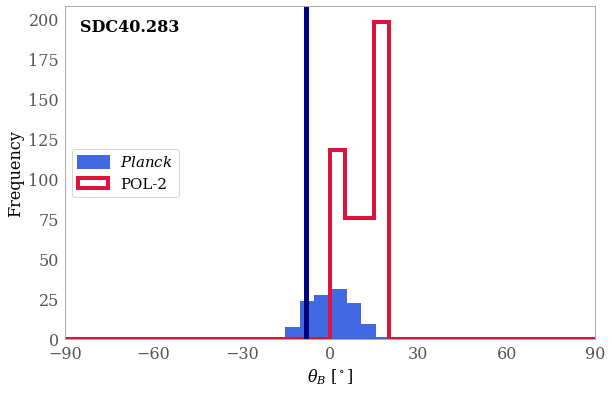}
	\caption{}
	\end{subfigure}
    \caption{Histograms of $\theta_B$ for each clump and its parent cloud, where the clump magnetic field measurements have been smoothed to {\it Planck} resolution. A vertical blue line marks the average angle of Planck pseudovectors falling within the
    boundary of the clump. }
    \label{fig:pseudovectorhistssmoothed}
\end{figure*}

\section{Velocity Spectra}
\label{sec:appendixvelspec}

Here we present the averaged $^{13}$CO(1-0) velocity spectra for the clouds, within their outer boundaries.

\begin{figure*}
	\begin{subfigure}[H]{0.45\textwidth}
	\includegraphics[width=\columnwidth, height=0.22\textheight]{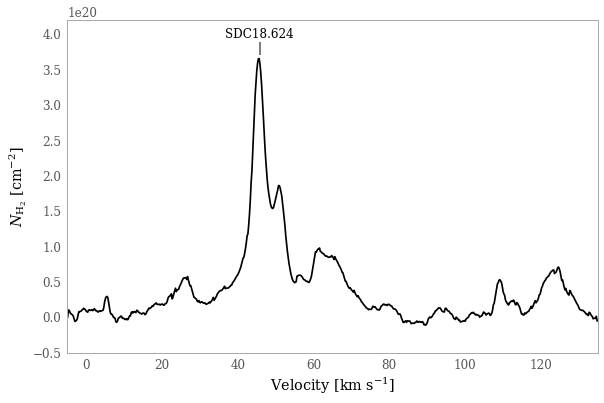}
	\caption{}
	\end{subfigure}
	\begin{subfigure}[H]{0.45\textwidth}
	\includegraphics[width=\columnwidth, height=0.22\textheight]{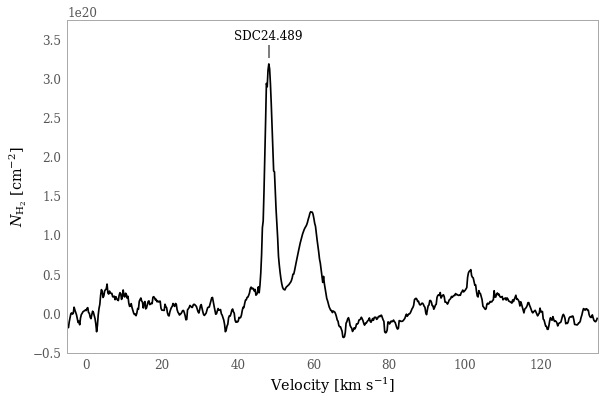}
	\caption{}
	\end{subfigure}
	\begin{subfigure}[H]{0.45\textwidth}
	\includegraphics[width=\columnwidth, height=0.22\textheight]{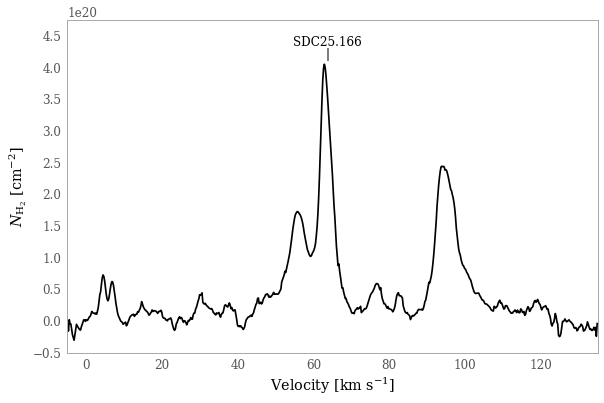}
	\caption{}
	\end{subfigure}
	\begin{subfigure}[H]{0.45\textwidth}
	\includegraphics[width=\columnwidth, height=0.22\textheight]{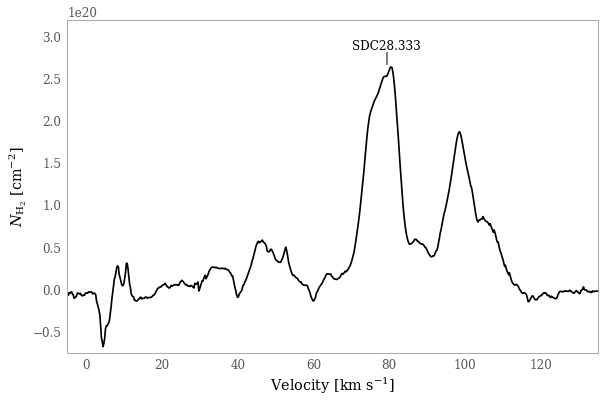}
	\caption{}
	\end{subfigure}
	\begin{subfigure}[H]{0.45\textwidth}
	\includegraphics[width=\columnwidth, height=0.22\textheight]{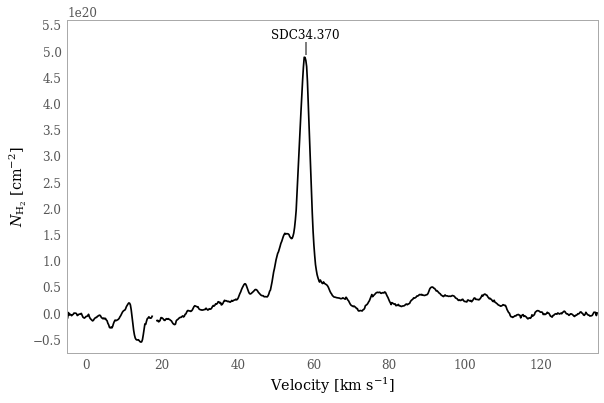}
	\caption{}
	\end{subfigure}
	\begin{subfigure}[H]{0.45\textwidth}
	\includegraphics[width=\columnwidth, height=0.22\textheight]{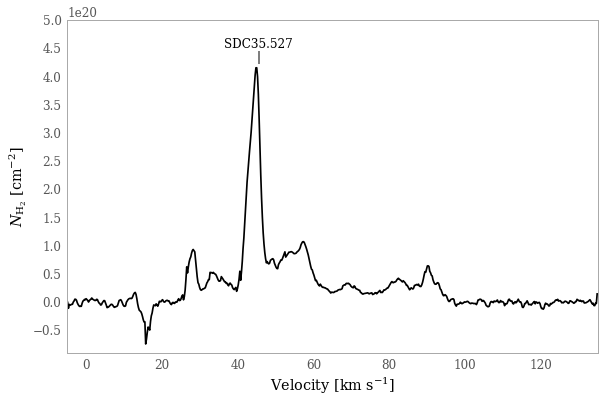}
	\caption{}
	\end{subfigure}
	\begin{subfigure}[H]{0.45\textwidth}
	\includegraphics[width=\columnwidth, height=0.22\textheight]{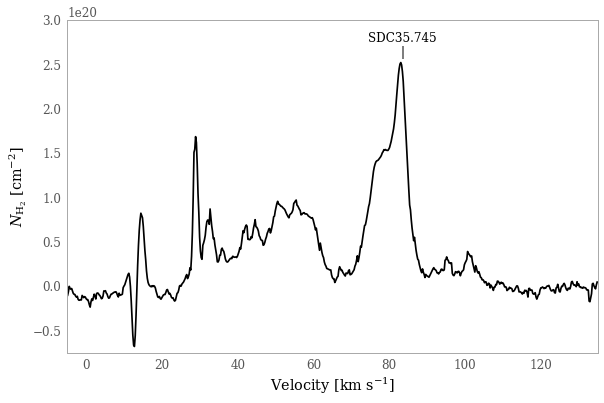}
	\caption{}
	\end{subfigure}
	\begin{subfigure}[H]{0.45\textwidth}
	\includegraphics[width=\columnwidth, height=0.22\textheight]{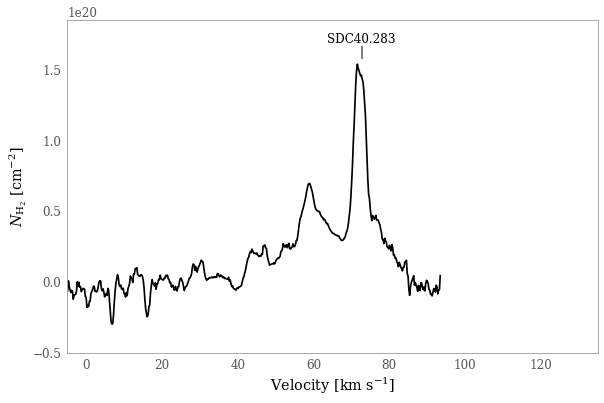}
	\caption{}
	\end{subfigure}
    \caption{Averaged $^{13}$CO(1-0) velocity spectra within the boundary of the main body of the cloud, over the full velocity range of the PPV cube, for: a) SDC18.624 b) SDC24.489 c) SDC25.166 d) SDC28.333 e) SDC34.370 f) SDC35.527 g) SDC35.745 h) SDC40.283.}
    \label{fig:13COvelocityspectra}
\end{figure*}

\section{VGT Images}
\label{sec:appendixVGTimages}

The figures presented here show the results of the application of the VGT to the remaining clouds in the sample, similarly to as shown in Figure \ref{fig:VGT} for SDC34.370.

\begin{figure*}
	\begin{subfigure}[H]{0.425\textwidth}
	\includegraphics[width=\columnwidth]{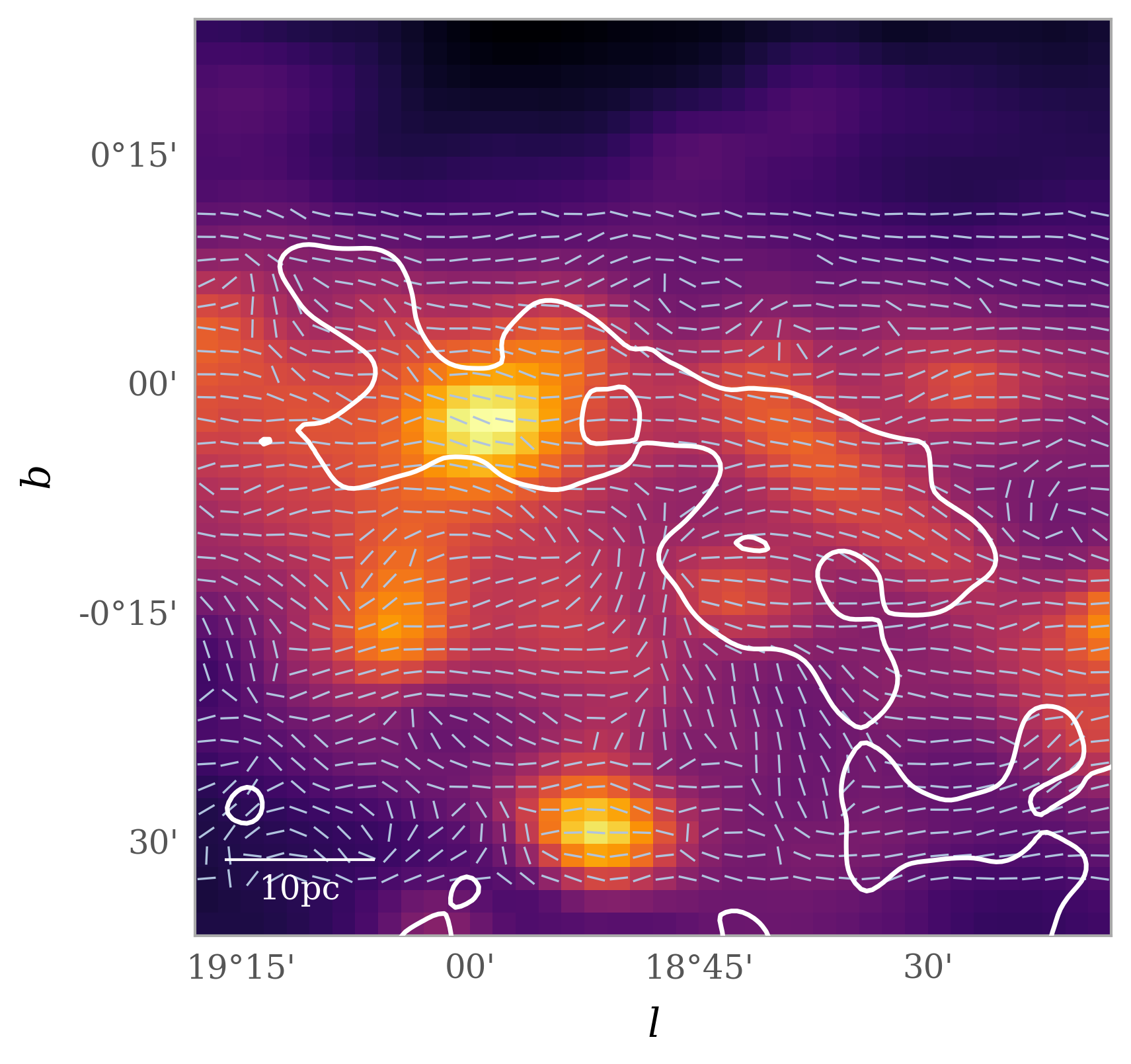}
	\caption{}
	\label{fig:VGT-a-SDC18p624}
	\end{subfigure}
	\begin{subfigure}[H]{0.475\textwidth}
	\includegraphics[width=\columnwidth]{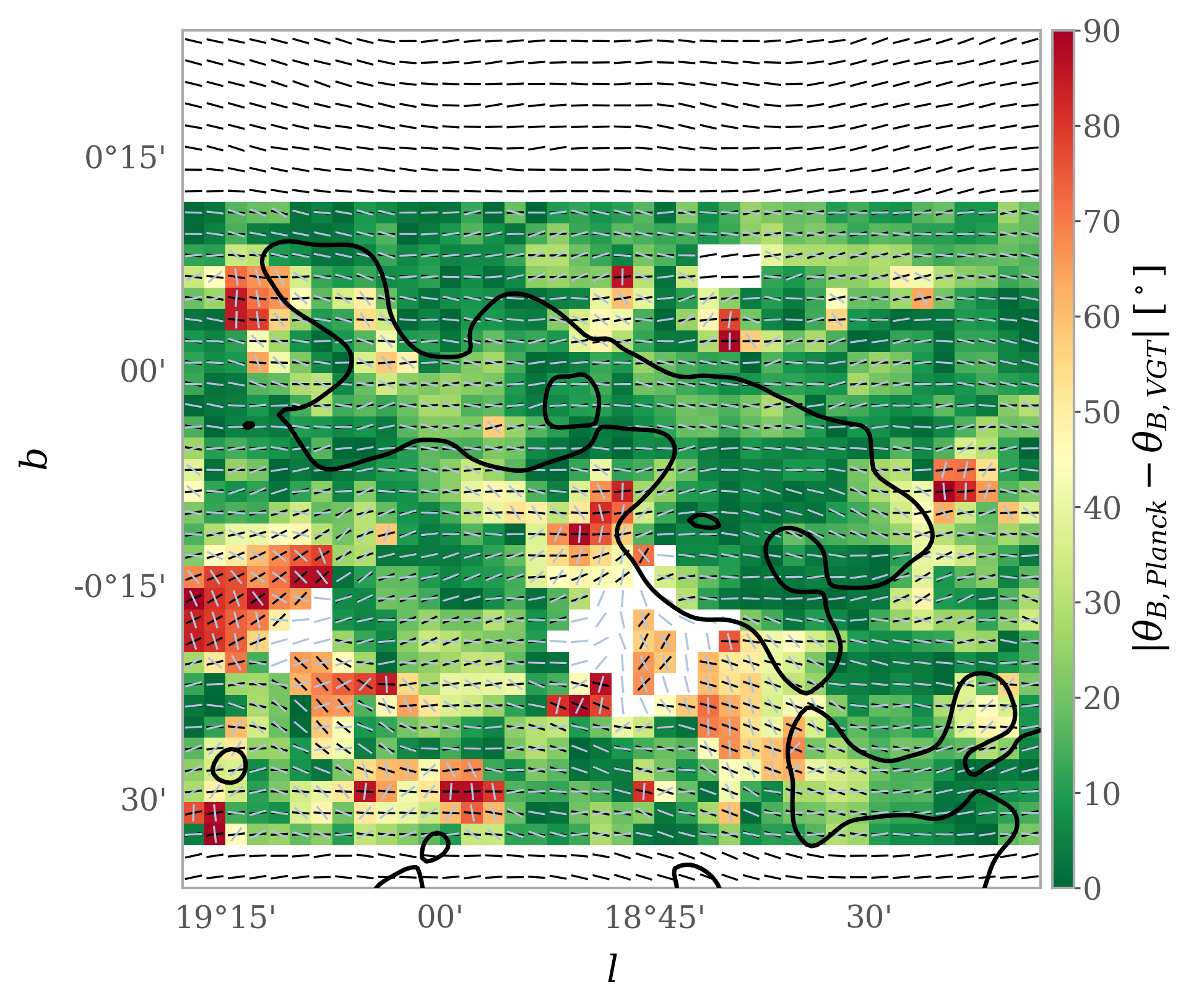}
	\caption{}
	\label{fig:VGT-b-SDC18p624}
	\end{subfigure}
    \caption{Same as Figure \ref{fig:VGT} for SDC18.624}
    \label{fig:VGT-SDC18p624}
\end{figure*}

\begin{figure*}
	\begin{subfigure}[H]{0.425\textwidth}
	\includegraphics[width=\columnwidth]{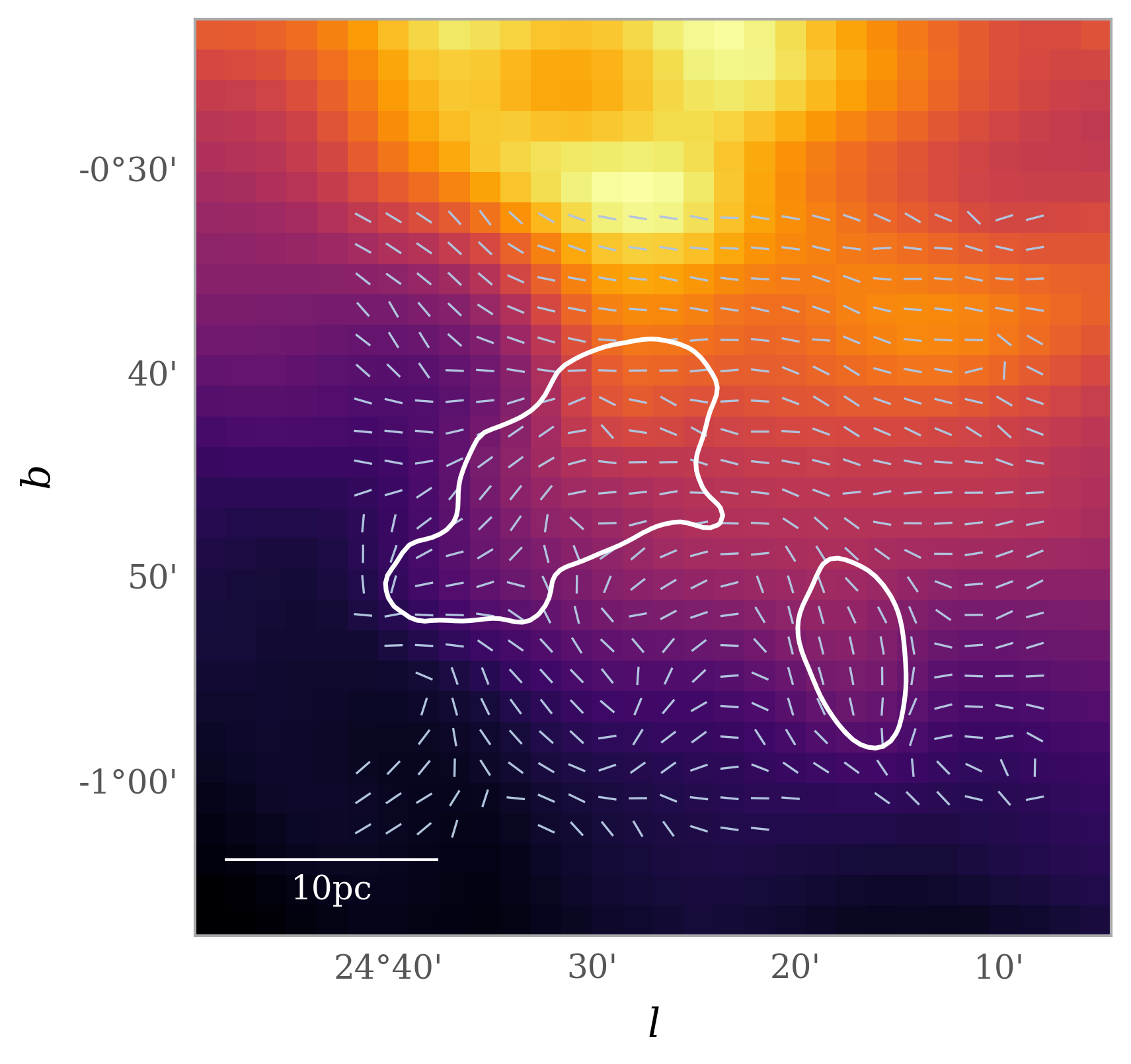}
	\caption{}
	\label{fig:VGT-a-SDC24p489}
	\end{subfigure}
	\begin{subfigure}[H]{0.475\textwidth}
	\includegraphics[width=\columnwidth]{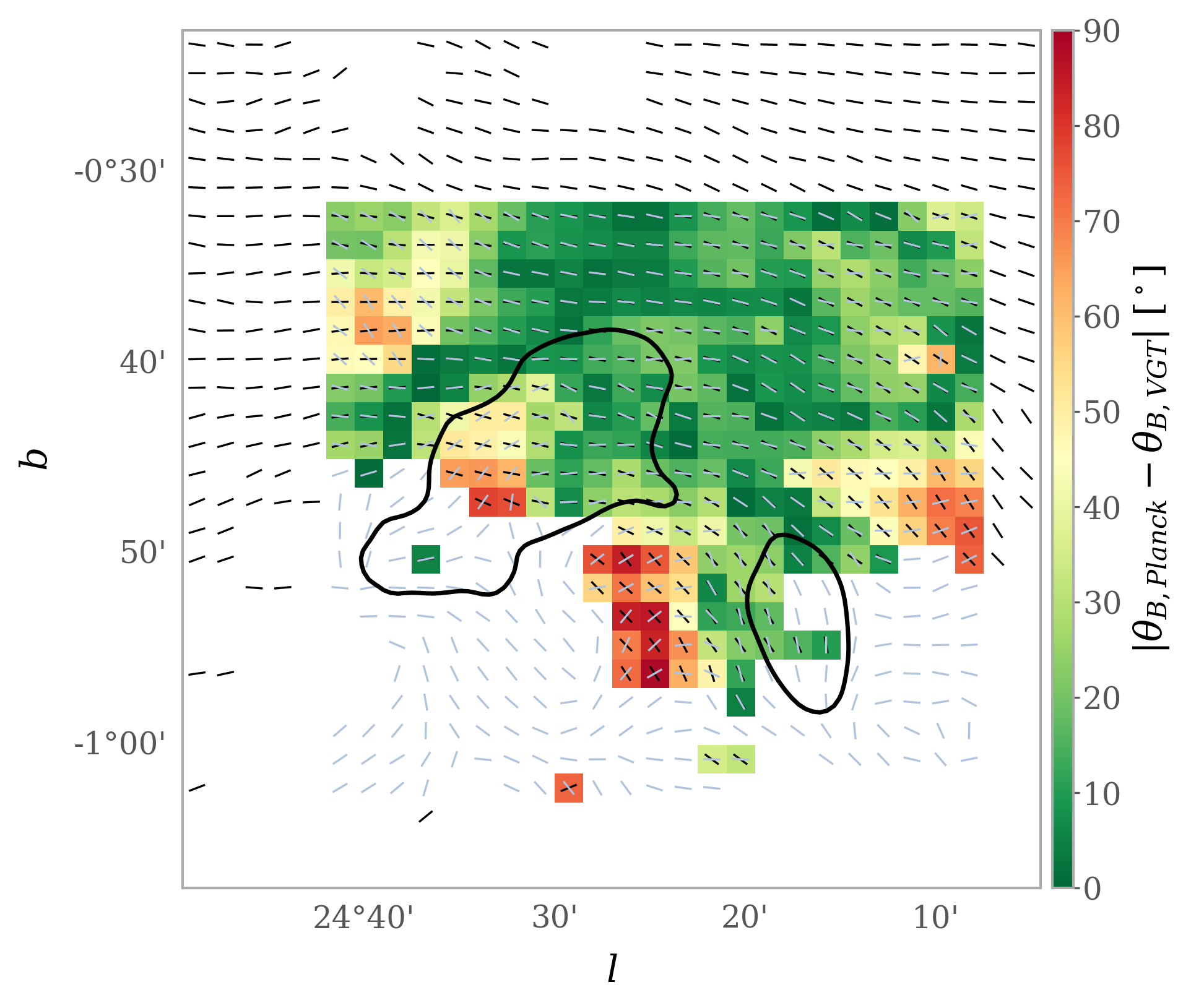}
	\caption{}
	\label{fig:VGT-b-SDC24p489}
	\end{subfigure}
    \caption{Same as Figure \ref{fig:VGT} for SDC24.489}
    \label{fig:VGT-SDC24p489}
\end{figure*}

\begin{figure*}
	\begin{subfigure}[H]{0.425\textwidth}
	\includegraphics[width=\columnwidth]{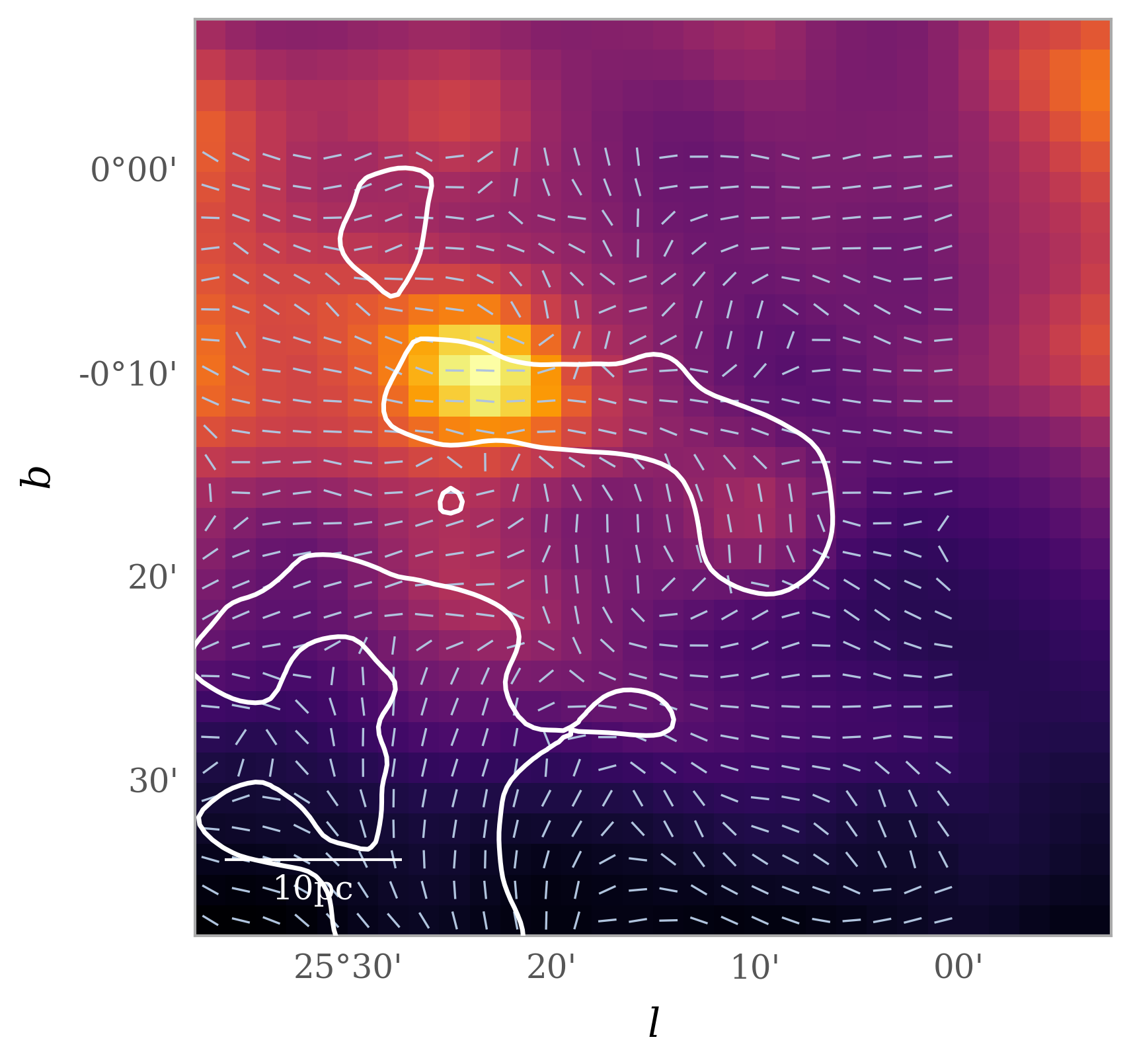}
	\caption{}
	\label{fig:VGT-a-SDC25p166}
	\end{subfigure}
	\begin{subfigure}[H]{0.475\textwidth}
	\includegraphics[width=\columnwidth]{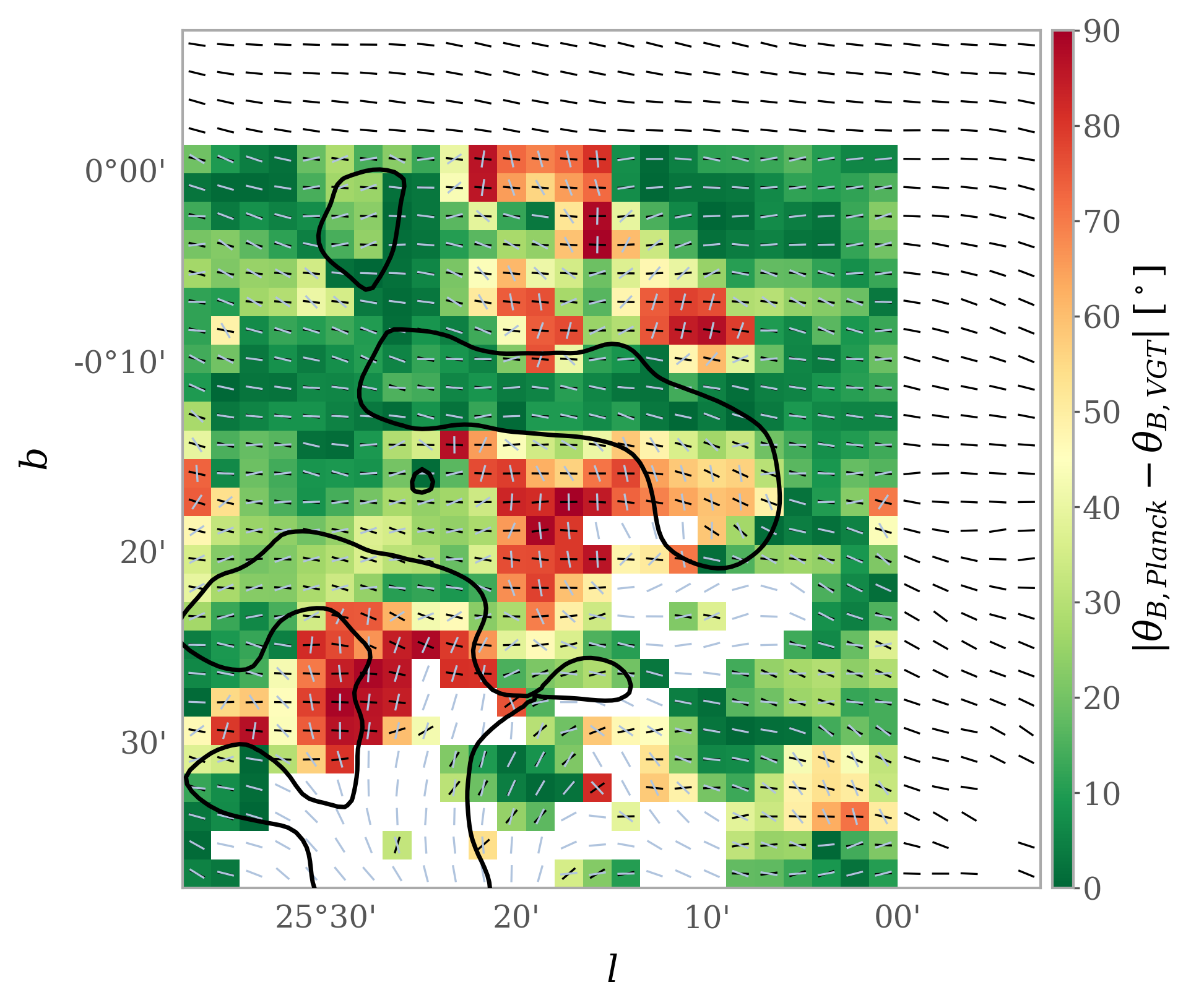}
	\caption{}
	\label{fig:VGT-b-SDC25p166}
	\end{subfigure}
    \caption{Same as Figure \ref{fig:VGT} for SDC25.166}
    \label{fig:VGT-SDC25p166}
\end{figure*}

\begin{figure*}
	\begin{subfigure}[H]{0.425\textwidth}
	\includegraphics[width=\columnwidth]{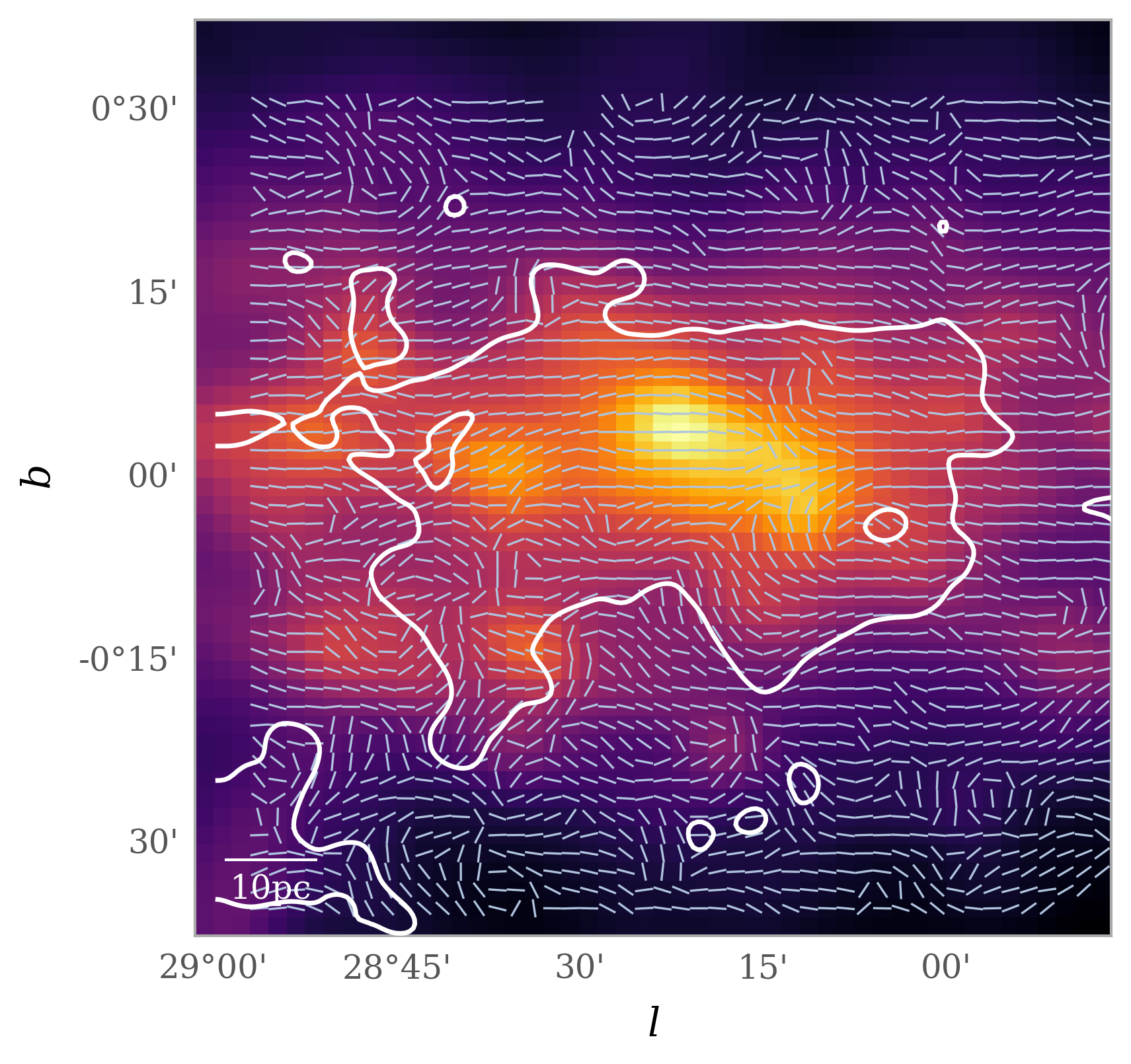}
	\caption{}
	\label{fig:VGT-a-SDC28p333}
	\end{subfigure}
	\begin{subfigure}[H]{0.475\textwidth}
	\includegraphics[width=\columnwidth]{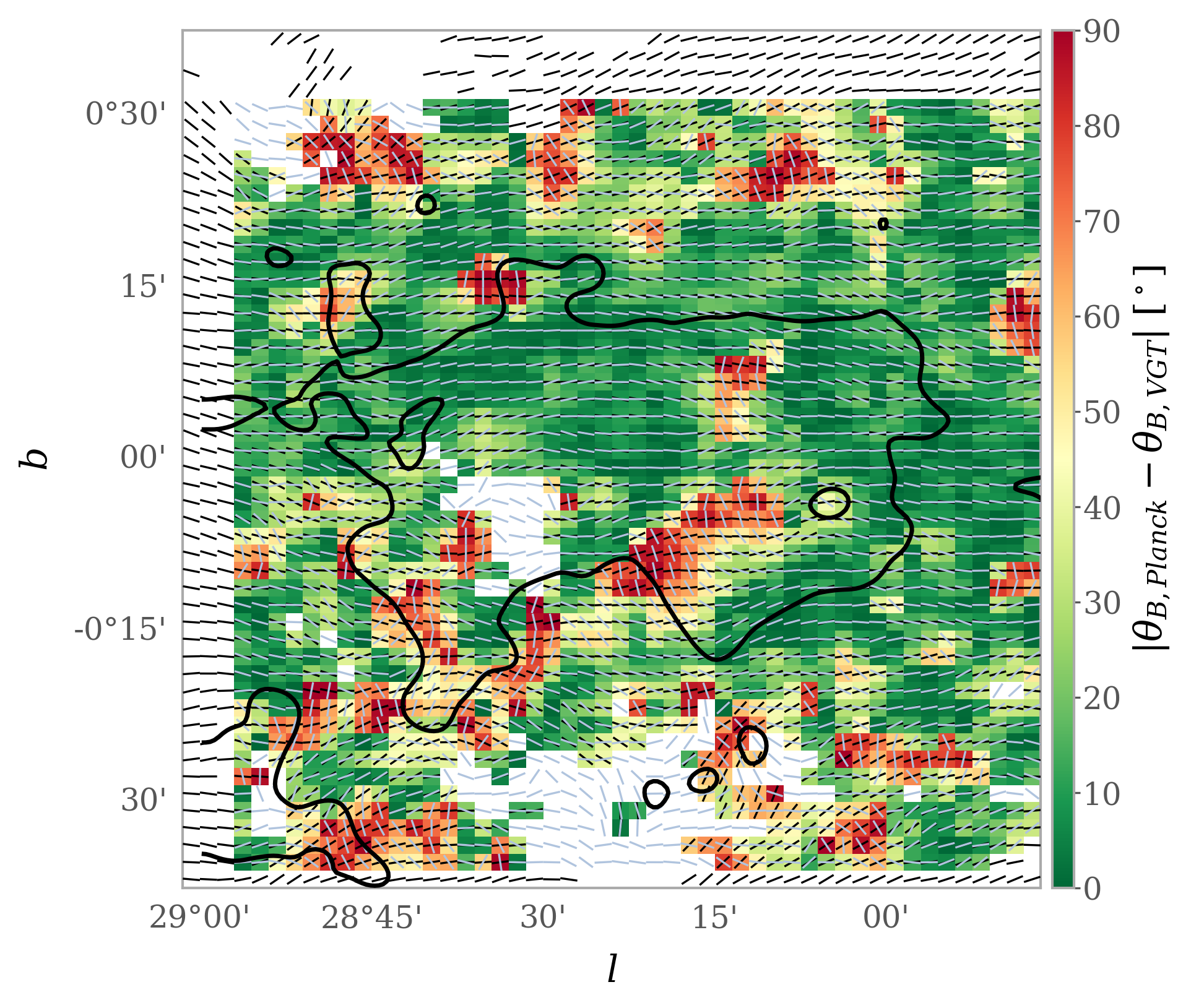}
	\caption{}
	\label{fig:VGT-b-SDC28p333}
	\end{subfigure}
    \caption{Same as Figure \ref{fig:VGT} for SDC28.333}
    \label{fig:VGT-SDC28p333}
\end{figure*}

\begin{figure*}
	\begin{subfigure}[H]{0.425\textwidth}
	\includegraphics[width=\columnwidth]{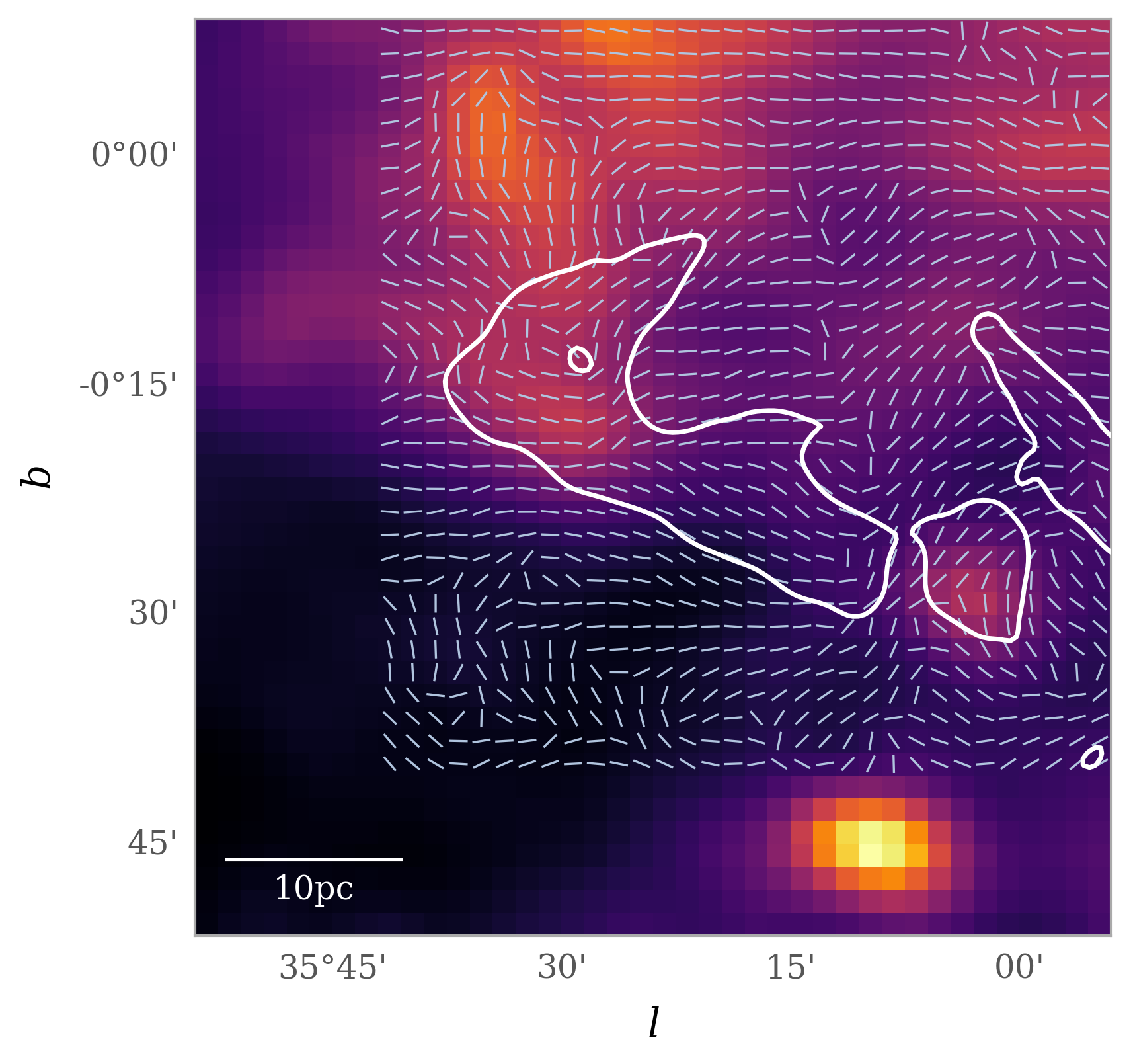}
	\caption{}
	\label{fig:VGT-a-SDC35p527}
	\end{subfigure}
	\begin{subfigure}[H]{0.475\textwidth}
	\includegraphics[width=\columnwidth]{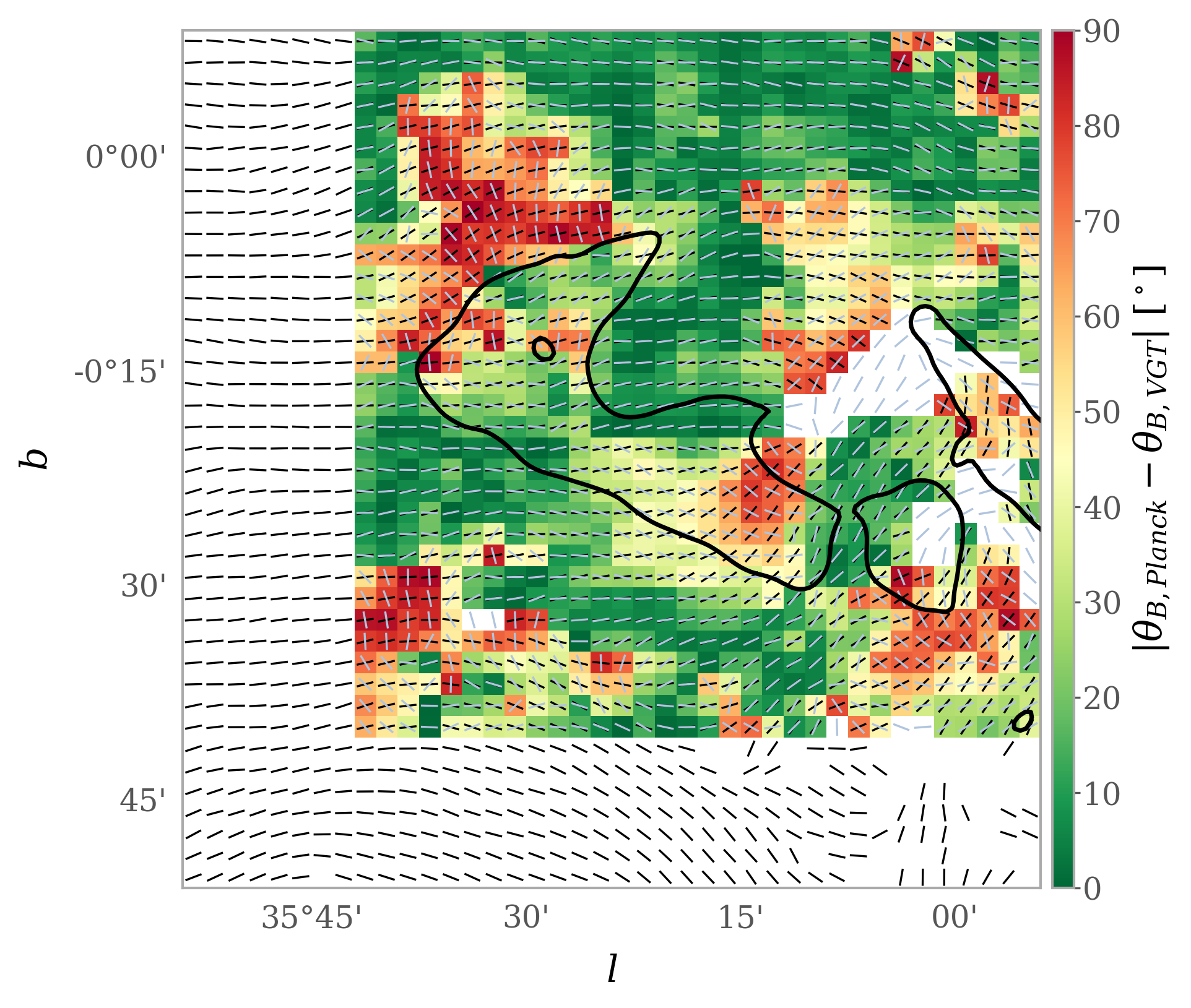}
	\caption{}
	\label{fig:VGT-b-SDC35p527}
	\end{subfigure}
    \caption{Same as Figure \ref{fig:VGT} for SDC35.527}
    \label{fig:VGT-SDC35p527}
\end{figure*}

\begin{figure*}
	\begin{subfigure}[H]{0.425\textwidth}
	\includegraphics[width=\columnwidth]{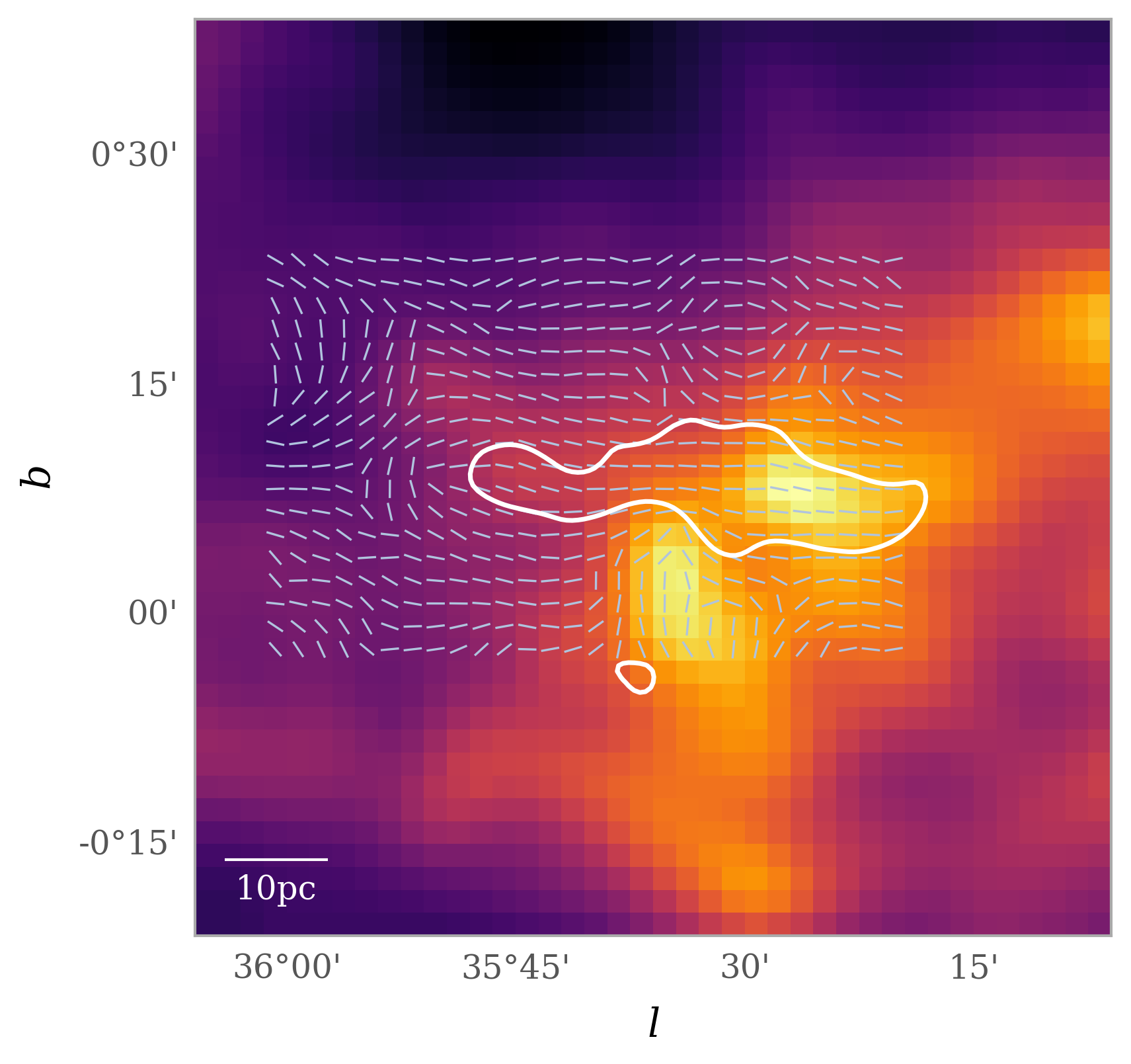}
	\caption{}
	\label{fig:VGT-a-SDC35p745}
	\end{subfigure}
	\begin{subfigure}[H]{0.475\textwidth}
	\includegraphics[width=\columnwidth]{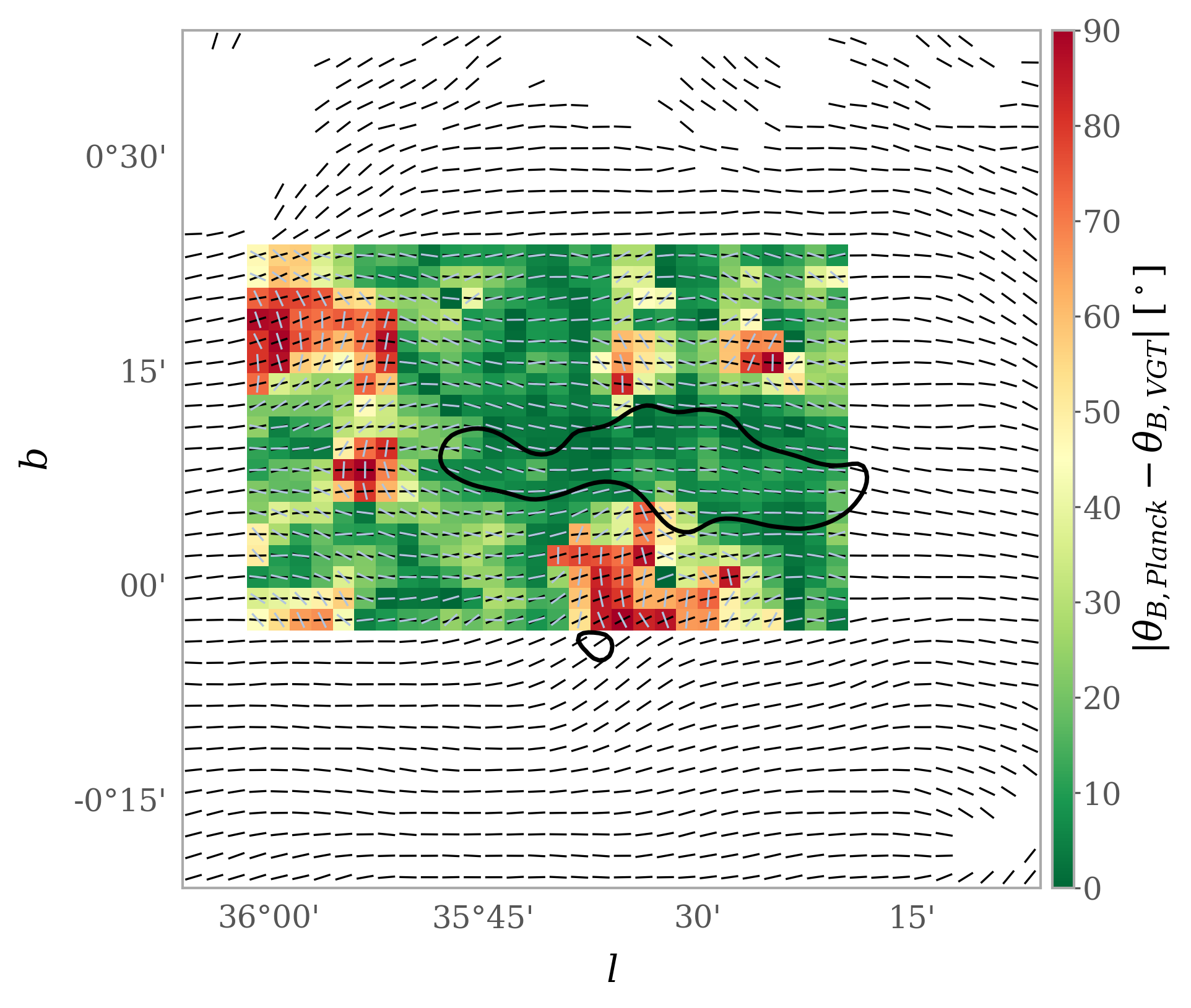}
	\caption{}
	\label{fig:VGT-b-SDC35p745}
	\end{subfigure}
    \caption{Same as Figure \ref{fig:VGT} for SDC35.745}
    \label{fig:VGT-SDC35p745}
\end{figure*}

\begin{figure*}
	\begin{subfigure}[H]{0.425\textwidth}
	\includegraphics[width=\columnwidth]{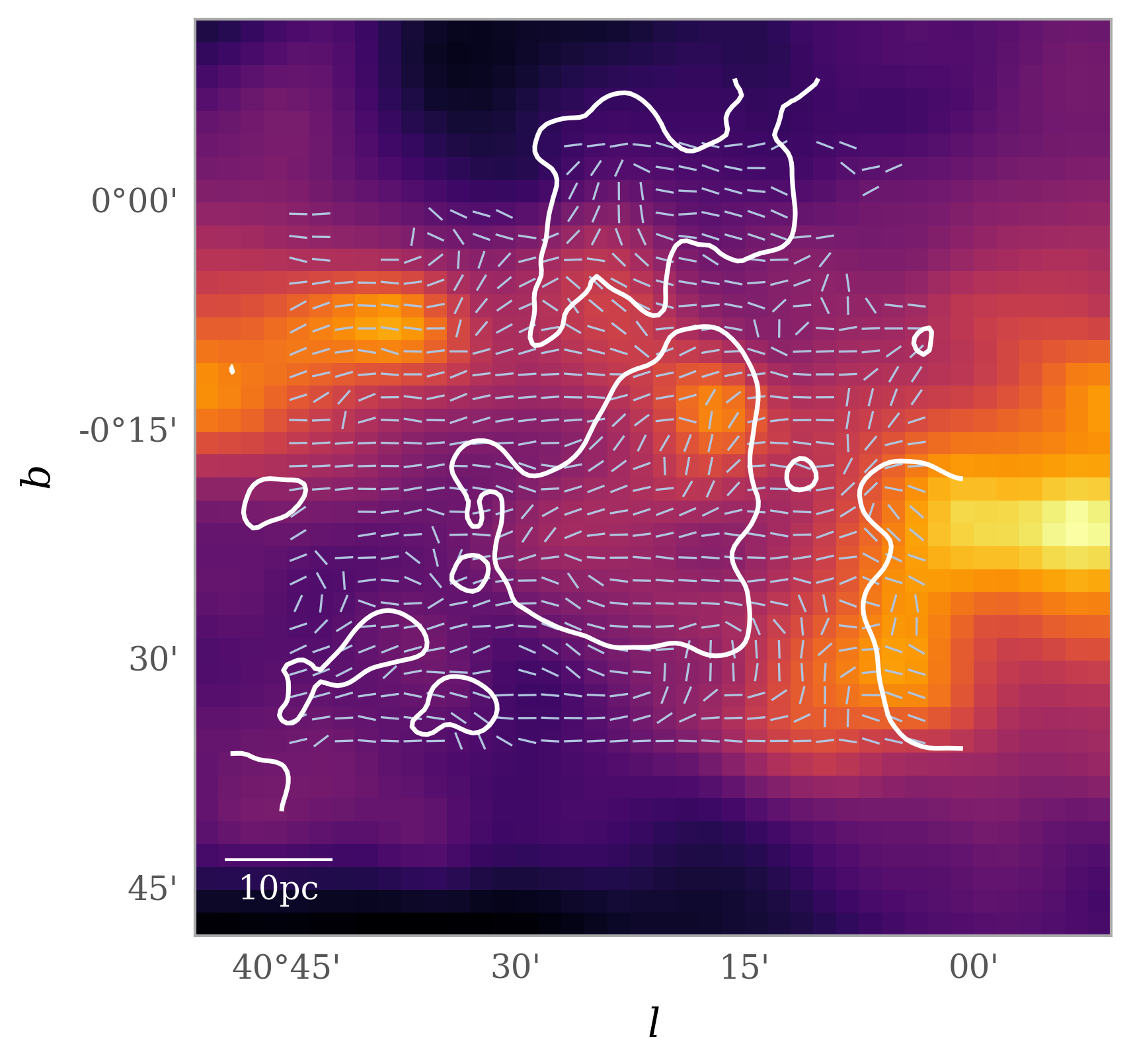}
	\caption{}
	\label{fig:VGT-a-SDC40p283}
	\end{subfigure}
	\begin{subfigure}[H]{0.475\textwidth}
	\includegraphics[width=\columnwidth]{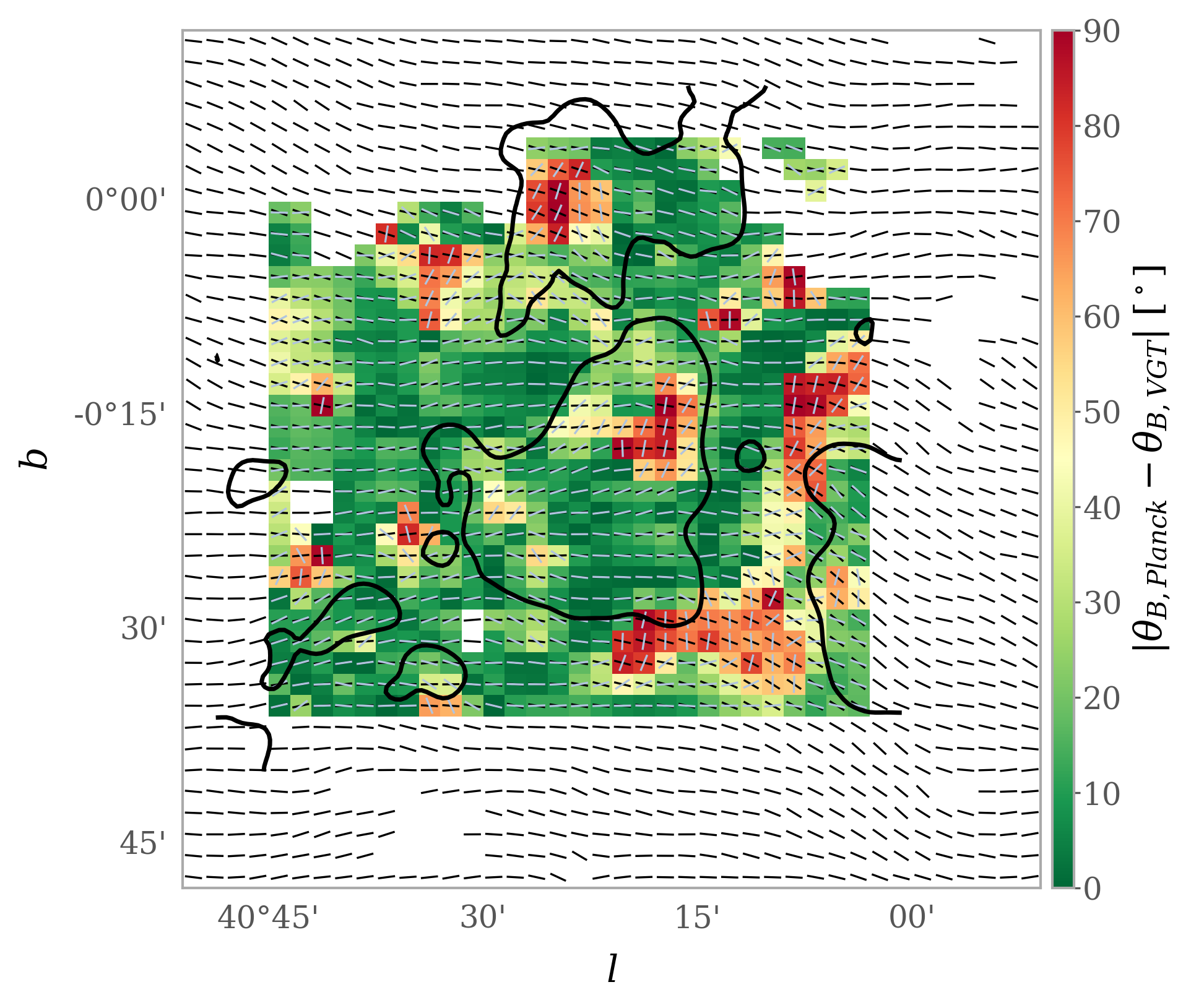}
	\caption{}
	\label{fig:VGT-b-SDC40p283}
	\end{subfigure}
    \caption{Same as Figure \ref{fig:VGT} for SDC40.283}
    \label{fig:VGT-SDC40p283}
\end{figure*}

\section{VGT Block Size Comparison}
\label{sec:appendixVGTblocksize}

\begin{figure*}
	\includegraphics[width=\textwidth]{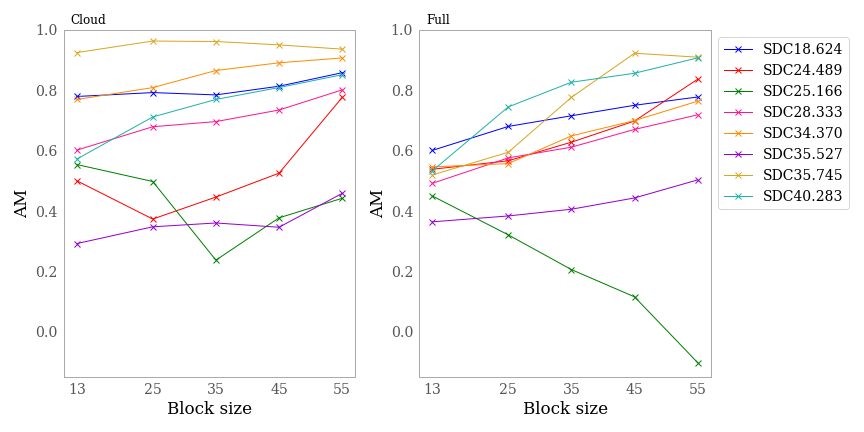}
	\caption{How the AM of the VGT changes with the block size used in the subblock averaging step within the outer contour of the cloud and for the full {\it Planck} image. The block size is given in units of pixels of the $^{13}$CO(1-0) data.}
    \label{fig:vgtblocksize}
\end{figure*}

Here we compare the results of the VGT applied across the full velocity range of the $^{13}$CO(1-0) PPV cube when using different block sizes for the subblock-averaging step. Figure \ref{fig:vgtblocksize} shows the results of this investigation. It should be noted that the total number of valid pseudovectors gained from the VGT may vary between different clouds and block sizes. We can see that for the full image the AM tends to increase as block size increases but usually not by a considerable amount. Within the boundary of the cloud the AM stays relatively constant as block size increases for most clouds, slightly increasing for some and slightly decreasing for others. Since increasing the block size does not seem to make a large difference to the results of the VGT in most cases, our chosen block size of 13$\times$13 (corresponding to a spatial scale of 4.8'$\times$4.8'), which is close to the {\it Planck} beam size, is adequate.

\section{ADF Fits}
\label{sec:appendixADF}

The ADF plots and their best-fitting functions for both {\it Planck} and POL-2 scales are shown here for all sources.

\begin{figure*}
	\begin{subfigure}[H]{0.49\textwidth}
	\includegraphics[width=\columnwidth]{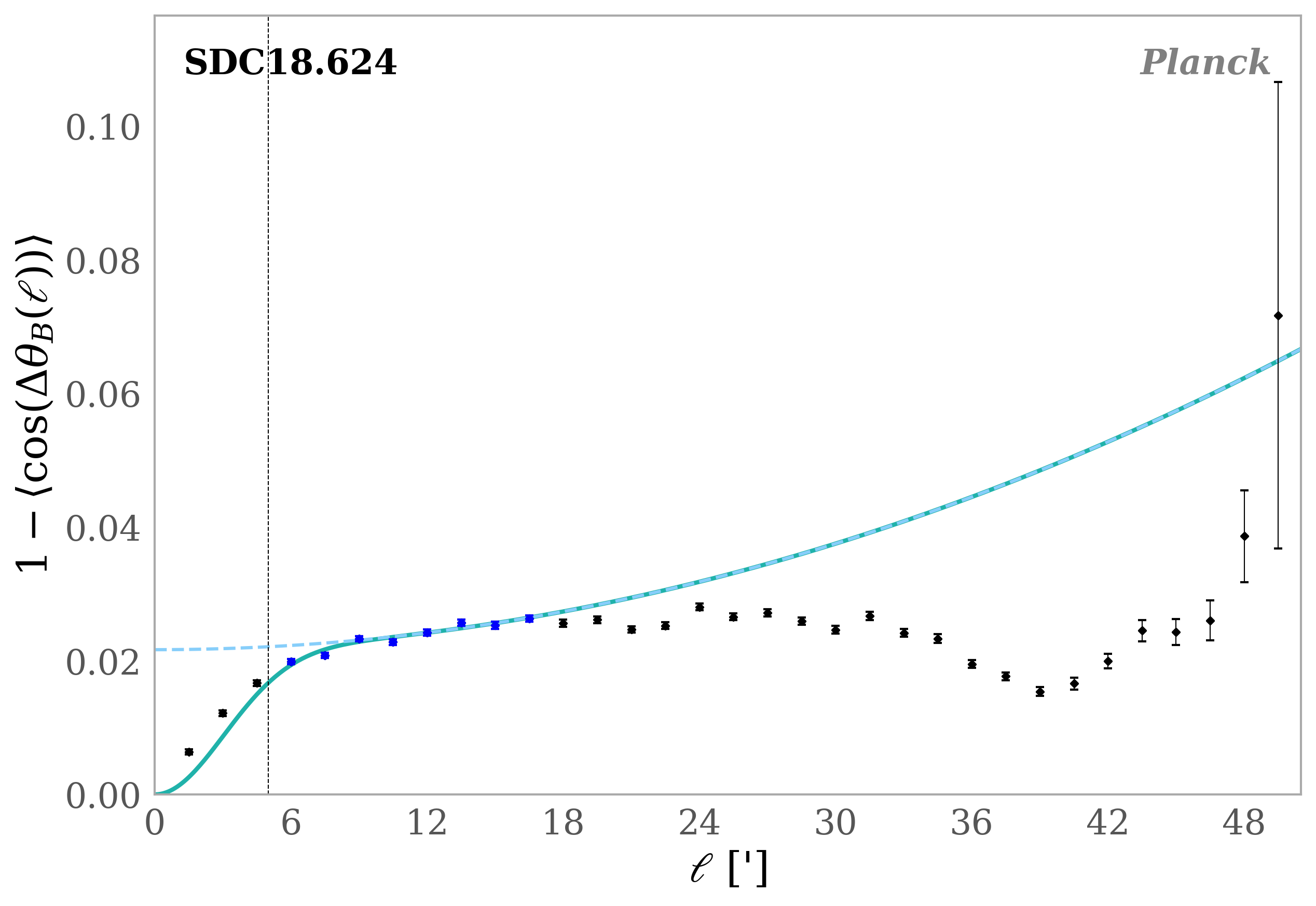}
	\caption{}
	\label{fig:ADF-Planck-SDC18p624}
	\end{subfigure}
	\begin{subfigure}[H]{0.49\textwidth}
	\includegraphics[width=\columnwidth]{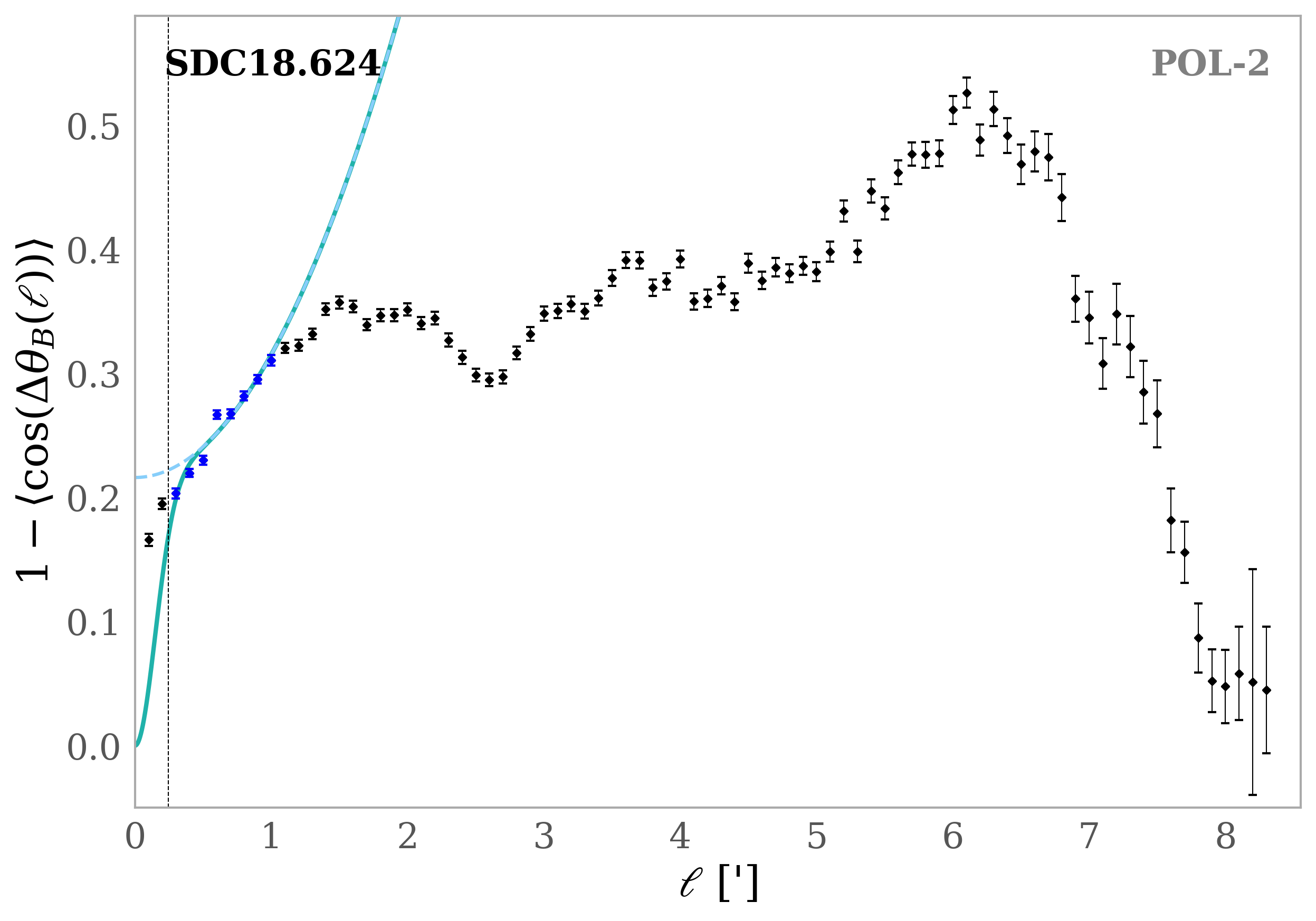}
	\caption{}
	\label{fig:ADF-POL-2-SDC18p624}
	\end{subfigure}
    \caption{Plots of the ADF for SDC18.624 for: a) {\it Planck} b) POL-2. A black dashed vertical line marks the beam size ($5'$ for {\it Planck} and $14.6''$ for POL-2). A solid turquoise curve depicts the best-fitting function for the case where all three parameters $\frac{1}{N}\frac{\langle B_t ^2 \rangle}{\langle B_o^2 \rangle}$, $\delta$, and $a$ were fit. A dashed light blue curve shows the large-scale component. The points in dark blue indicate which points were included in the illustrated fit.}
    \label{fig:ADF-SDC18p624}
\end{figure*}

\begin{figure*}
	\begin{subfigure}[H]{0.49\textwidth}
	\includegraphics[width=\columnwidth]{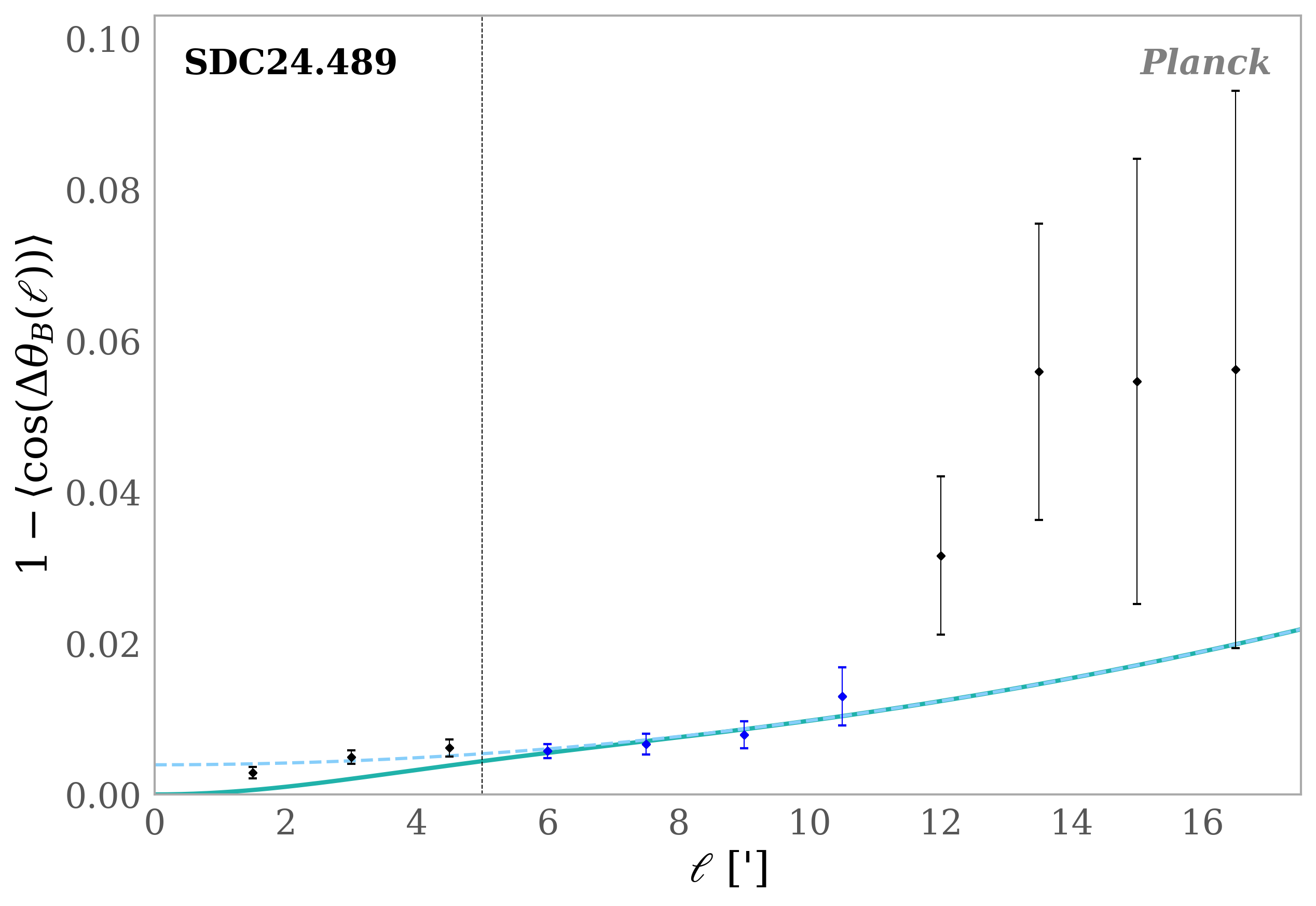}
	\caption{}
	\label{fig:ADF-Planck-SDC24p489}
	\end{subfigure}
	\begin{subfigure}[H]{0.49\textwidth}
	\includegraphics[width=\columnwidth]{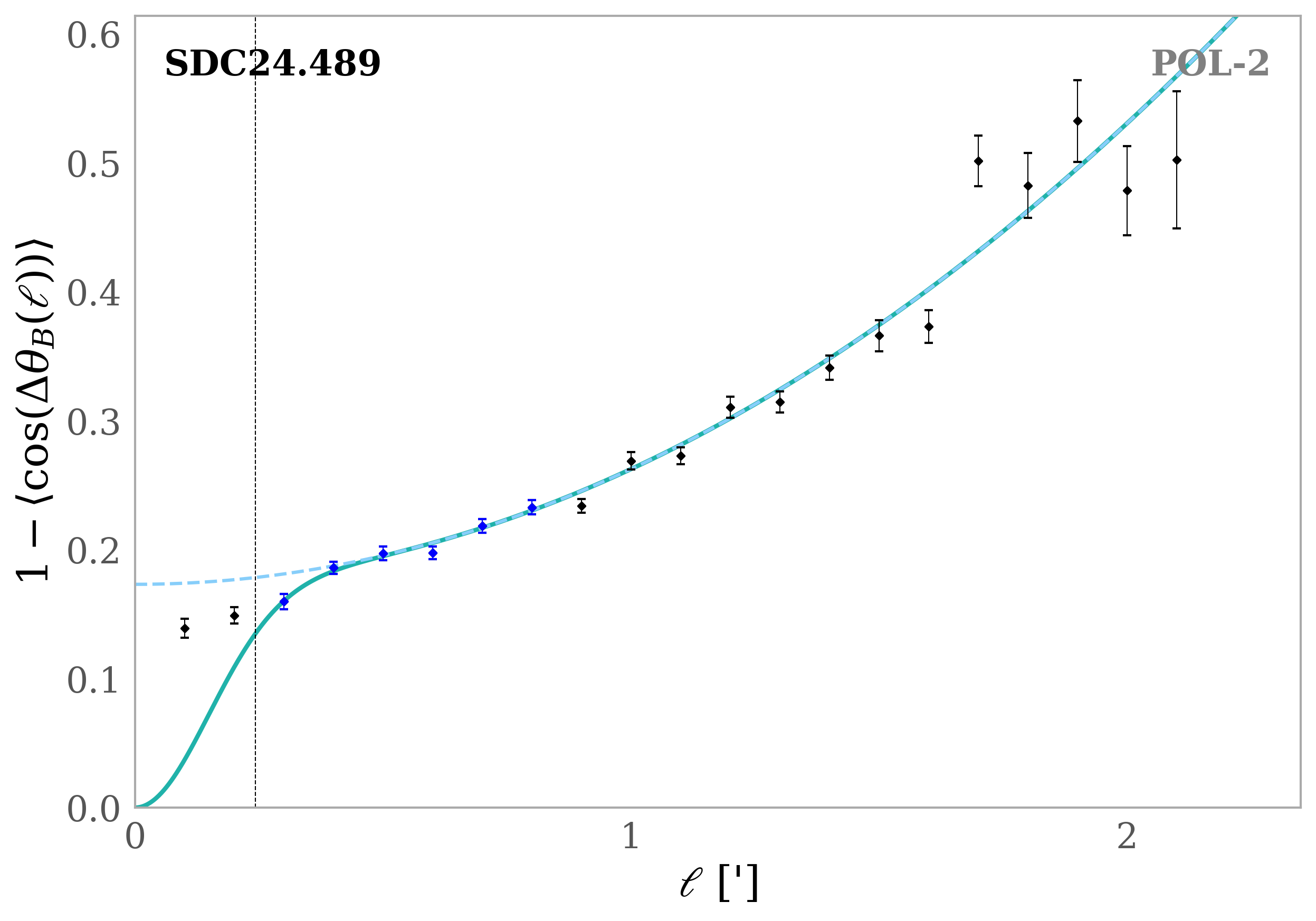}
	\caption{}
	\label{fig:ADF-POL-2-SDC24p489}
	\end{subfigure}
    \caption{Same as Figure \ref{fig:ADF-SDC18p624} for SDC24.489.}
    \label{fig:ADF-SDC24p489}
\end{figure*}

\begin{figure*}
	\begin{subfigure}[H]{0.49\textwidth}
	\includegraphics[width=\columnwidth]{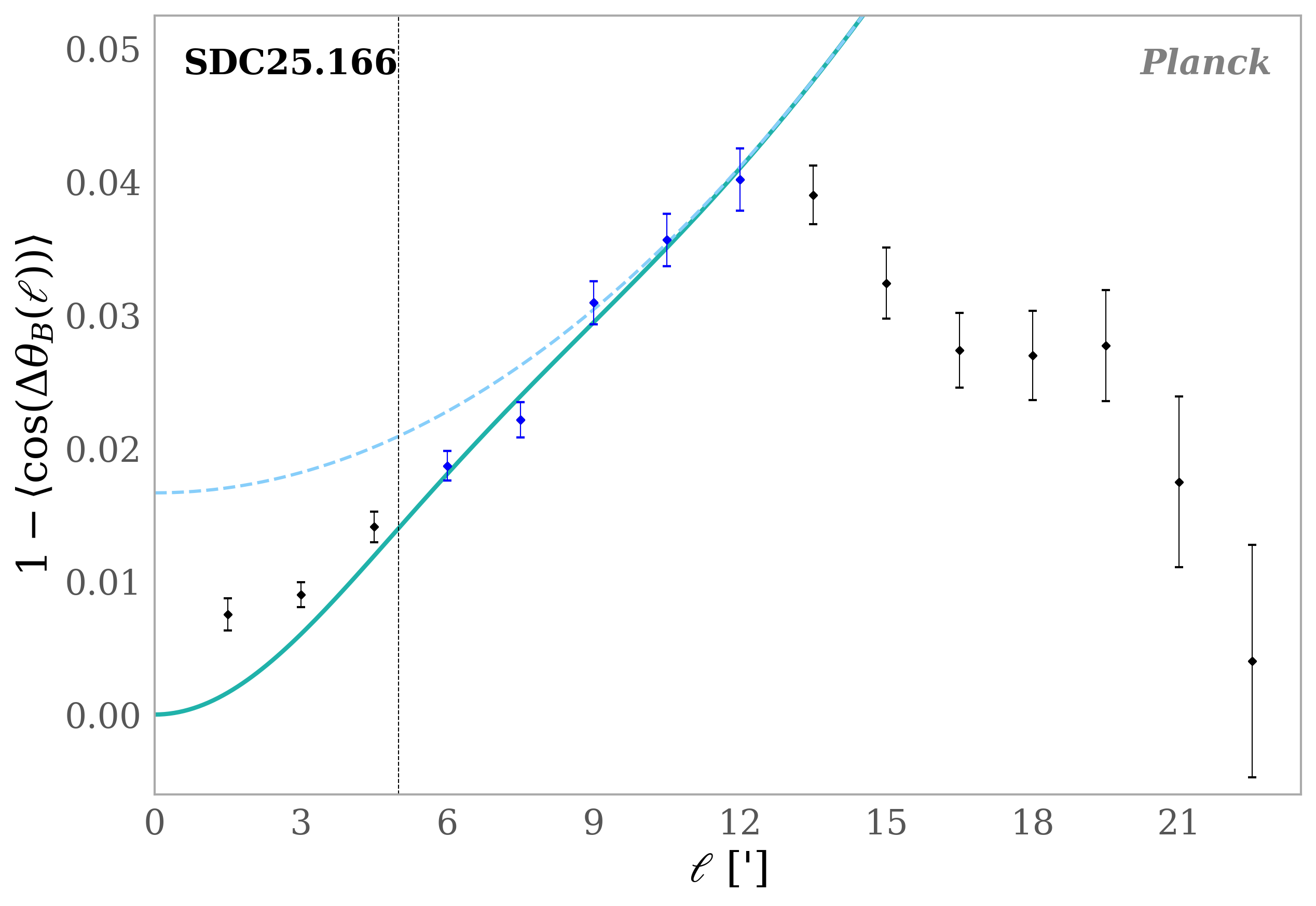}
	\caption{}
	\label{fig:ADF-Planck-SDC25p166}
	\end{subfigure}
	\begin{subfigure}[H]{0.49\textwidth}
	\includegraphics[width=\columnwidth]{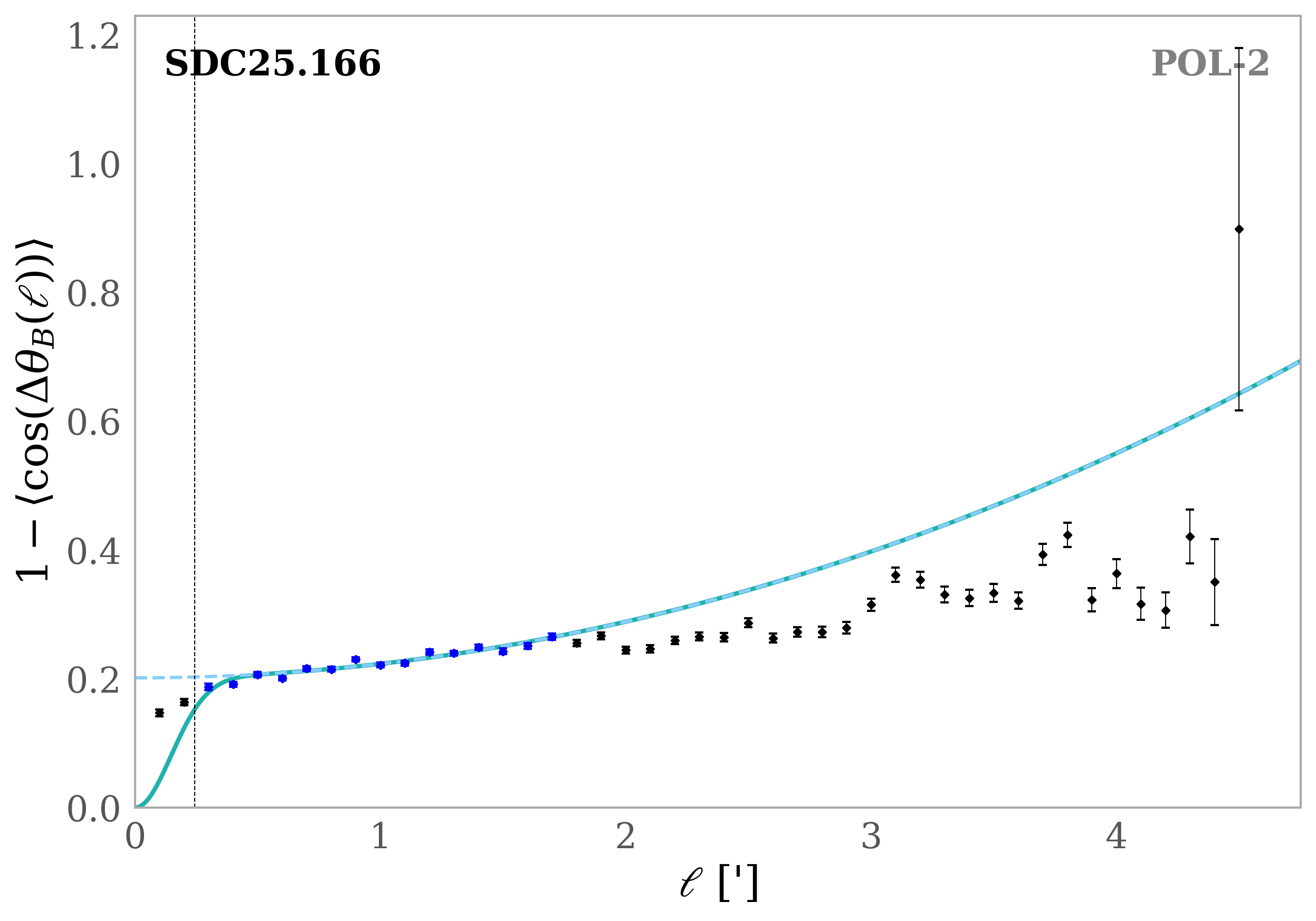}
	\caption{}
	\label{fig:ADF-POL-2-SDC25p166}
	\end{subfigure}
    \caption{Same as Figure \ref{fig:ADF-SDC18p624} for SDC25.166.}
    \label{fig:ADF-SDC25p166}
\end{figure*}

\begin{figure*}
	\begin{subfigure}[H]{0.49\textwidth}
	\includegraphics[width=\columnwidth]{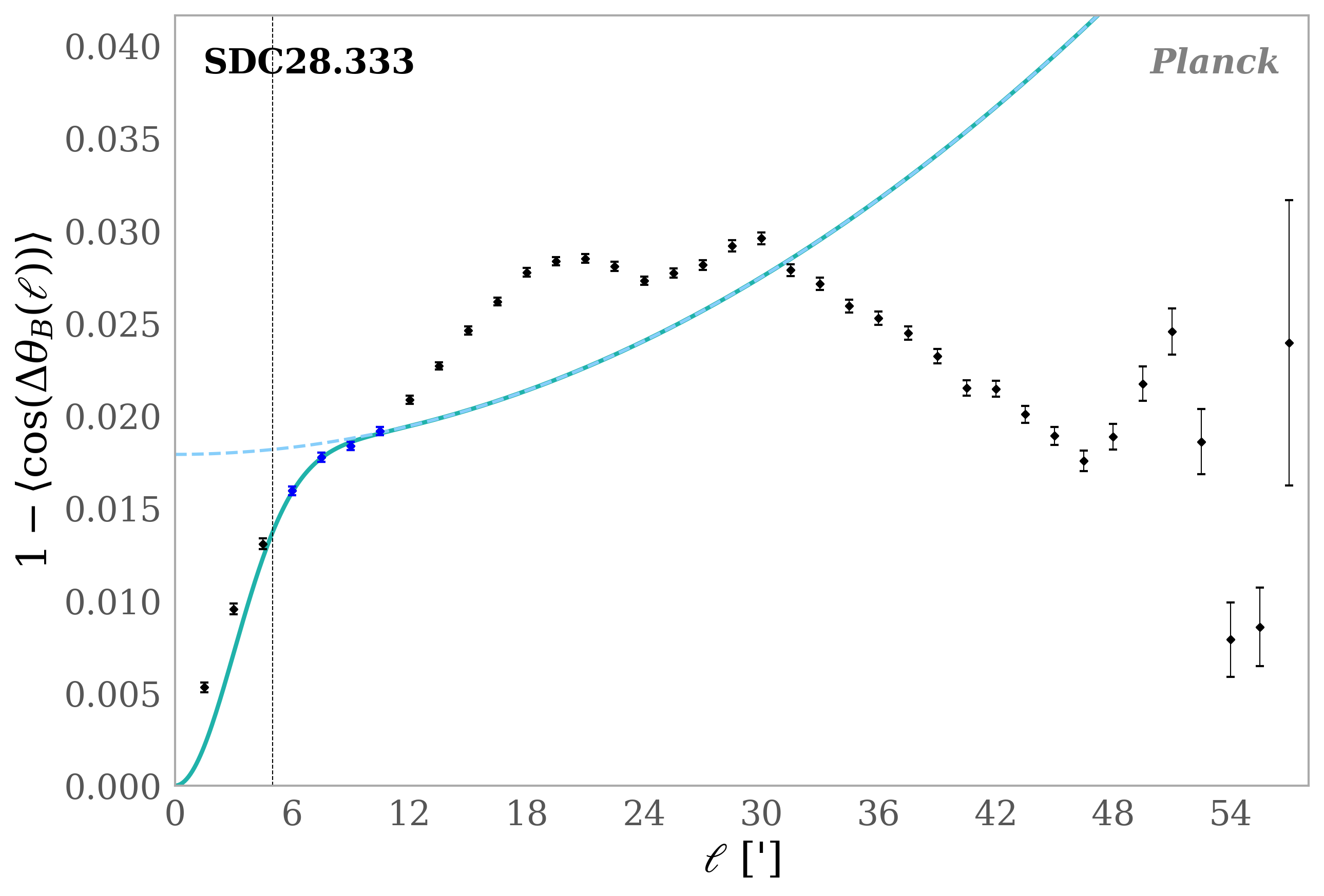}
	\caption{}
	\label{fig:ADF-Planck-SDC28p333}
	\end{subfigure}
	\begin{subfigure}[H]{0.48\textwidth}
	\includegraphics[width=\columnwidth]{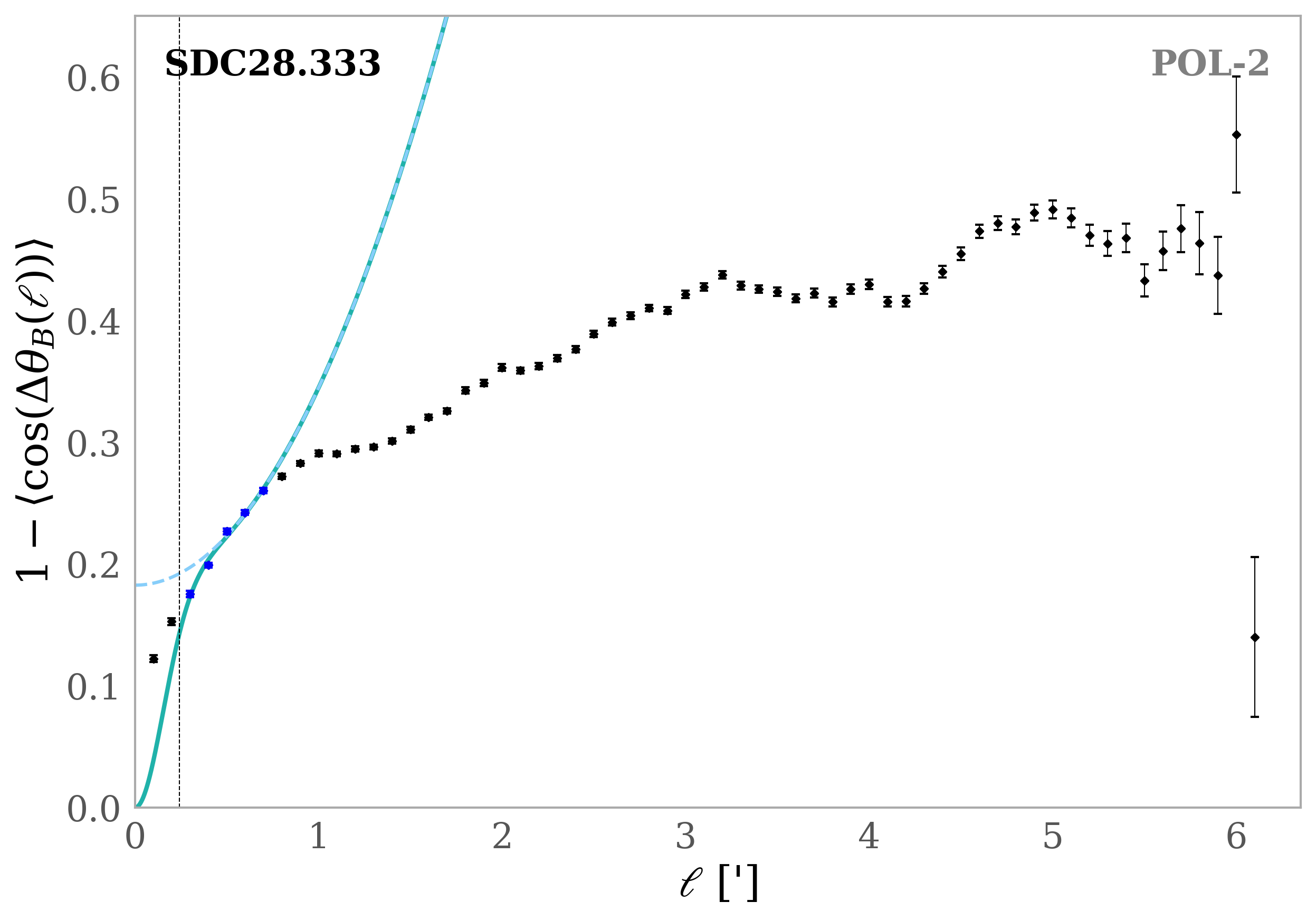}
	\caption{}
	\label{fig:ADF-POL-2-SDC28p333}
	\end{subfigure}
    \caption{Same as Figure \ref{fig:ADF-SDC18p624} for SDC28.333.}
    \label{fig:ADF-SDC28p333}
\end{figure*}

\begin{figure*}
	\begin{subfigure}[H]{0.49\textwidth}
	\includegraphics[width=\columnwidth]{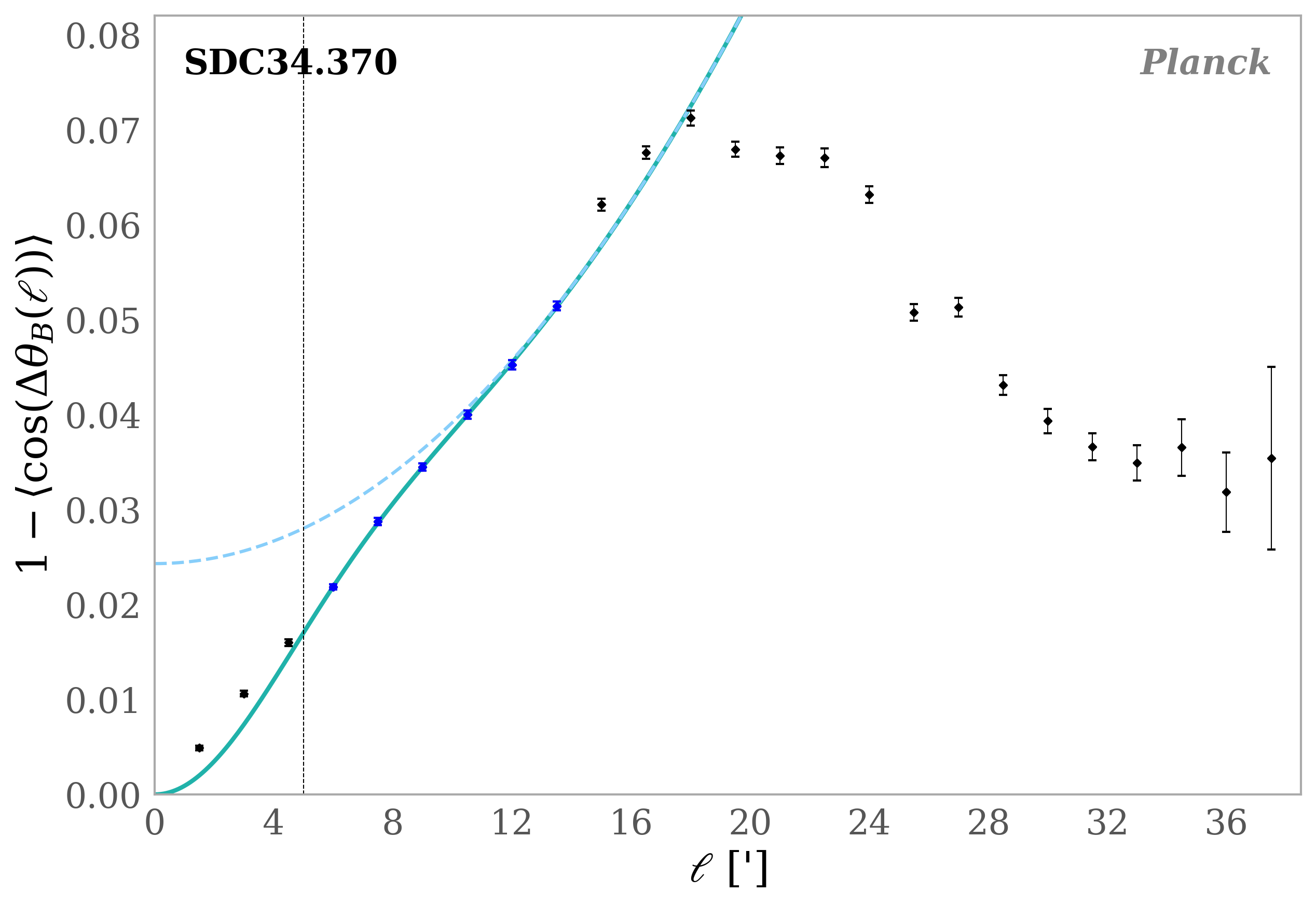}
	\caption{}
	\label{fig:ADF-Planck-SDC34p370}
	\end{subfigure}
	\begin{subfigure}[H]{0.49\textwidth}
	\includegraphics[width=\columnwidth]{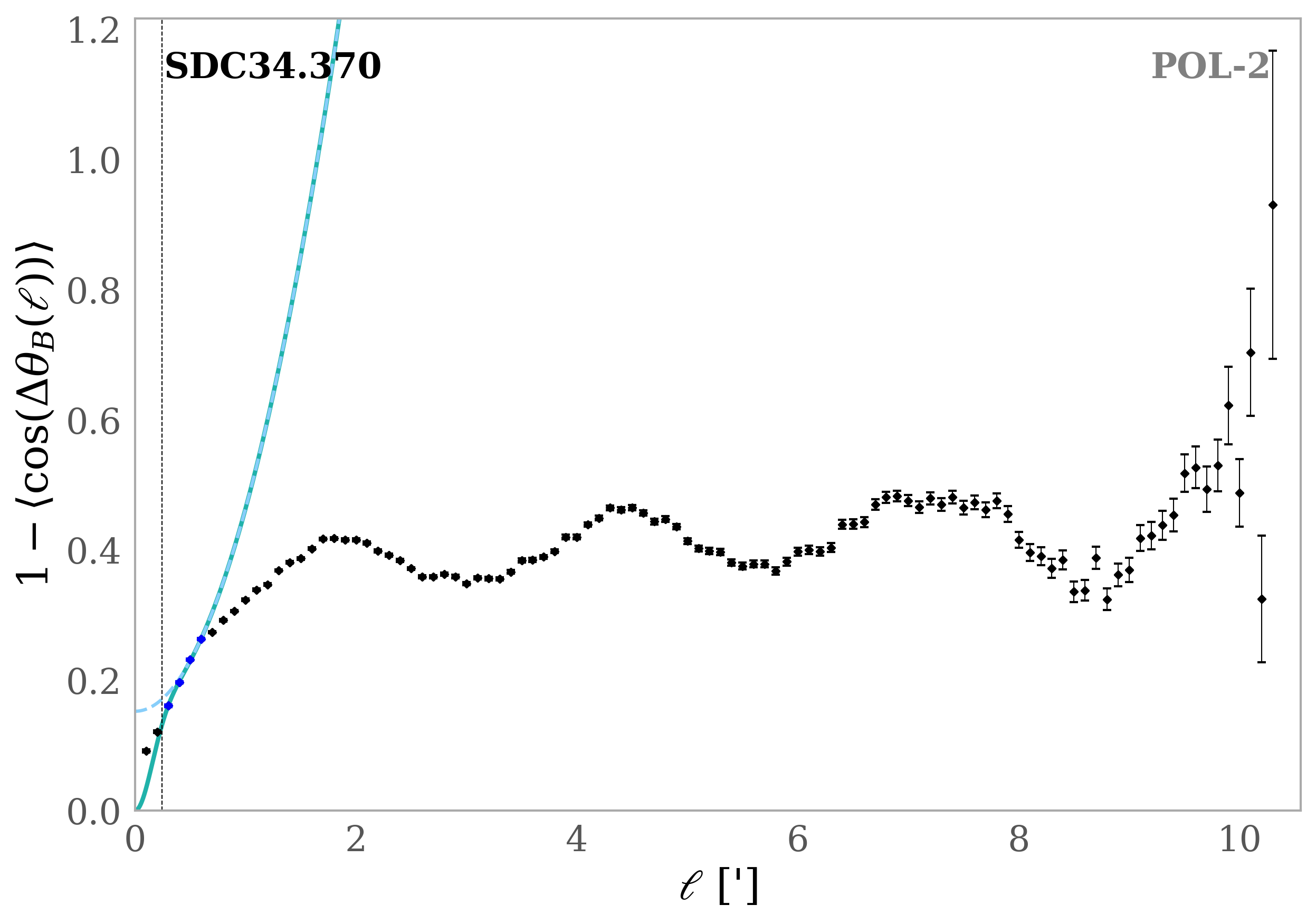}
	\caption{}
	\label{fig:ADF-POL-2-SDC34p370}
	\end{subfigure}
    \caption{Same as Figure \ref{fig:ADF-SDC18p624} for SDC34.370.}
    \label{fig:ADF-SDC34p370}
\end{figure*}

\begin{figure*}
	\begin{subfigure}[H]{0.49\textwidth}
	\includegraphics[width=\columnwidth]{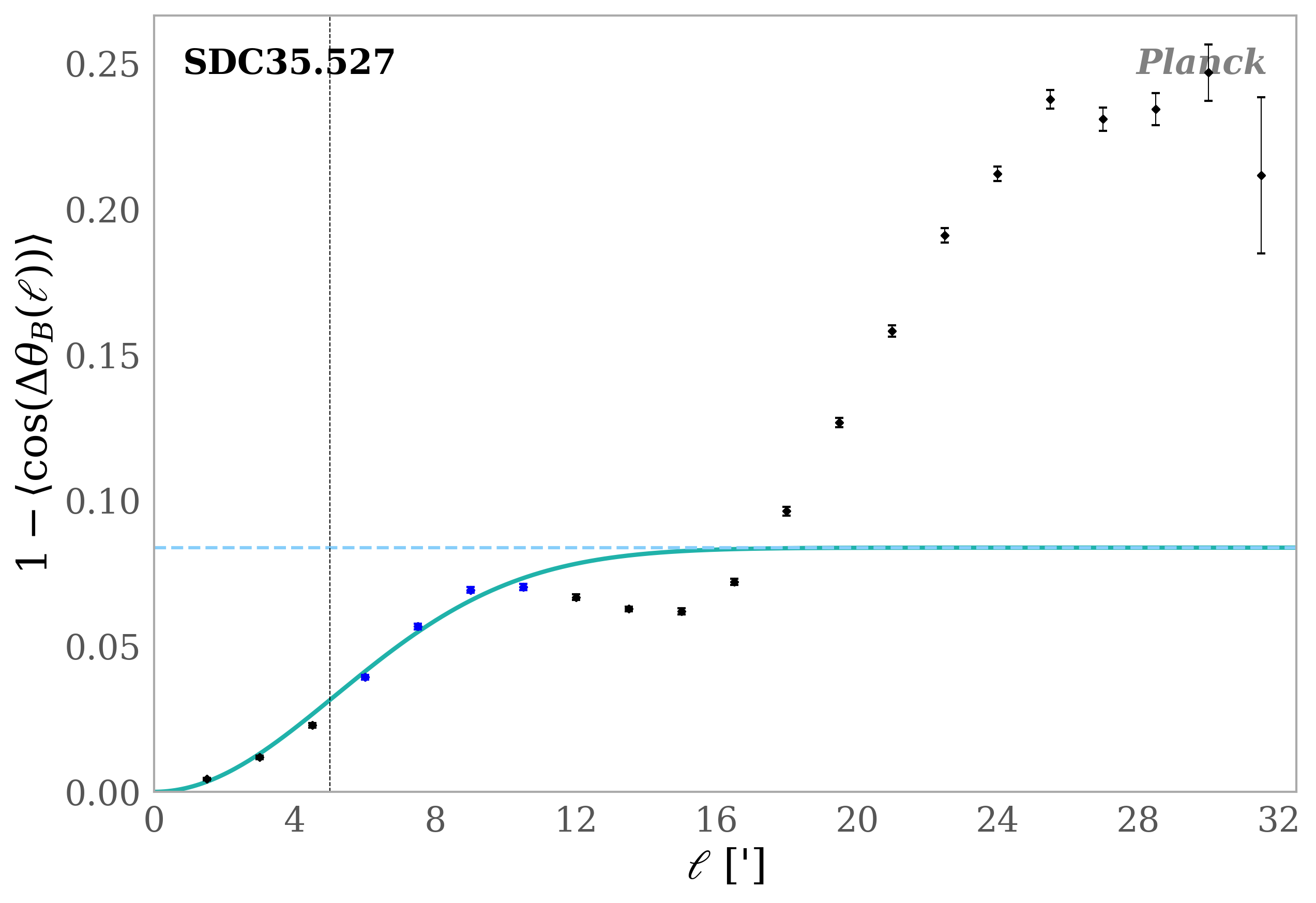}
	\caption{}
	\label{fig:ADF-Planck-SDC35p527}
	\end{subfigure}
	\begin{subfigure}[H]{0.49\textwidth}
	\includegraphics[width=\columnwidth]{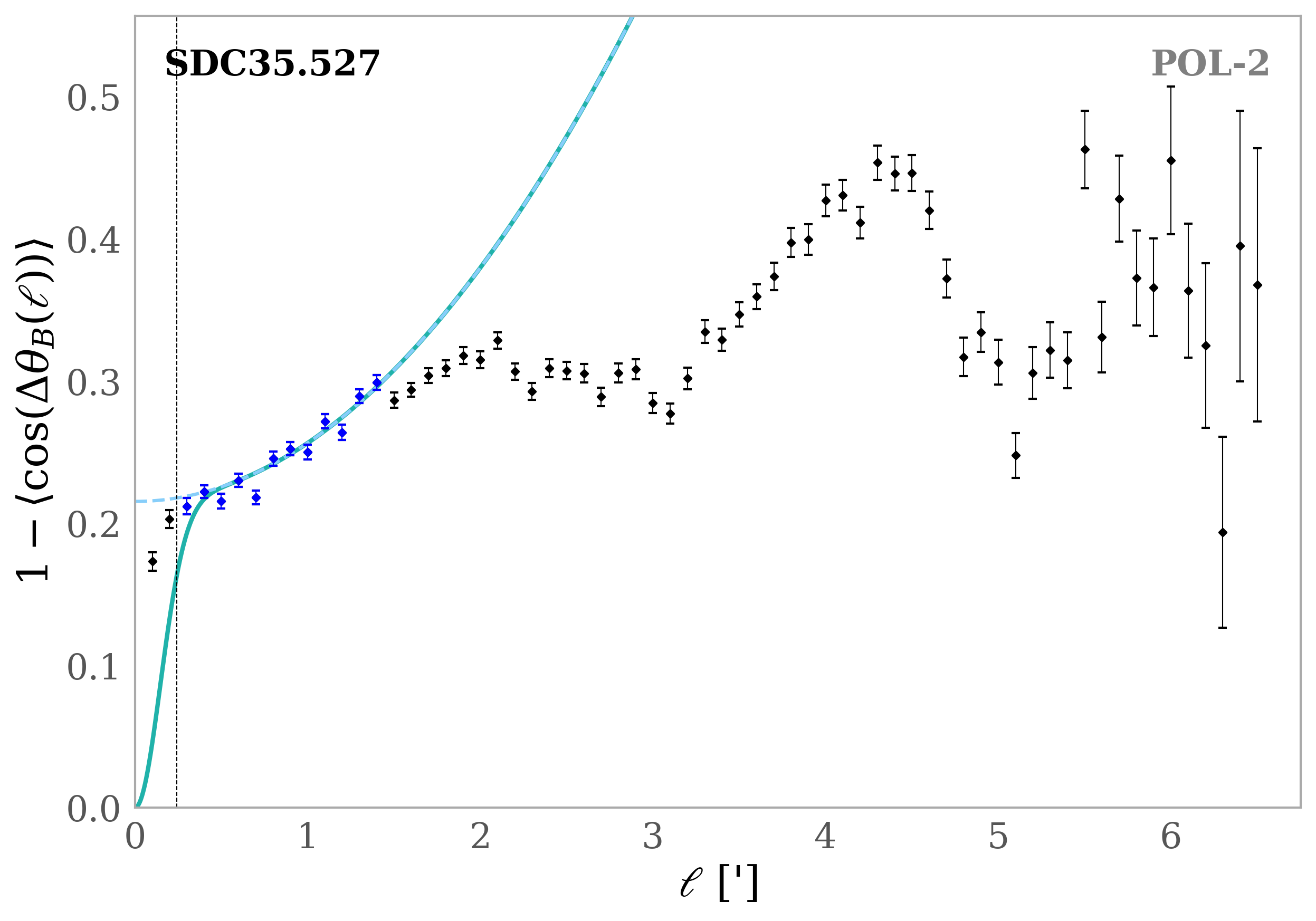}
	\caption{}
	\label{fig:ADF-POL-2-SDC35p527}
	\end{subfigure}
    \caption{Same as Figure \ref{fig:ADF-SDC18p624} for SDC35.527.}
    \label{fig:ADF-SDC35p527}
\end{figure*}

\begin{figure*}
	\begin{subfigure}[H]{0.49\textwidth}
	\includegraphics[width=\columnwidth]{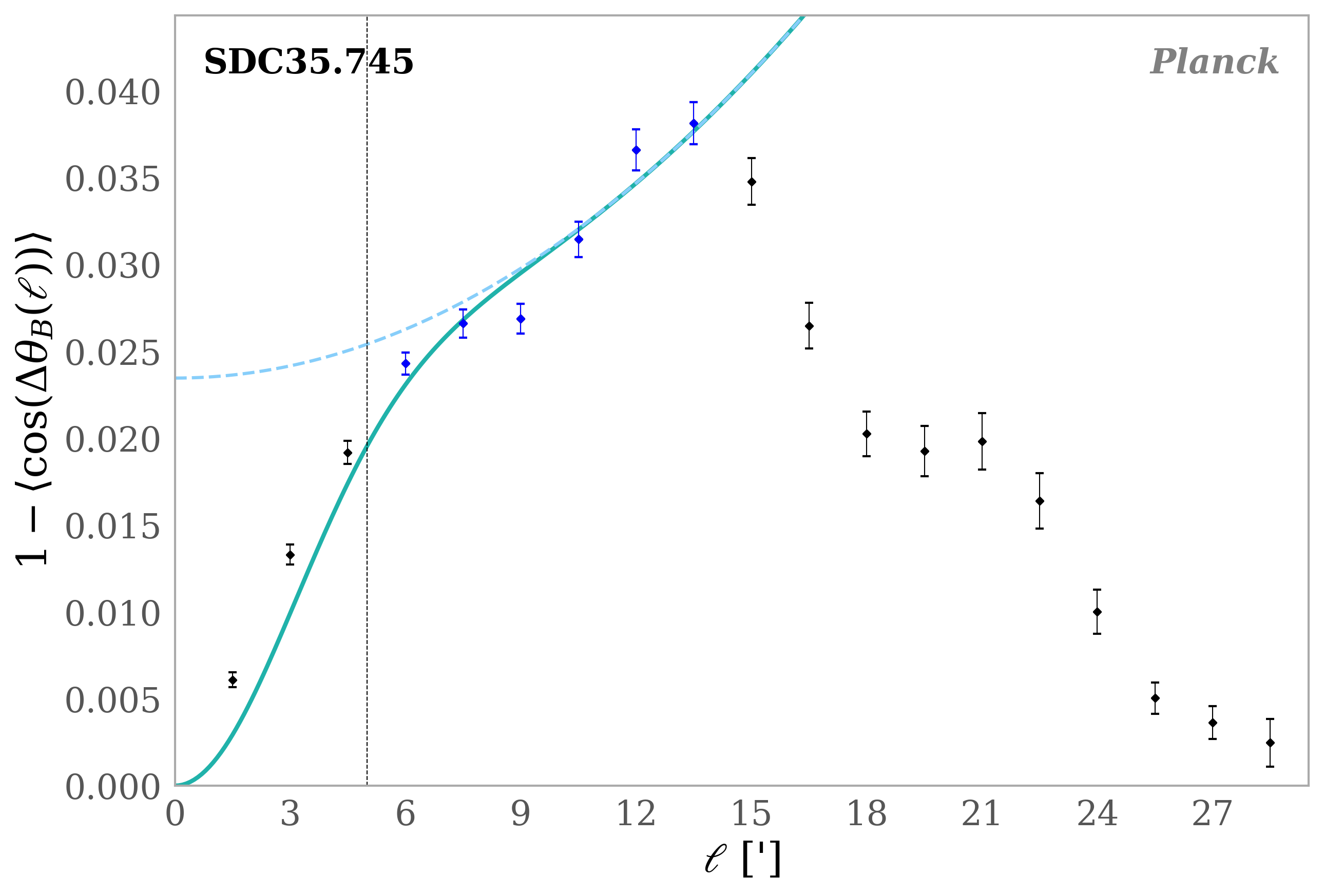}
	\caption{}
	\label{fig:ADF-Planck-SDC35p745}
	\end{subfigure}
	\begin{subfigure}[H]{0.48\textwidth}
	\includegraphics[width=\columnwidth]{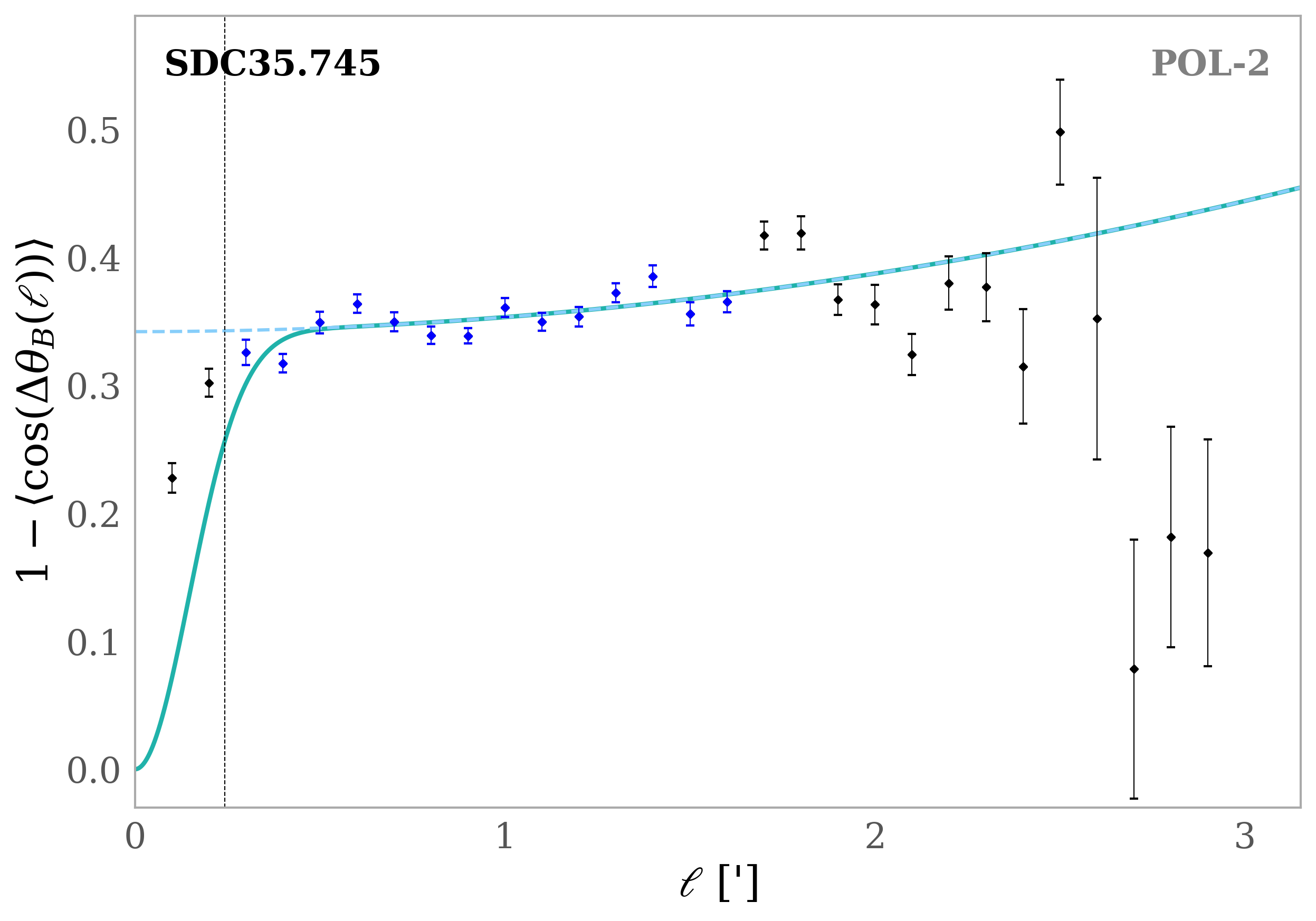}
	\caption{}
	\label{fig:ADF-POL-2-SDC35p745}
	\end{subfigure}
    \caption{Same as Figure \ref{fig:ADF-SDC18p624} for SDC35.745.}
    \label{fig:ADF-SDC35p745}
\end{figure*}

\begin{figure*}
	\begin{subfigure}[H]{0.49\textwidth}
	\includegraphics[width=\columnwidth]{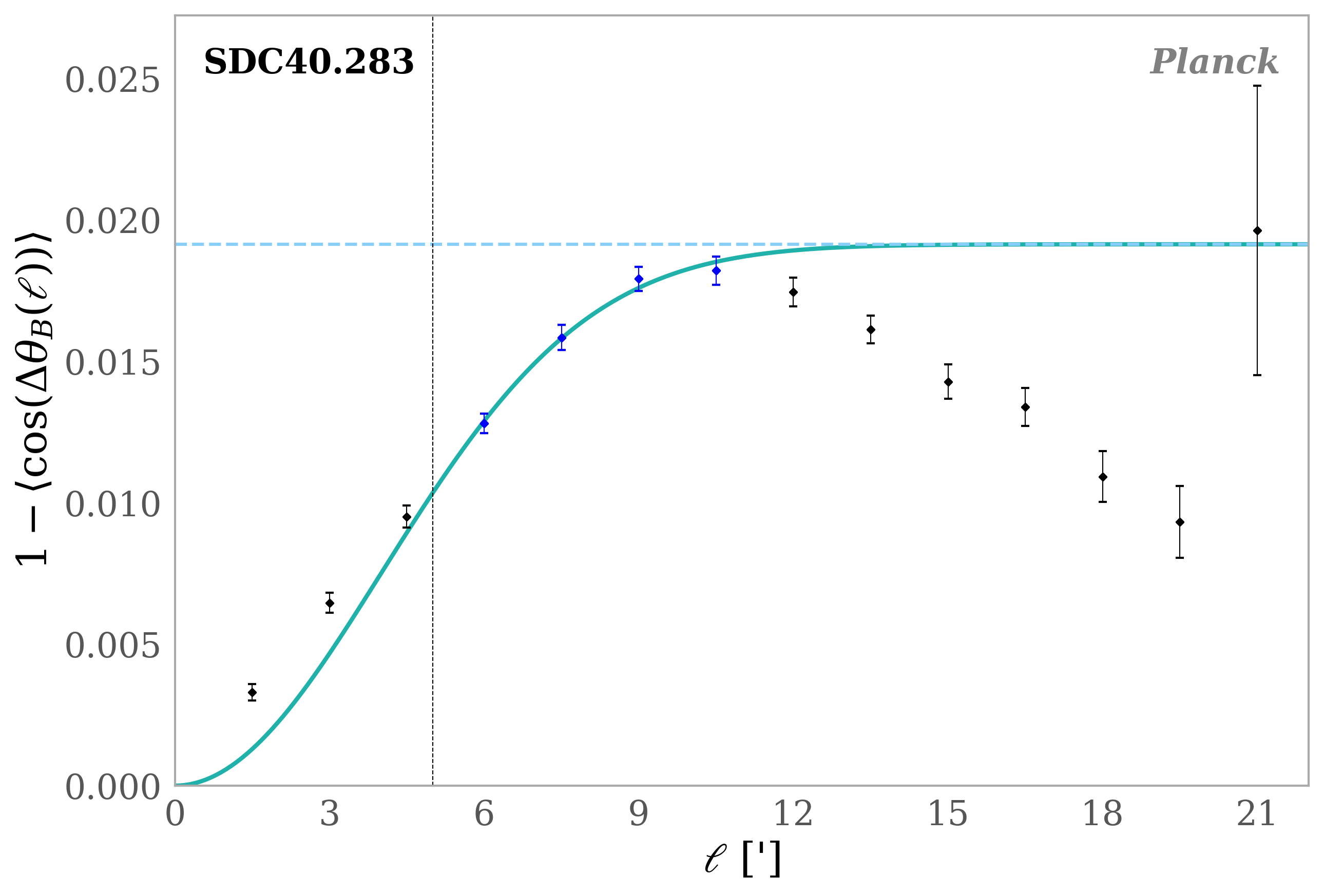}
	\caption{}
	\label{fig:ADF-Planck-SDC40p283}
	\end{subfigure}
	\begin{subfigure}[H]{0.49\textwidth}
	\includegraphics[width=\columnwidth]{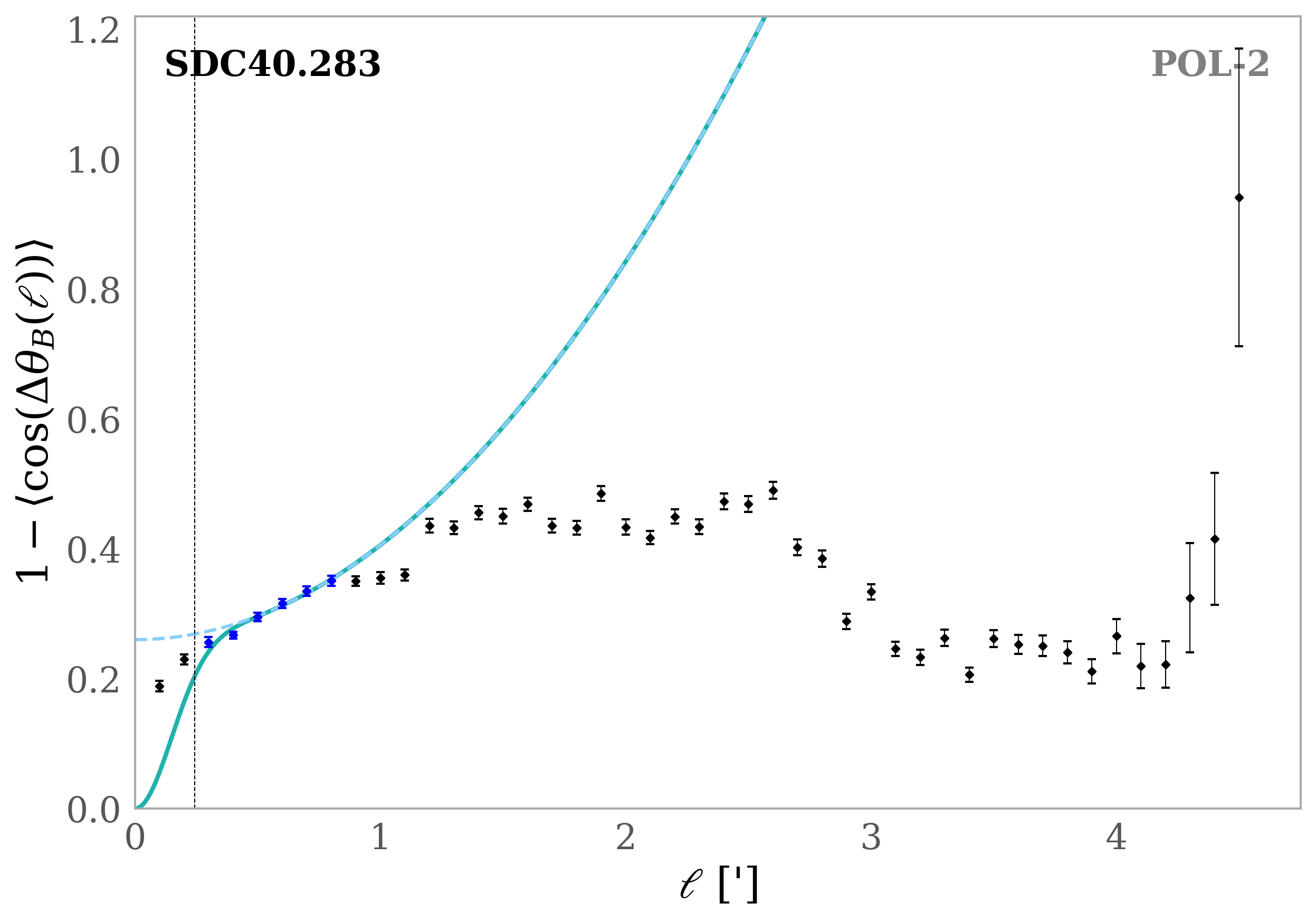}
	\caption{}
	\label{fig:ADF-POL-2-SDC40p283}
	\end{subfigure}
    \caption{Same as Figure \ref{fig:ADF-SDC18p624} for SDC40.283.}
    \label{fig:ADF-SDC40p283}
\end{figure*}
%%%%%%%%%%%%%%%%%%%%%%%%%%%%%%%%%%%%%%%%%%%%%%%%%%

% Don't change these lines
\bsp	% typesetting comment
\label{lastpage}
\end{document}